\documentclass[final,3p,times,twocolumn]{elsarticle}

\usepackage{amssymb}
\usepackage{amsmath}

\journal{International Journal of Multiphase Flow}

\usepackage{hyperref}
\usepackage{color}
\usepackage{bm}		
\usepackage{subcaption}	
\usepackage{booktabs}	

\hypersetup{
	colorlinks=true,    
	linkcolor=blue,
	citecolor=blue,
	urlcolor=blue,
}

\definecolor{corr1}{rgb}{0,0,0}
\definecolor{rev1}{rgb}{0,0,0}

\begin{document}

\begin{frontmatter}


\title{Interphase coupling for gas-droplet flows using the fully Lagrangian approach\tnoteref{label1}}
\author{C. P. Stafford}
\ead{chris.stafford@manchester.ac.uk}
\affiliation{organization={Department of Mechanical and Aerospace Engineering, University of Manchester},
             city={Manchester},
             country={UK}}

\title{}


\author{O. Rybdylova} 

\affiliation{organization={Advanced Engineering Centre, School of Architecture, Technology and Engineering, University of Brighton},
            city={Brighton},
            country={UK}}

\begin{abstract}
A novel method combining the fully Lagrangian approach (FLA) and kernel regression has been developed for two-way coupled simulations of evaporating sprays. The carrier phase is incompressible viscous flow described by the Navier-Stokes equations. The admixture is considered to be a cloud of monodisperse evaporating droplets, which is treated as a continuum in the FLA. All droplet parameters are calculated along selected trajectories with the number density calculated using the Lagrangian form of the continuity equation. To enable two-way coupling, the momentum and mass phase exchange terms must be calculated in each volume element of an Eulerian mesh. This is achieved by using kernel regression in conjunction with the FLA trajectory data, which retains the detail of complex structures in droplet clouds by adaptively scaling the kernel support according to the local droplet field deformation. In this work, the mass and momentum coupling source terms obtained using the FLA are assessed against reference values calculated using a standard Lagrangian particle tracking simulation that incorporates a PSI-CELL box-counting method. It is shown that the FLA retains the same level of fidelity and smoothness as the reference PSI-CELL case, whilst also providing a computational speedup factor of around 100 times due to the decreased droplet seeding.
\end{abstract}

\begin{graphicalabstract}
	\includegraphics[width=\textwidth]{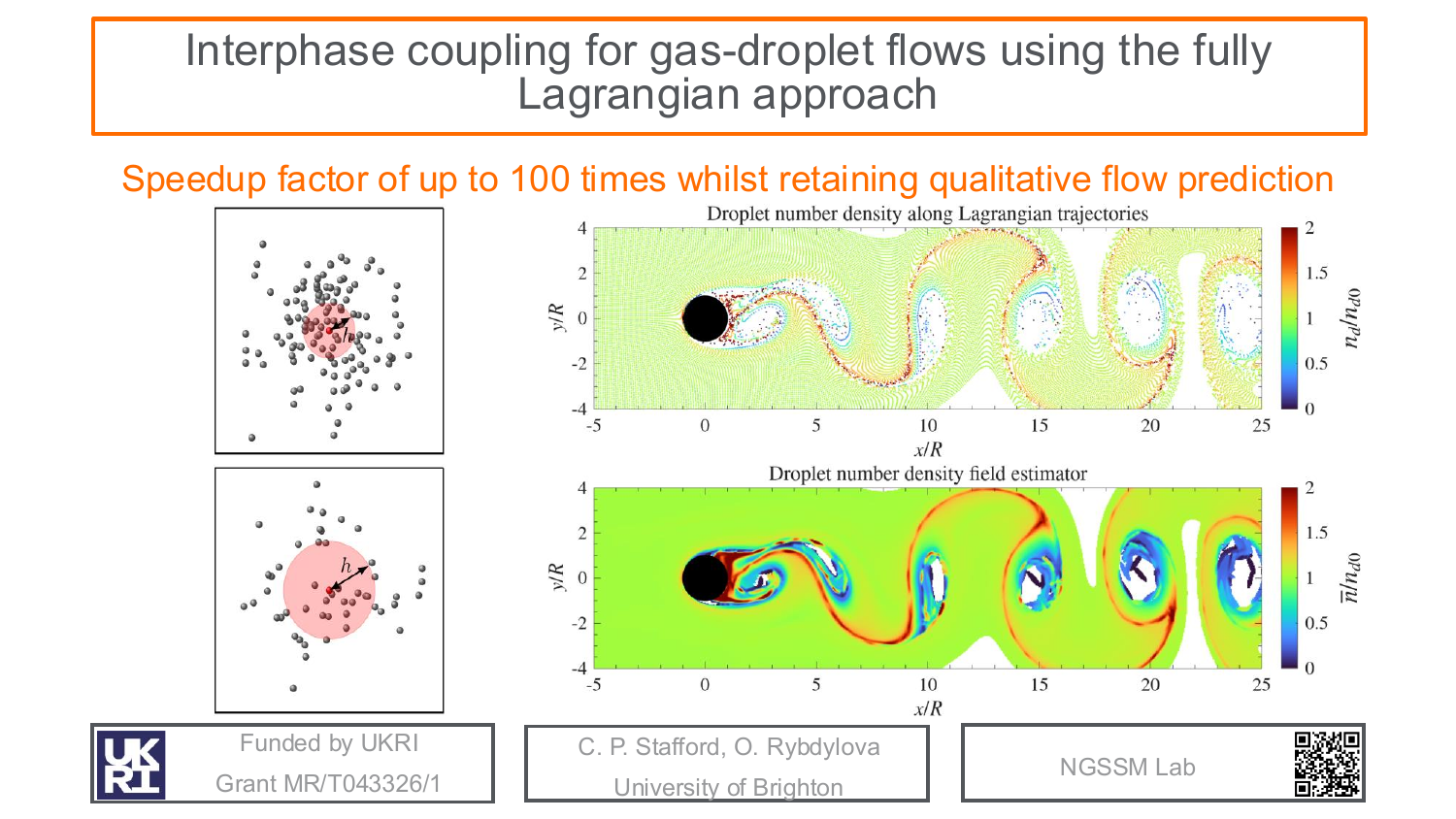}
\end{graphicalabstract}

\begin{highlights}
\item {\color{rev1}Formulation of two-way coupling source terms in an Eulerian-Lagrangian framework}
\item {\color{rev1}Kernel regression reconstruction procedure using the fully Lagrangian approach (FLA)}
\item {\color{corr1}Implementation within OpenFOAM, and verification against a PSI-CELL solver}
\item {\color{rev1}Demonstrated for the test case of steady-state and unsteady flow past a cylinder}
\item {\color{rev1}Speedup factor of up to 100 times whilst retaining qualitative flow prediction}
\end{highlights}

\begin{keyword}
droplets and sprays
\sep
continuum modelling
\sep
fully Lagrangian approach
\sep
OpenFOAM


\end{keyword}

\end{frontmatter}



\section{Introduction}
\label{sec:introduction}

The ubiquity of gas-droplet flows in engineering and environmental applications highlights the relevance of understanding and controlling the behaviour of such systems. Important examples include a variety of spray configurations \cite{ref:Sazhin_2014}, such as exhalation of respiratory
aerosols after coughing or sneezing \cite{ref:Bourouiba2014}, aerosol drug delivery using inhalers \cite{ref:Dolovich2011}, and fuel injection in internal combustion engines \cite{ref:Begg01082009}. This motivates both the interest in the physical phenomena that are manifest in the flow behaviour, as well as the need for accurate simulations to gain insight and
for design optimisation. Such systems involve interaction between a dispersed phase of droplets {\color{corr1}suspended in a background carrier fluid}, and can be subject to a range of physical effects. These include phenomena such as body forces, turbulence and evaporation that all affect the droplet behaviour to varying extents, the significance of which is dependent on the behaviour of the carrier flow. As a result, the close interplay between the local carrier flow and droplets is crucial to represent within computational fluid dynamics models of gas-droplet flows, in order to accurately capture within simulations the transient physics of how such systems evolve.

The effect of the carrier flow on droplets is generally well documented, and is key to determining how the droplets respond to the flow. Since this is the principal determinant of the individual droplet trajectories and collective motion, these one-way coupling effects are accounted for through established models which represent the various forces experienced by droplets, for instance drag, added mass and lift \cite{crowe2014multiphase}. By contrast, the effect of the droplets on the carrier flow {\color{corr1}varies with the nature of the droplet phase, and in some cases can be a major influence on the evolution of the flow. In particular,} flows with a high loading of suspended droplets can experience a notable change in the carrier flow behaviour due to the presence of the droplets \cite{Elghobashi1993}. Accounting for this two-way coupling through the inclusion of momentum and mass exchange terms is therefore necessary to guarantee a realistic modelling framework for simulation purposes.

This need for two-way coupling has been the focus of several previous works for both laminar and turbulent flows {\color{rev1}in an Eulerian-Lagrangian framework} \cite{Eaton2009}. The standard procedure for this which is now implemented in most computational fluid dynamics codes is the Particle-Source-In Cell (PSI-CELL) method \cite{Crowe1977}, which accumulates particle contributions using a simple box-counting approach of the particles within each cell of the computational mesh. The associated interpolation function used for particle contributions in this case is simply a delta function for whether the particle is in the cell or not, meaning that the method only provides $C^0$ continuity, with the location of particles within a cell not being accounted for \cite{fehske2007computational}. Consequently, the PSI-CELL method requires a high number of particles to achieve accuracy and ensure statistical noise from the mesh-based accumulation process does not confound the resultant interphase coupling terms. Furthermore, the PSI-CELL method is known to not be numerically convergent \cite{Garg2009}. An improvement is offered by the Cloud-In-Cell (CIC) method \cite{Laux1996}, which accounts for the particle location by interpolating carried particle quantities to the corners of the cell in the computational mesh which is occupied by the particle. This is accomplished using an inverse distance-based kernel, resulting in a linear interpolation procedure which is $C^1$ continuous. The CIC method had become the workhorse of interphase coupling in codes where the spatial distribution of particles must be precisely accounted for in order to retain the accuracy of simulations, for instance the Direct Numerical Simulation of turbulence \cite{Eaton2009}. Despite the additional detail offered, the CIC method still however requires a high number of particles to achieve statistical convergence, adding to the computational expense of simulations.

More recent work has focused on more advanced procedures for constructing the coupling source terms. An assessment of the numerical accuracy for several interpolation procedures used for both mesh to particle forward interpolation, as well as accumulation of the particle contributions into the interphase coupling terms, observed that for spatially non-uniform particle distributions the estimate for the mean interphase momentum transfer term was poor in regions with fewer particles \cite{Garg2007}. A suitable remedy to this was found to be ensuring that the number density of computational particles is maintained at a relatively uniform level by introducing more computational particles in regions of low number density, resulting in the statistical error remaining uniformly low across the entire domain. For developing more numerically stable implementations of the interphase momentum transfer term, it was proposed to use variable statistical weights for each particle which are evolved using the fractional rate of change of each weight \cite{Garg2009}. This procedure encapsulates some of the underlying physical behaviour, however the evolution of the weights is based upon a more empirical measure, and so becomes limited in its applicability for more extreme values of number density. Whilst the statistical nature of this approach is able to offset the high number of particles required for statistical convergence of the PSI-CELL and CIC methods, it is not as robust as these more primitive approaches, leading to the ongoing need for the development of {\color{rev1}Eulerian-Lagrangian} two-way coupling approaches which are both accurate and efficient.

The objective of the present work is to address this need by developing a robust numerical scheme for calculation of the interphase coupling source terms in dilute mixtures that incorporates the fully Lagrangian approach (FLA) \cite{ref:Osiptsov2000}. This methodology has been extensively applied to monodisperse and one-way coupled gas-particle flows \cite{Osiptsov2024}, and it has been demonstrated that the FLA is an efficient method for calculating the distribution of particles in comparison to other Lagrangian approaches \cite{ref:HealyYoung2005}. {\color{rev1}Furthermore, the FLA has been to shown to offer insights into the underlying statistical properties of particle segregation in particle-laden turbulent flows \cite{Meneguz2011}, highlighting the versatility of the method.} More recently, the application of a generalised FLA (gFLA) to model polydisperse evaporating droplets showed that while the gFLA maintains the computational economy afforded by the original FLA, a robust method for interpolation of Lagrangian droplet data to an Eulerian mesh is required \cite{ref:LiRybdylova2021}. This is achieved through the use of kernel regression \cite{ref:StaffordRybdylova2022}. The procedure retains the detail of complex structures in droplet clouds, caustics and voids, by adaptively scaling the kernel domain of influence using information about the local droplet field deformation provided by the FLA. The present paper is concerned with the next step in the advancement of the FLA, and focuses upon using the reconstructed droplet contributions on the mesh to develop a two-way coupling approach incorporating the FLA, and its implementation into the open-source computational fluid dynamics software OpenFOAM. The essence of this work is therefore in developing a continuum model for disperse particle-laden flow that retains a Lagrangian core, {\color{corr1}with use of kernel regression to reconstruct Eulerian field representations} for the interphase coupling source terms {\color{corr1}that can be appended directly to the carrier flow transport equations}.

{\color{rev1}
It should be noted that whilst the present work focuses on interphase coupling in an Eulerian-Lagrangian framework, the coupling in dispersed particle-laden flows can also be incorporated within Eulerian continuum models for the particle phase \cite{crowe2014multiphase}. In particular kinetic-based approaches offer an appropriate means of achieving this \cite{Reeks2021}, since they are able to capture the crossing trajectory effect of inertial particles in unsteady flows \cite{Yudine1959} within the continuum description. The key difference in Eulerian formulations is their averaged approach however, with additional closure models required to account for correlations between continuum variables and detail of the flow structure. Therefore ultimately this remains a separate modelling paradigm, and is not discussed further in this work.
}

The remainder of the paper is structured as follows; Section \ref{sec:mathematical-modelling} describes the mathematical model formulation of the FLA for dispersed gas-droplet flow; Section \ref{sec:coupling} outlines the proposed coupling procedure and details the specifics of the source term reconstruction; {\color{rev1} Section \ref{sec:numerics} focuses on the numerical implementation of the source term reconstruction in OpenFOAM;} Section \ref{sec:results} presents and discusses the results of the application of the methodology to the analysis of a dispersed gas-droplet flow around a cylinder; and Section \ref{sec:conclusions} summarises the main results of the paper.

\section{Mathematical model for dispersed droplet flow}
\label{sec:mathematical-modelling}

{\color{corr1}The FLA models a fluid-droplet system as a dispersed multiphase flow, and is applicable to a general carrier flow which can be compressible or incompressible and viscous or inviscid \cite{ref:Osiptsov2000}. In this work the carrier flow is taken to be an incompressible viscous gas, with its momentum evolution and continuity described by the Navier-Stokes equations}
\begin{subequations}
	\label{eq:NS}
	\begin{align}
		\nabla \cdot \bm{u} & = 0 \, , \label{eq:fluid-continuity} \\
		\frac{\partial \bm{u}}{\partial t} + \left( \bm{u} \cdot \nabla \right) \bm{u} & = - \nabla p + \frac{1}{Re} \nabla^2 \bm{u} + \bm{S}_{\text{Mom}} \, , \label{eq:fluid-momentum}
	\end{align}
\end{subequations}
where $\bm{u}$ is the gas phase velocity, $p$ is the static pressure, $Re$ is the flow Reynolds number, and $\bm{S}_{\text{Mom}}$ is the momentum exchange source from the droplet phase. To account for phase change, the behaviour of the vapour accumulated from evaporated droplets is governed by the passive scalar transport equation
\begin{equation} \label{eq:vapour-transport}
	\frac{\partial \rho_v}{\partial t} + \nabla \cdot \left( D \nabla \rho_v \right) + \nabla \cdot \left( \bm{u} \rho_v \right) = S_{\text{Mass}} \, ,
\end{equation}
where $\rho_v$ is the droplet vapour concentration, $D$ is the vapour diffusion coefficient, and $S_{\text{Mass}}$ is the mass exchange term from the droplet phase. The source terms $\bm{S}_{\text{Mom}}$ and $S_{\text{Mass}}$ therefore represent the effect of droplets on the {\color{corr1}carrier phase flow}, {\color{corr1}and require specification in terms of the droplet phase variables.} {\color{corr1}Note that the following methodology in this work only concerns the specifics of the droplet phase, and so is applicable to a wider class of more general carrier flows.}

The droplet phase is represented through the FLA framework, which considers the evolution of droplets along trajectories in a Lagrangian sense. {\color{corr1}In the present work the dispersed phase is a treated as a cloud of monodisperse evaporating droplets.} The FLA is based on the assumption that the dispersed phase can be represented as a {\color{corr1}pressureless continuum (i.e. an inert gas in which there is no Brownian motion or interaction between the droplets)} \cite{ref:Osiptsov2000}, and this restricts its application to dilute suspensions. In this regime, since the droplet liquid density is much greater than the carrier flow gas density, effects such as buoyancy, added-mass, and Basset-Boussinesq forces can be neglected, and further assuming a low droplet Reynolds number enables the momentum of spherical droplets to be modelled using Stokes drag law
\begin{align} \label{eq:Stokes-drag}
	& \ddot{\bm{x}}_d (t) = \frac{1}{St} \left( \bm{u}(\bm{x}_d (t), t) - \dot{\bm{x}}_d (t) \right) \, , \\
	& \bm{x}_d (t_0) = \bm{x}_0 \, , \dot{\bm{x}}_d (t_0) = \bm{v}_0 \, , \nonumber
\end{align}
where $\bm{x}_d$ is the droplet trajectory, $St$ is the droplet Stokes number describing the droplet inertia, and $\bm{x}_0$ and $\bm{v}_0$ are respectively the initial droplet position and velocity at time $t_0$. The subscript $d$ denotes that the relevant variables are Lagrangian quantities which are considered along droplet trajectories.
In terms of phase change, an idealised physical model of evaporation is employed in which all heat at the droplet surface is assumed to be spent on evaporation. In this case, the droplet evaporation rate then depends only upon the droplet radius, and takes the form
\begin{equation} \label{eq:evap-model}
	\dot{r}_d (t) = \frac{\delta}{2 r_d} \, ,
	\qquad \qquad
	r_d (t_0) = r_0 \, ,
\end{equation}
in which $r_d$ is the droplet radius, $\delta$ is the rate of change of the droplet surface area, and $r_0$ is the initial droplet radius at time $t_0$. The nondimensional parameters $St$ and $\delta$ are expressed in terms of physical quantities as
\begin{equation}
	St = \frac{2 \rho_d r_d^2 U}{9 \mu L} \, , \qquad
	\delta = -\frac{4 k}{9 \mu H} (T - T_d) \, ,
\end{equation}
where $\rho_d$ is the droplet density, $\mu$ is the gas phase dynamic viscosity, $k$ is the droplet {\color{corr1}liquid} thermal conductivity, $H$ is the droplet {\color{corr1}liquid} latent heat of vaporization, $T$ and $T_d$ are the temperatures of the gas phase and droplet respectively, and $L$ and $U$ are respectively reference length and {\color{corr1}velocity} scales. The variation in droplet size governed by Eq.~\eqref{eq:evap-model} and the droplet momentum equation \eqref{eq:Stokes-drag} are therefore coupled through the Stokes number dependence on $r_d$.

The central concept of the FLA lies in expression of the droplet phase continuity equation in Lagrangian form \cite{ref:Osiptsov2000}
\begin{equation} \label{eq:FLA-COM}
	n_d (\bm{x}_d,t) = \frac{n_d (\bm{x}_0,t_0)}{\lvert \det \left( \bm{J} (\bm{x}_0,t) \right) \rvert} \, ,
\end{equation}
where $n_d$ is the {\color{corr1}droplet} number density field as sampled by individual Lagrangian droplets $\bm{x}_d$, and $\bm{J} (\bm{x}_0,t)$ is the Jacobian matrix defined as
\begin{equation}
	\bm{J} (\bm{x}_0,t) = \frac{\partial \bm{x}_d}{\partial \bm{x}_0} \, ,
\end{equation}
which represents the Eulerian-Lagrangian transformation along a droplet trajectory. The governing equation for the evolution of $\bm{J}$ is found by taking the partial derivative of the droplet momentum equation \eqref{eq:Stokes-drag} {\color{corr1}with respect to the initial position $\bm{x}_0$}, yielding
\begin{align} \label{eq:Jacobian-evolution}
	& \ddot{\bm{J}} = \frac{1}{St} \nabla \bm{u} (\bm{x}_d,t) \cdot \bm{J} - \frac{1}{St} \dot{\bm{J}} \, , \\
	& \bm{J} (\bm{x}_0,t_0) = \bm{I} \, , \dot{\bm{J}} (\bm{x}_0,t_0) = \frac{\partial \bm{v}_0}{\partial \bm{x}_0} \, . \nonumber
\end{align}
Eq.~\eqref{eq:Jacobian-evolution} can be numerically solved along trajectories in conjunction with Eq.~\eqref{eq:Stokes-drag} during a particle tracking simulation, providing a means of evaluating $n_d (\bm{x}_d,t)$ via Eq.~\eqref{eq:FLA-COM} for a given seeding of droplets. The FLA is thus able to describe the evolution of the continuum number density in terms of the contribution from a single droplet, and offers an attractive method of evaluating the droplet mean-field statistics without resorting to the conventional brute force averaging procedures used in the majority of Lagrangian particle tracking simulations.

In this work, the aforementioned FLA framework is applied to consider the behaviour of monodisperse evaporating droplets, in order to provide a straightforward setup for investigation of the interphase mass and momentum coupling source terms. In this simplified case, since the droplets are initially a monodisperse seeding, and their evaporation is governed by the idealised model presented in Eq.~\eqref{eq:evap-model} which is only time dependent, the droplets will all evaporate at the same rate. Consequently, the droplet phase can be approximated as being locally monodisperse, which in practice is a reasonable assumption for a steady state flow, but less so for a transient flow. This permits usage of the FLA without having to recourse to the more involved gFLA framework that explicitly accounts for the droplet size distribution within the methodology \cite{ref:LiRybdylova2021}, and therefore enables the focus to remain on the coupling procedure proposed in this work.

\section{Interphase coupling source term reconstruction}
\label{sec:coupling}

For calculation of the interphase mass and momentum exchange from droplets to the gas phase, the coupling source terms $S_{\text{Mass}}$ and $\bm{S}_{\text{Mom}}$ in the scalar transport and carrier flow momentum equations \eqref{eq:vapour-transport} and \eqref{eq:fluid-momentum} respectively need to be specified. {\color{corr1}These terms consist of the appropriate contributions from the droplet phase, and are stated in general form as \cite{Osiptsov2024}}
{\color{corr1}
\begin{subequations}
	\label{eq:coupling-terms-specification}
	\begin{align}
		S_{\text{Mass}} (\bm{x},t) & = - \overline{\dot{m}_d} (\bm{x},t) \, , \label{eq:mass-coupling-specification} \\
		\bm{S}_{\text{Mom}} (\bm{x},t) & = - \overline{\bm{F}_d} (\bm{x},t) - \overline{\dot{m}_d \bm{v}_d} (\bm{x},t) \, , \label{eq:momentum-coupling-specification}
	\end{align}
\end{subequations}
where $\bm{F}_d = m_d \ddot{\bm{x}}_d$ is the hydrodynamic force produced by the continuous phase on the droplet phase, $m_d$ and $\dot{m}_d$ are the mass and mass rate of change of the droplet phase respectively, $\bm{v}_d = \dot{\bm{x}}_d$ is the velocity of the droplet phase, and $\overline{(\cdot)}$ is a volume averaging procedure applied to individual Lagrangian droplets to obtain these quantities as spatially dependent fields. It is thus seen that the mass source term comes purely from the droplet evaporation, whilst the momentum source term has contributions from both the hydrodynamic force and droplet evaporation. To evaluate these expressions, the droplet acceleration $\ddot{\bm{x}}_d$ is given by Eq.~\eqref{eq:Stokes-drag}, whilst for spherical droplets expressions for $m_d$ and $\dot{m}_d$ can be obtained from the formula for the volume of a sphere and the droplet evaporation model \eqref{eq:evap-model}.

In the treatment of disperse particle-laden flows using standard Lagrangian methods, these source terms are defined using the PSI-CELL approach, in which the volume average is specified simply as a discrete sum over all the droplets contributing to an averaging volume $V$, specifically
\begin{equation} \label{eq:vol-avg-discrete}
	\overline{\phi}(\bm{x},t) = \frac{1}{V} \sum_i \phi_d^i \, ,
\end{equation}
in which the contributions of a Lagrangian quantity $\phi_d$ from individual droplets with index $i$ are accumulated to form an averaged Eulerian field description $\overline{\phi}(\bm{x},t)$.
When applied to Eq.~\eqref{eq:coupling-terms-specification}, this yields the familiar representations \cite{crowe2014multiphase}
\begin{subequations}
	\label{eq:coupling-terms-definition}
	\begin{align}
		S_{\text{Mass}} (\bm{x},t) & = - \frac{1}{V} \sum_i \dot{m}_d^i \, , \label{eq:mass-coupling} \\
		\bm{S}_{\text{Mom}} (\bm{x},t) & = - \frac{1}{V} \sum_i \bm{F}_d^i - \frac{1}{V} \sum_i \dot{m}_d^i \bm{v}_d^i \, . \label{eq:momentum-coupling}
	\end{align}
\end{subequations}
Eqs.~\eqref{eq:coupling-terms-definition} make clear the interpretation of the interphase coupling source terms as continuum representations of the individual droplets.
Whilst conceptually simple, the major disadvantage to this direct approach is that the averaging procedure implicit in Eqs.~\eqref{eq:coupling-terms-definition} is zeroth-order due to the box-counting approach that is employed {\color{rev1}not accounting for the droplet location in the accumulation process}, and therefore requiring a high number of droplet realisations to ensure statistical convergence of the resultant volume averaging procedure.

In this work, the FLA is used to specify the interphase coupling source terms given in Eqs.~\eqref{eq:coupling-terms-specification} in order to realise the computational efficiency advantage associated with the reduced droplet seeding offered by the method. To achieve a result that is comparable to standard Lagrangian simulations, the FLA number density $n_d (\bm{x}_d,t)$ defined in Eq.~\eqref{eq:FLA-COM} is used as a multiplicative factor in each source term to account for the contribution of each seed droplet. Accordingly, to incorporate the FLA into the procedure Eqs.~\eqref{eq:coupling-terms-specification}, the individual averages are written with the FLA droplet number density also included as
\begin{subequations}
	\label{eq:coupling-terms-averaged}
	\begin{align}
		S_{\text{Mass}} (\bm{x},t) & = - \overline{n_d \dot{m}_d} (\bm{x},t) \, , \\
		\bm{S}_{\text{Mom}} (\bm{x},t) & = - \overline{n_d \bm{F}_d} (\bm{x},t) - \overline{n_d  \dot{m}_d \bm{v}_d} (\bm{x},t) \, ,
	\end{align}
\end{subequations}
where $n_d$ is the FLA number density field for the droplet phase.
%
%
In practice, to utilise the FLA effectively in the framework of Eqs.~\eqref{eq:coupling-terms-averaged}, an approach to the volume averaging procedure of the individual droplet contributions that reconstructs a smooth Eulerian field description is desired, as opposed to the elementary box counting inherent in Eq.~\eqref{eq:vol-avg-discrete}.} Previous work has addressed this using a kernel regression approach to effectively interpolate the known values of $n_d (\bm{x}_d,t)$ along droplet trajectories onto the required Eulerian grid \cite{ref:StaffordRybdylova2022}. Specifically, the Nadaraya-Watson estimator is used for this procedure, which applied to the droplet number density $n_d (\bm{x}_d,t)$ is given by \cite{ref:hastie2009elements}
\begin{equation} \label{eq:kernel-estimator}
\overline{n}(\bm{x},t) = \frac{\sum_{i=1}^N K_{h}(\bm{x},\bm{x}_d^i) \, n_d^i (\bm{x}_d^i,t)}{\sum_{j=1}^N K_{h}(\bm{x},\bm{x}_d^j)} \, ,
\end{equation}
where $K_{h}(\bm{x},\bm{x}_d^i)$ is a kernel that provides the weighting of the contribution for the droplet $\bm{x}_d^i$ to the estimator at the location $\bm{x}$, and $N$ is the number of droplets which contribute towards the estimator at a given location. In this work a spherical Gaussian kernel is used, specified by
\begin{equation} \label{eq:kernel}
	K_{h}(\bm{x},\bm{x}_d^i) = \frac{1}{h} \exp \left[ - \frac{1}{2h^2} \lvert\vert \bm{x} - \bm{x}_d^i \rvert\rvert^2 \right] \, ,
\end{equation}
in which $h$ is the smoothing length of the kernel, which is defined in terms of the Jacobian for a given droplet as
\begin{equation} \label{eq:smoothing-length}
	h(t) = h_0 \lvert \det \left( \bm{J} (\bm{x}_0,t) \right) \rvert \, ,
\end{equation}
where $h_0$ is the initial smoothing length for the droplet at time $t_0$. This specification of the smoothing length allows the kernel to adaptively scale according to the associated droplet number density field, resulting in a reconstruction which can more accurately capture the structures within the droplet phase \cite{ref:StaffordRybdylova2022}.

{\color{rev1}
The specific choice of a Gaussian kernel is made in this work due to its possession of $C^{\infty}$ continuity and straightforward geometrical interpretation. This ensures the reconstruction procedure remains realisable for any quantity, since the kernel is strictly non-negative, and that incorporation of the Jacobian into the procedure using Eq.~\eqref{eq:smoothing-length} is done in a physically meaningful manner. The caveat of a Gaussian kernel having an infinite range of support in practice is then addressed by imposing a cutoff length for the locations $\bm{x}$ which receive contributions from a given droplet $\bm{x}_d^i$. This cutoff is specified by employing the 3-sigma rule of the Gaussian distribution, thereby stipulating that only locations such that $(\bm{x} - \bm{x}_d^i) \le 3h$ fall within this artificial compact support.
}

\begin{figure}[!ht]
	\begin{subfigure}[c]{0.495\columnwidth}
		\fbox{\includegraphics[width=0.9\columnwidth,trim={40 10 10 10},clip] {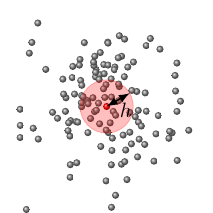}}
		\caption{}
		\label{fig:partFieldrhohigh}
	\end{subfigure}
	\begin{subfigure}[c]{0.495\columnwidth}
		\fbox{\includegraphics[width=0.9\columnwidth,trim={40 10 10 10},clip] {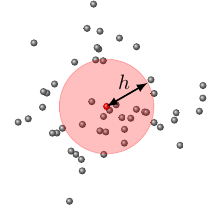}}
		\caption{}
		\label{fig:partFieldrholow}
	\end{subfigure}
	\caption{{\color{rev1}Variation of the kernel smoothing length $h$ defined by Eq.~\eqref{eq:smoothing-length}, for the cases of : \eqref{fig:partFieldrhohigh} High droplet concentration; \eqref{fig:partFieldrholow} Lower droplet concentration. The range of support of the central droplet (coloured in red) is shown as the shaded red circle, with $h$ being proportional to its radius.}}
	\label{fig:partFieldrho}
\end{figure}

{\color{rev1}
An illustration of how the smoothing length $h$ of the kernel changes with the droplet concentration according to Eq.~\eqref{eq:smoothing-length} is provided in Figure \ref{fig:partFieldrho}.
}
{\color{rev1}
This depicts the magnitude of the Jacobian determinant $\lvert \det \left( \bm{J} (\bm{x}_0,t) \right) \rvert$ associated with the central droplet (coloured in red) as the shaded red circle that encompasses the local neighbourhood of that droplet, and this is the region covered by the range of support of the kernel in Eq.~\eqref{eq:kernel}. The smoothing length $h$ is proportional to the radius of the support. This illustrates that in the case of a high droplet concentration, as seen in Figure \eqref{fig:partFieldrhohigh}, $\lvert \det \left( \bm{J} (\bm{x}_0,t) \right) \rvert$ and $h$ reduce in size. In contrast, for a lower droplet concentration, as seen in Figure \eqref{fig:partFieldrholow}, $\lvert \det \left( \bm{J} (\bm{x}_0,t) \right) \rvert$ and $h$ increase in size. The effect is such that a similar number of droplets is maintained within the support of the kernel associated with a droplet at a given time as the local droplet concentration varies, in order to provide a consistent estimator in Eq.~\eqref{eq:kernel-estimator}.
}

{\color{rev1}
The estimator in Eq.~\eqref{eq:kernel-estimator} is such that while the droplet contributions sum to unity at each location $\bm{x}$, it does not inherently preserve physical conservation laws during the mapping from the discrete Lagrangian quantities to the reconstructed continuum field in its given form. Whilst the normalisation of the kernel in Eq.~\eqref{eq:kernel} is necessary for a meaningful representation of the coupling source term fields, an additional correction procedure must be applied to ensure conservation of mass and momentum during the reconstruction process. This is performed at the individual droplet level by ensuring that the sum of the weights at all locations $\bm{x}^k$ for a given droplet $\bm{x}_d^i$ is unity. This is enforced by first calculating the weights $K_{h}(\bm{x}^k,\bm{x}_d^i)$ at all locations $\bm{x}^k$ within the kernel support for the droplet $\bm{x}_d^i$, and then normalising these weights by their sum $\sum_{k=1}^N K_{h}(\bm{x}^k,\bm{x}_d^i)$ before adding the contributions from the droplet $\bm{x}_d^i$ to the estimator in Eq.~\eqref{eq:kernel-estimator}. This then guarantees that the Eulerian mass and momentum source terms produced are an accurate physical representation of the carried Lagrangian droplet quantities from which they are reconstructed.
}

{\color{rev1}
To illustrate the efficacy of the kernel regression procedure in reconstructing the droplet number density field, Figure \ref{fig:part-estimator-comp} depicts both the droplet number density obtained from Lagrangian trajectories and the kernel estimator at time $t = 50$. Figure \eqref{fig:plot-part-numden} shows the individual droplet locations given by Eq.~\eqref{eq:Stokes-drag}, with droplets coloured by the number density calculated from Eq.~\eqref{eq:FLA-COM}. Figure \eqref{fig:plot-estimator} shows the Eulerian reconstruction for the droplet number density field given by Eq.~\eqref{eq:kernel-estimator}. It is seen that the use of the smoothing kernel in Eq.~\eqref{eq:kernel} produces a smooth number density field which effectively captures the regions of high and low concentration where droplet clusters and voids form. That this procedure is applied to the droplet number density data in Eq.~\eqref{eq:FLA-COM} is the feature which provides the method with its improvement in computational efficiency, since the information about the droplet concentration is known from the FLA without having to evaluate it separately.
\begin{figure}[!ht]
	\begin{subfigure}[c]{\columnwidth}
		\includegraphics[width=\columnwidth,trim={0 5 0 0},clip] {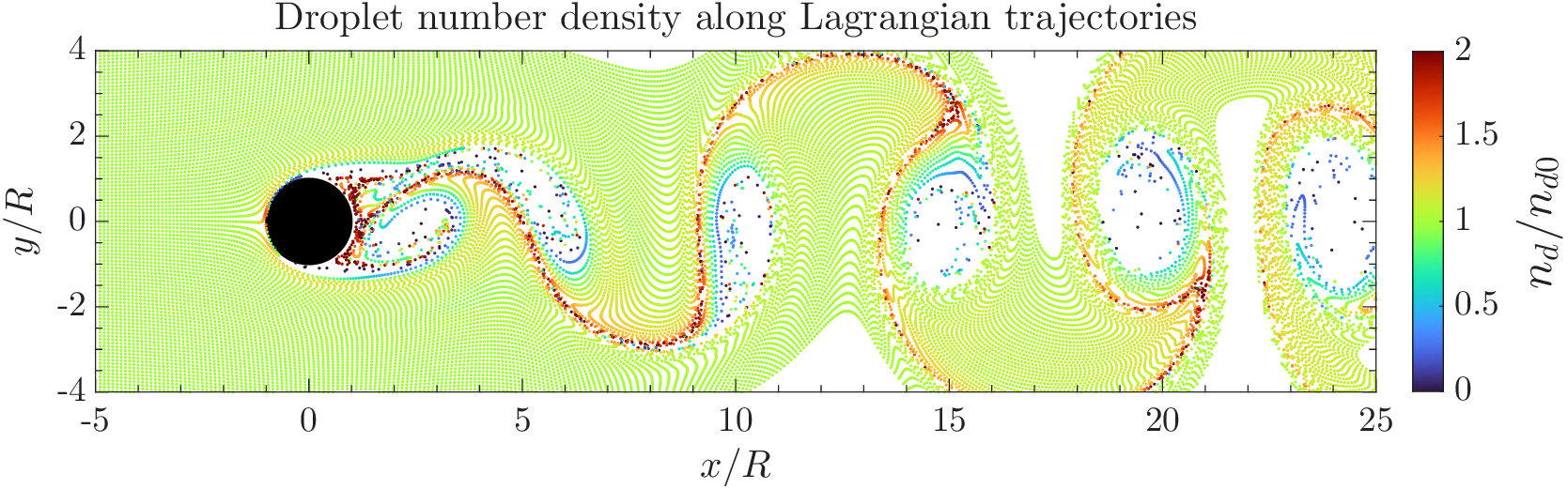}
		\caption{}
		\label{fig:plot-part-numden}
	\end{subfigure}
	\begin{subfigure}[c]{\columnwidth}
		\includegraphics[width=\columnwidth,trim={0 15 0 40},clip] {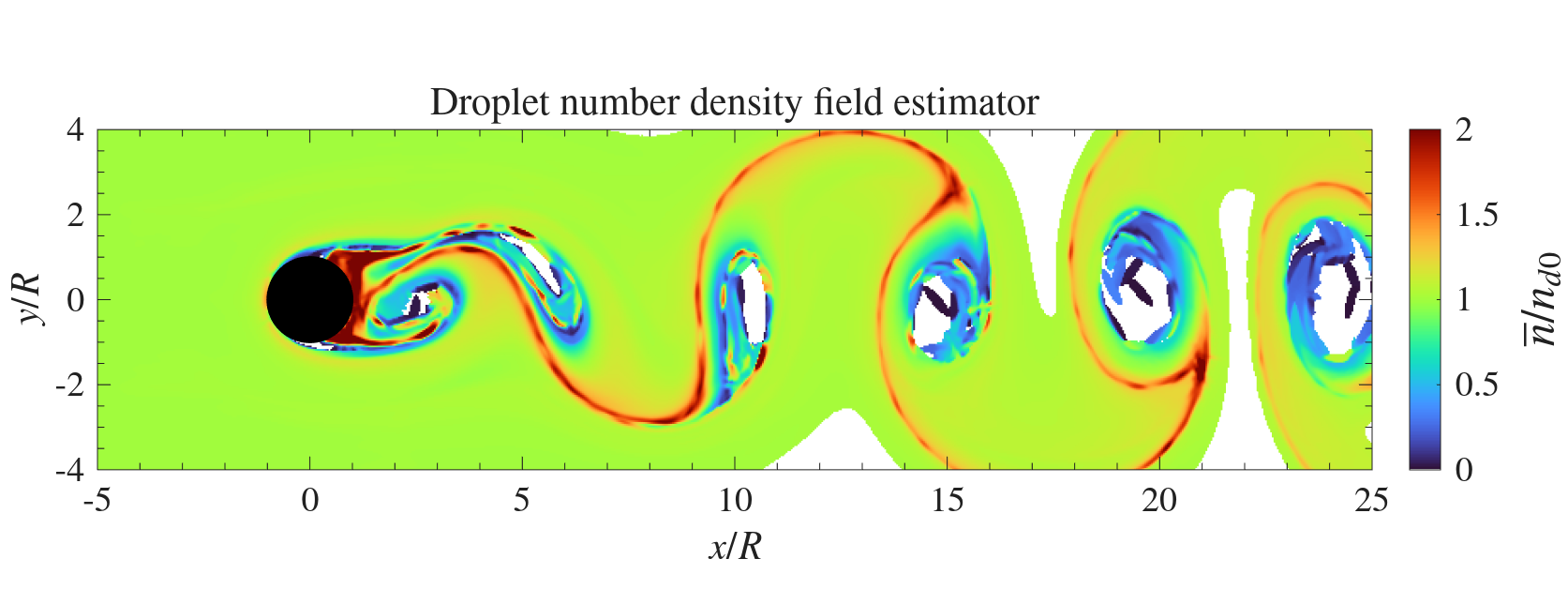}
		\caption{}
		\label{fig:plot-estimator}
	\end{subfigure}
	\caption{{\color{rev1}Comparison of the droplet number density for $Re = 100$ and $St = 0.1$ without evaporation at time $t = 50$ for: \eqref{fig:plot-part-numden} Lagrangian trajectories; \eqref{fig:plot-estimator} kernel estimator reconstruction.}}
	\label{fig:part-estimator-comp}
\end{figure}
For further information on the kernel regression procedure, the reader is referred to prior work focusing on its development \cite{ref:StaffordRybdylova2022}.
}

The estimator \eqref{eq:kernel-estimator} can be employed to evaluate the interphase coupling source terms as specified using the FLA in Eqs.~\eqref{eq:coupling-terms-averaged} by applying the procedure to the terms describing the change in mass and momentum experienced by droplets, resulting in
\begin{subequations}
	\label{eq:coupling-terms-kernel}
	\begin{align}
		S_{\text{Mass}}^{\text{FLA}} (\bm{x},t) & = 
		\frac{\sum_{i=1}^N K_{h}(\bm{x},\bm{x}_d^i) \, n_d^i (\bm{x}_d^i,t) \, \dot{m}_d^i (t)}{\sum_{j=1}^N K_{h}(\bm{x},\bm{x}_d^j)} \, , \label{eq:mass-coupling_FLA} \\
		\bm{S}_{\text{Mom}}^{\text{FLA}} (\bm{x},t) & = 
		\frac{\sum_{i=1}^N K_{h}(\bm{x},\bm{x}_d^i) \, n_d^i (\bm{x}_d^i,t) \, {\color{corr1} \bm{F}_d^i (t)}}{\sum_{j=1}^N K_{h}(\bm{x},\bm{x}_d^j)} \nonumber \\
		& + \frac{\sum_{i=1}^N K_{h}(\bm{x},\bm{x}_d^i) \, n_d^i (\bm{x}_d^i,t) \, \dot{m}_d^i (t) \, \bm{v}_d^i (t)}{\sum_{j=1}^N K_{h}(\bm{x},\bm{x}_d^j)} \, . \label{eq:momentum-coupling_FLA}
	\end{align}
\end{subequations}
Eqs.~\eqref{eq:coupling-terms-kernel} present a novel methodology for the construction of coupling terms compared to those used in standard Lagrangian particle tracking simulations, with the distinction here being that the accumulation of droplet contributions utilises the meshfree procedure offered by kernel regression as opposed to the conventional grid-based volume averaging methods.

\section{Numerical implementation}
\label{sec:numerics}

The above methodology for constructing the interphase coupling source terms has been implemented into the open source computational fluid dynamics package OpenFOAM as part of a custom solver within the Lagrangian library. The source terms are then included in the numerical solution of the carrier flow phase, and can be applied within an arbitrary flow solver in the OpenFOAM package. In this work the pimpleFOAM solver is used to simulate general transient flow, and the Lagrangian momentum parcel-based solver has been extended to include the FLA Jacobian governing equations \eqref{eq:Jacobian-evolution}, droplet evaporation by means of Eq.~\eqref{eq:evap-model}, and the kernel regression procedure for constructing the interphase coupling source terms in Eqs.~\eqref{eq:coupling-terms-kernel}.

To ensure that the {\color{corr1}representation provided by Eqs.~\eqref{eq:coupling-terms-kernel} is comparable to the reference PSI-CELL case in Eqs.~\eqref{eq:coupling-terms-definition}} despite the different droplet seedings and injection rates used, the initial FLA number density must be carefully specified so that the mass loading across the injection face is {\color{corr1}physically consistent}. Additionally, since the FLA number density is calculated with respect to a reference volume, the correct mass loading within a given mesh cell at a certain point in time during the simulation must be obtained by scaling the number density with respect to the volume of that mesh cell. The details of the scaling factors needed to account for the mass loading and reference volume are outlined in \ref{sec:initial-number-density}.

The computational procedure used to reconstruct the interphase coupling source terms in Eqs.~\eqref{eq:coupling-terms-kernel} within the OpenFOAM framework is delineated in \ref{sec:solution-coupling}.

\section{Results and discussion}
\label{sec:results}

For the present study, the configuration used is the prototypical test case of two-dimensional flow past a cylinder of radius $R$. Steady-state at $Re = 20$ and unsteady flow at $Re = 100$ are investigated, and the cases for both non-evaporating and evaporating droplets are considered {\color{corr1}in each flow regime}. In the case of {\color{corr1}non-evaporating droplets}, this enables the momentum coupling source term to be considered in isolation since no phase change is taking place, meaning that only the first term in Eq.~\eqref{eq:momentum-coupling-specification} involving the hydrodynamic force contribution exists. This simplified scenario is ideal for illustrating the applicability of the FLA-based solver before considering the additional effects from evaporation.
Simulations using this new methodology are compared against a reference solver provided by the existing built-in OpenFOAM Lagrangian momentum parcel-based library that uses the PSI-CELL method to accumulate the interphase coupling source terms. The FLA-based reconstruction of the interphase coupling source terms is calculated using Eqs.~\eqref{eq:coupling-terms-kernel}, whilst the corresponding reconstruction from the reference PSI-CELL simulations is evaluated using Eqs.~\eqref{eq:coupling-terms-definition}.

For simulations using the FLA, a uniformly spaced seeding of {\color{rev1}$NP$} monodisperse droplets is injected across the inlet at $x/R = -5$, with an injection interval of {\color{rev1}$\Delta t_i$} seconds. {\color{rev1}It is necessary to vary $\Delta t_i$ in addition to $NP$, since $NP$ only determines the droplet concentration across the inlet cross-section, whilst $\Delta t_i$ accounts for the droplet concentration in the inlet-normal direction. In order to demonstrate the applicability of the kernel regression procedure for reconstructing the coupling source terms, cases are presented for $NP = 100$ and $\Delta t_i = 0.1$.} In contrast, to achieve statistical convergence on the mesh, the reference PSI-CELL simulations use 1000 droplets injected with an interval of $0.01$ seconds, representing a hundredfold increase in the number of droplets injected per second of simulation time. This is correspondingly reflected in the amount of CPU time needed to execute the respective methods, with the FLA case needing 27 CPU hours compared to the built-in solver case requiring 4030 CPU hours.

\begin{figure}[!ht]
	\includegraphics[width=\columnwidth,trim={0 0 0 5},clip] {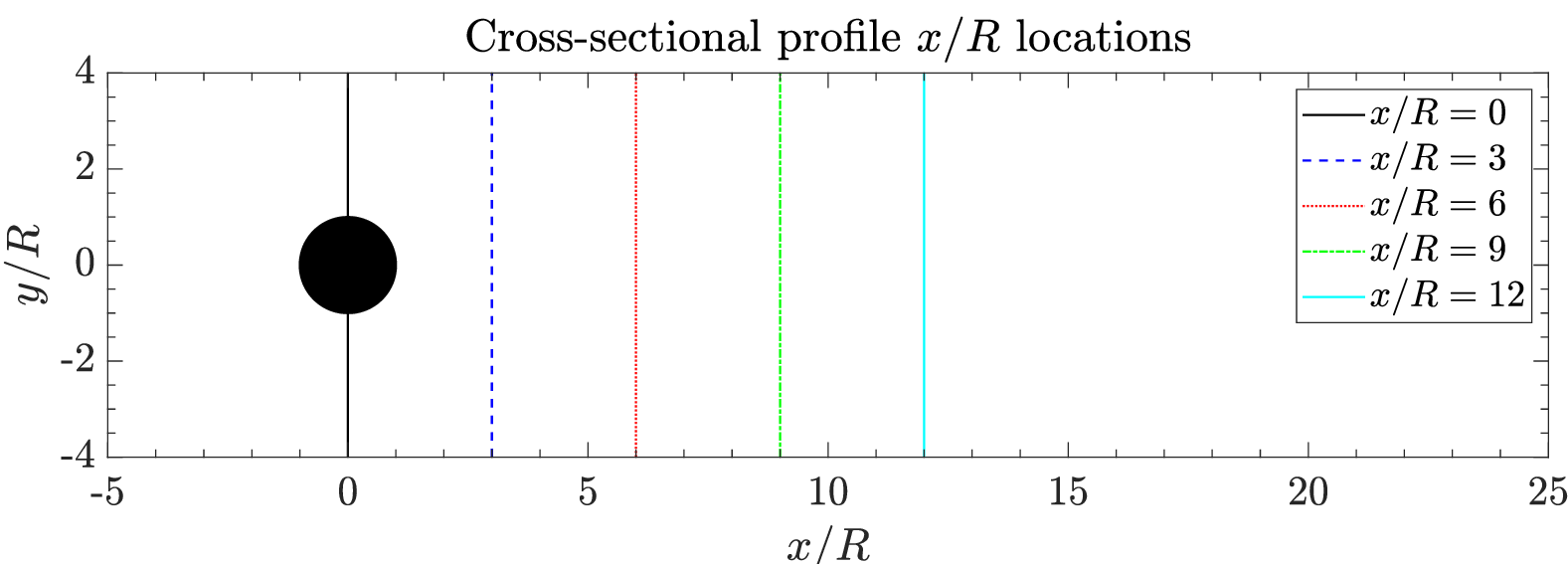}
	\caption{Schematic diagram of the computational domain showing the cross-sectional profile $x / R$ locations:
		{\color{black}-----} ${x} / R = 0$, {\color{blue}- - -} ${x} / R = 3$,
		{\color{red}$\cdots\cdots$} ${x} / R = 6$,
		{\color{green}-$\cdot$-$\cdot$-$\cdot$} ${x} / R = 9$,
		{\color{cyan}-----} ${x} / R = 12$
		.}
	\label{fig:FPC_prof_schematic}
\end{figure}

All statistics in the following analysis are presented at $t = 50$ seconds after the start of the droplet injection, to enable the carrier flow and droplet field to evolve and adjust according to the droplet-gas interaction effects that are introduced by the two-way coupling mechanism. Further to this, before the introduction of droplets the carrier flow is allowed to equilibrate into either the steady-state or periodic configurations being considered. Comparison of the FLA against the reference PSI-CELL simulations is made at the selected cross-sectional profile $x / R$ locations illustrated in Figure \ref{fig:FPC_prof_schematic}, with these being firstly at the location of the cylinder, then each subsequent profile at a distance of 3 cylinder radii downstream.

The parameters used across the different simulations are given in Table~\ref{table:param}. For $Re = 20$, the parameters for both the evaporating and non-evaporating cases are identical, whilst for $Re = 100$ the droplet inertia is increased for the evaporating case in order to elucidate the effects of this on the coupling mechanism (see Section \ref{sec:transient-evap-flow}). In all cases the carrier flow is taken to be air with density $\rho_f = 1$. The desired value of $Re$ for steady-state and transient flow behaviour is obtained by altering the flow kinematic viscosity $\nu$, however the droplet density or size must then also be changed to compensate for this in order to maintain a given value of $St$. At the end of the simulation at $t = 50$, the droplet volume fraction $\alpha_d$ reaches $1.41 \times 10^{-3}$, apart from in the non-evaporating case for $Re = 100$ when it is $1.77 \times 10^{-3}$. These values correspond to the {\color{corr1}global} upper limit for which the FLA is valid as a continuum representation {\color{corr1}of a dilute droplet loading}. The associated mass fraction $\omega_d$ is $77.9\%$ at $t = 50$, again with the exception of the non-evaporating case for $Re = 100$ where it is $26.1\%$. This signifies that despite the low droplet volume fraction, the mass loading of the droplet phase is significant, and in most cases accounts for the majority of mass in the two-phase system by the end of the simulation. Under such conditions this non-negligible mass loading can be expected to have a notable effect upon the behaviour of the underlying carrier flow, and it is this aspect which the following cases are built up to investigate. 
\begin{table}[!ht]
	\begin{center}
		\begin{tabular}{ccccc}
			\toprule
			& \multicolumn{2}{c}{Non-evaporating} &  \multicolumn{2}{c}{Evaporating} \\
			\cmidrule{2-3}
			\cmidrule{4-5}
			& $Re = 20$ & $Re = 100$ & $Re = 20$ & $Re = 100$ \\
			\midrule
			$\rho_d$ & 2500 & 2000 & 2500 & 2500 \\
			$r_d$ & 0.003 & 0.0015 & 0.003 & 0.003 \\
			$\nu$ & 0.05 & 0.01 & 0.05 & 0.01 \\
			$St$ & 0.1 & 0.1 & 0.1 & 0.5 \\
			$\alpha_d$ & $0.00141$ & $0.00177$ & $0.00141$ & $0.00141$ \\
			$\omega_d$ & 0.779 & 0.261 & 0.779 & 0.779 \\
			\bottomrule
		\end{tabular} 
	\end{center}
	\caption{Simulation parameters. Values of $\rho_d$ and $r_d$ are given in SI units.}
	\label{table:param}
\end{table}

{\color{rev1}
Numerically, the only free parameter within the kernel regression procedure is the initial smoothing length $h_0$. This can be altered depending on the fidelity required in the droplet field reconstruction. A smaller value of $h_0$ will provide better resolved features, however will require a greater number of droplets to achieve a smooth reconstruction. Conversely, a larger value of $h_0$ provides more smoothing to the reconstruction and will not resolve features in as much detail, however will require fewer droplets in order to do so. In this manner, it is appropriate to set $h_0$ as inversely proportional to the injection droplet seeding $NP$, as then the same number of droplets fall within the initial kernel support for cases run with different values of $NP$. In this work, $h_0$ is set as the inter-droplet spacing at the point of injection, such that
\begin{equation}
	h_0 = \frac{(y/R)_{\text{max}} - (y/R)_{\text{min}}}{NP}
\end{equation}
This ensures that a sufficient number of droplets fall within the kernel support for different values of $NP$ to provide a smooth reconstruction of the mass and momentum coupling source terms in Eqs.~\eqref{eq:coupling-terms-kernel}.
}

{\color{rev1}
To quantify the accuracy associated with the kernel regression procedure more formally, the relative $l^2$-norm is employed as a global error metric for comparing the Eulerian coupling source term fields constructed from the FLA data against those from the reference PSI-CELL data. For this purpose, the appropriate $l^2$-norm for a given spatial field $(\cdot)$ is defined as
\begin{equation}
	\text{Error}_{l^2}[(\cdot)] = \frac{\sqrt{\sum_{i=1}^N [(.)_i^{\text{FLA}} - (.)_i^{\text{PSI-CELL}}]^2}}{\sqrt{\sum_{i=1}^N [(.)_i^{\text{PSI-CELL}}]^2}}
\end{equation}
where $N$ is the number of computational mesh points in the global domain, $(.)_i^{\text{FLA}}$ is the FLA estimator for the spatial field at point $i$, and $(.)_i^{\text{PSI-CELL}}$ is the PSI-CELL estimator for the spatial field at point $i$. This is calculated for an injection droplet seeding of $NP = 50$, $100$, and $200$ with respective injection intervals of $\Delta t_i = 0.2$, $0.1$, and $0.05$ for these seedings, in order to illustrate the effect of the number of droplets used in FLA simulations on the relative error.
}

{\color{rev1}
A comparison of the droplet counts at the point of injection and simulation runtime for the FLA and PSI-CELL cases used is provided in Table \ref{table:comp-spec}.
\begin{table}[!ht]
	\begin{center}
		\begin{tabular}{ccccc}
			\toprule
			Method & $NP$ & $\Delta t_i [s]$ & $NP/\Delta t_i$ & CPU hours \\
			\midrule
			FLA & 50 & 0.2 & 250 & 7.55 \\
			FLA & 100 & 0.1 & 1000 & 26.82 \\
			FLA & 200 & 0.05 & 4000 & 98.91 \\
			PSI-CELL & 1000 & 0.01 & 100000 & 4030.23 \\
			\bottomrule
		\end{tabular} 
	\end{center}
	\caption{{\color{rev1}Comparison of the droplets injected per second ($NP / \Delta t_i$) and simulation runtimes (CPU hours) for the different FLA and PSI-CELL cases run.}}
	\label{table:comp-spec}
\end{table}
The simulation cost for the droplet phase is proportional to the number of droplets injected per second, given by the quantity $NP/\Delta t_i$. The number of CPU hours is the total time taken for the simulation to run until time $t = 50$ across all processors used. It is seen that the FLA case with $NP = 100$ has an injection droplet count that is 100 times less than that of the PSI-CELL case, and a simulation runtime that is over 100 times faster. Decreasing the droplet seeding to $NP = 50$ results in the simulation runtime decreasing by a factor of around $3.5$ compared to that for $NP = 100$, whilst increasing the seeding to $NP = 200$ increases the simulation runtime by a similar factor. This quantifies the efficiency gain that can be achieved using the kernel regression procedure for the coupling source term reconstruction in the cases that have been run in this study.
}

In transient flow solvers, an additional source of {\color{corr1}numerical ambiguity} arising from interphase coupling methodologies that do not iterate to a converged solution can be present. Instead, such solvers simply apply a correction to the velocity field to account for the droplet phase. Different methods can only be compared exactly in the limit of converged solutions, so comparing the evolution of the simulation may not be a rigorous comparison of different coupling approaches. In this case, a more suitable discriminator is the source terms constructed from the droplet phase, which can be assessed both qualitatively and quantitatively in order to gauge the performance of the respective numerical procedures.

\subsection{Steady-state non-evaporating flow}
\label{sec:steady-state-non-evap-flow}

\begin{figure}[!ht]
	\begin{subfigure}[c]{\columnwidth}
		\includegraphics[width=\columnwidth,trim={0 10 5 15},clip] {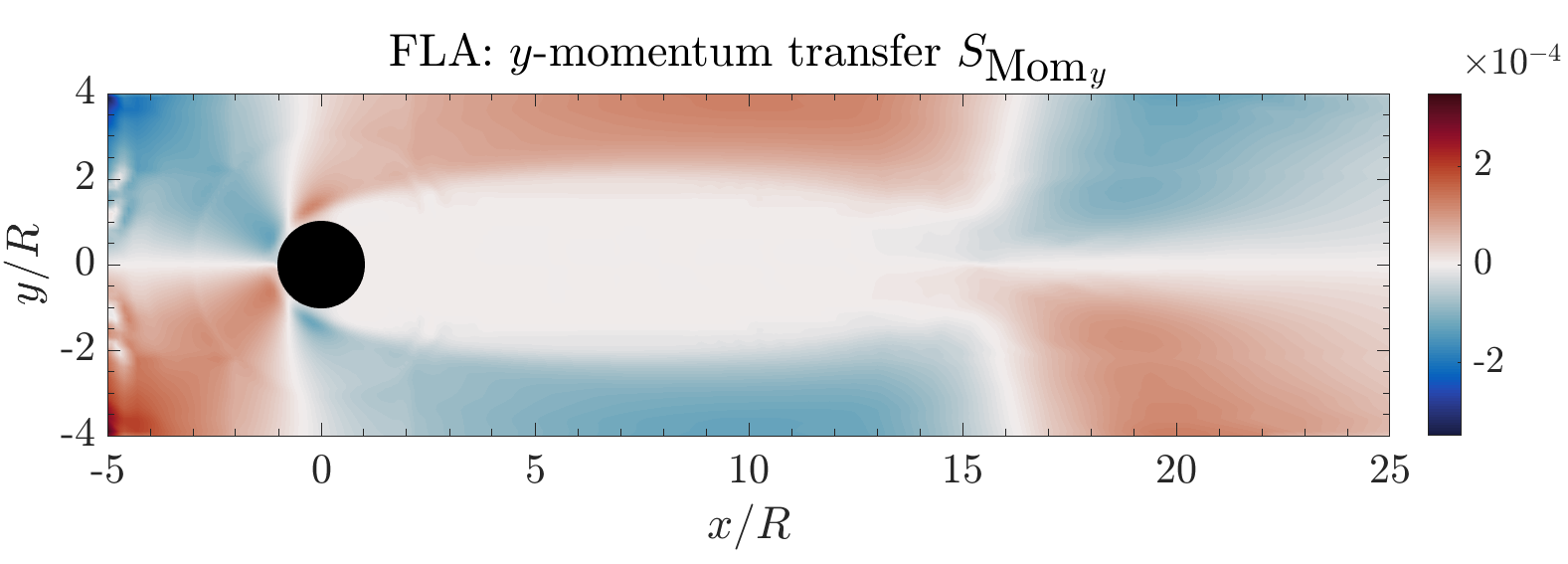}
		\caption{}
		\label{fig:fieldPlotFLA_nonevap_Re20_UTransY}
	\end{subfigure}
	\begin{subfigure}[c]{\columnwidth}
		\includegraphics[width=\columnwidth,trim={0 10 5 15},clip] {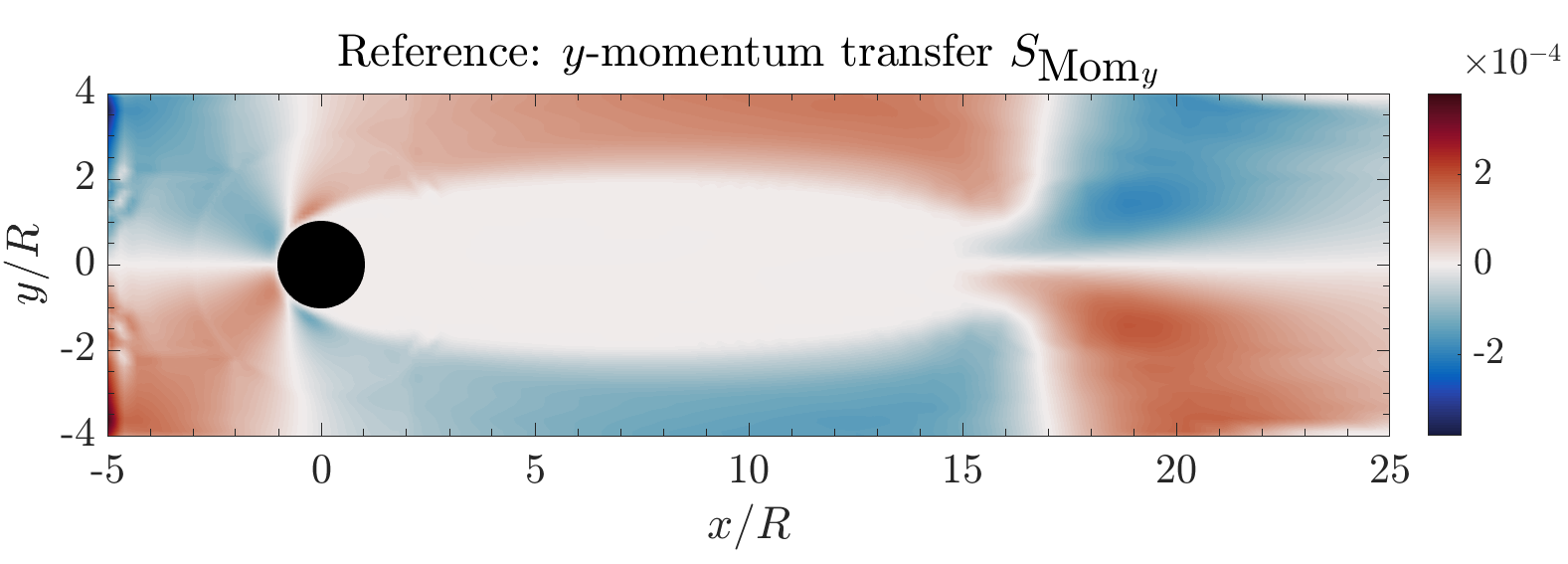}
		\caption{}
		\label{fig:fieldPlotCIC_nonevap_Re20_UTransY}
	\end{subfigure}
	\caption{Distribution of the $y-$component of the momentum transfer source term $\bm{S}_{\text{Mom}}$ for $Re = 20$ and $St = 0.1$ without evaporation at time $t = 50$ for: \eqref{fig:fieldPlotFLA_nonevap_Re20_UTransY} FLA solver; \eqref{fig:fieldPlotCIC_nonevap_Re20_UTransY} reference PSI-CELL solver.}
	\label{fig:fieldPlot_nonevap_Re20_UTransY}
\end{figure}

\begin{figure}[!ht]
	\begin{subfigure}[c]{\columnwidth}
		\includegraphics[width=\columnwidth,trim={0 0 0 0},clip] {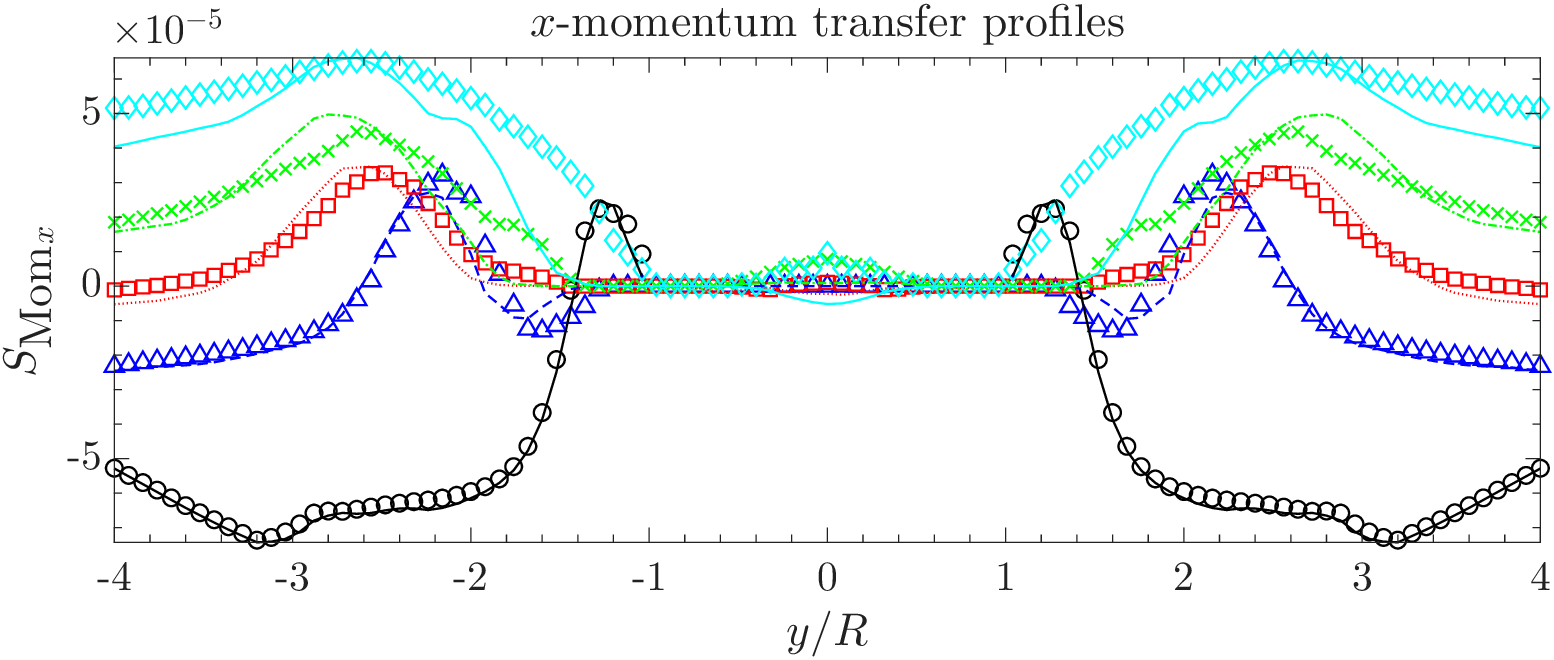}
		\caption{}
		\label{fig:plotprofCompFLACIC_nonevap_Re20_UTransX}
	\end{subfigure}
	\begin{subfigure}[c]{\columnwidth}
		\includegraphics[width=\columnwidth,trim={0 0 0 0},clip] {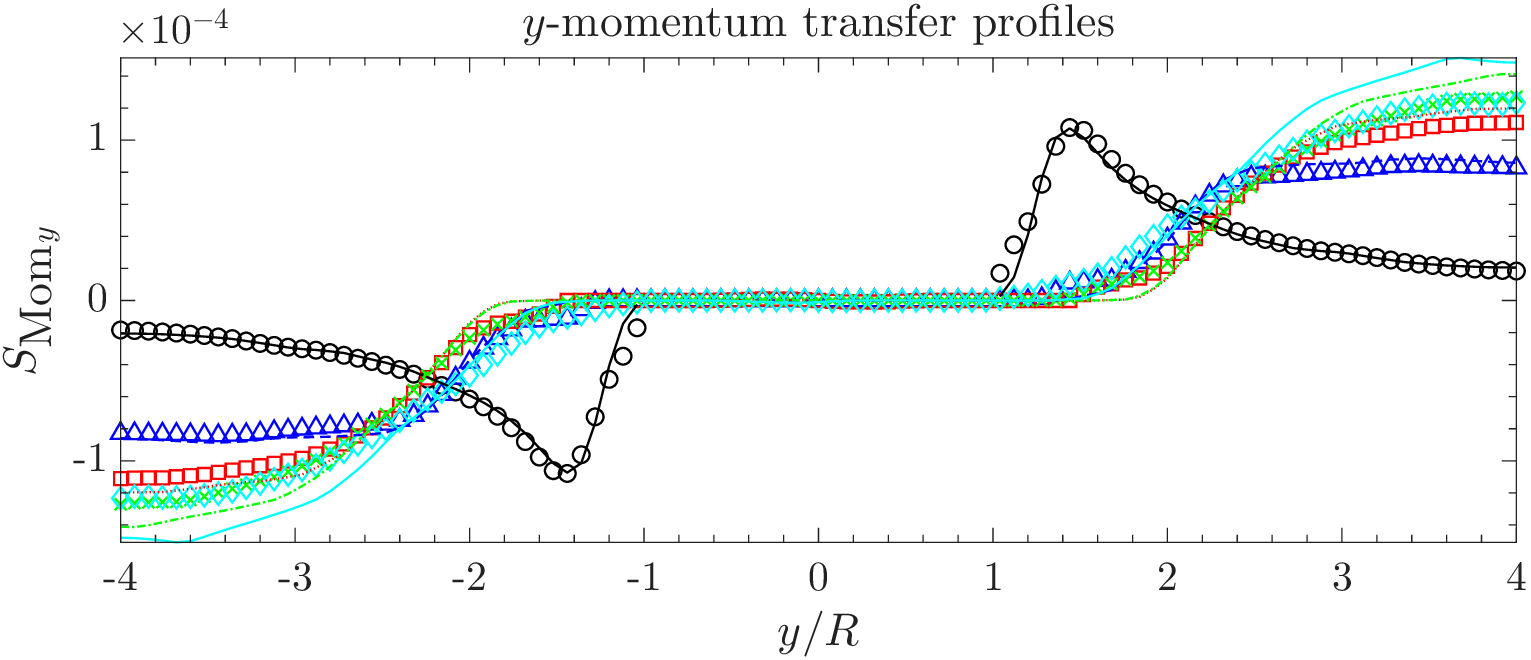}
		\caption{}
		\label{fig:plotprofCompFLACIC_nonevap_Re20_UTransY}
	\end{subfigure}
	\caption{Momentum transfer source term $\bm{S}_{\text{Mom}}$ profiles for the FLA solver (symbols) and the reference PSI-CELL solver (lines) at cross-sections
	{\color{black}$\boldsymbol{\bigcirc}$} ${x} / R = 0$, {\color{blue}$\boldsymbol{\bigtriangleup}$} ${x} / R = 3$,
	{\color{red}$\boldsymbol{\Box}$} ${x} / R = 6$,
	{\color{green}$\boldsymbol{\times}$} ${x} / R = 9$,
	{\color{cyan}$\boldsymbol{\diamond}$} ${x} / R = 12$
	for $Re = 20$ and $St = 0.1$ without evaporation at time $t = 50$ for: \eqref{fig:plotprofCompFLACIC_nonevap_Re20_UTransX} $x$-momentum; \eqref{fig:plotprofCompFLACIC_nonevap_Re20_UTransY} $y$-momentum.}
	\label{fig:plotprofCompFLACIC_nonevap_Re20}
\end{figure}

Initially the simplest case of non-evaporating flow for $Re = 20$ is considered. In this scenario the droplet field separates into two distinct regions as it passes around the cylinder, and equilibrates into its steady-state by the end of the simulation at $t = 50$. The spatial distribution for the $y-$component of the momentum transfer source term is shown in Figure \ref{fig:fieldPlot_nonevap_Re20_UTransY}.
Figure \eqref{fig:fieldPlotFLA_nonevap_Re20_UTransY} demonstrates that despite the reduced droplet seeding and simulation runtime of the FLA case, the qualitative agreement with the statistically-converged reference PSI-CELL case in Figure \eqref{fig:fieldPlotCIC_nonevap_Re20_UTransY} is high, with the only notable difference being a slightly larger magnitude of the coupling term in some locations for the PSI-CELL case, along with the wake behind the cylinder being marginally longer. This is detailed further in Figure \ref{fig:plotprofCompFLACIC_nonevap_Re20}, which depicts a comparison between cross-sectional profiles of the momentum transfer fields located at different positions downstream of the cylinder, and is shown for both the $x-$component in Figure \eqref{fig:plotprofCompFLACIC_nonevap_Re20_UTransX} and the $y-$component in Figure \eqref{fig:plotprofCompFLACIC_nonevap_Re20_UTransY}.
It is seen that in general both components of the momentum transfer field are well recovered by the FLA reconstruction procedure when compared to the reference PSI-CELL solver, however the error in the FLA values increases with distance from the cylinder. This is attributable to the kernel regression process smoothing the momentum transfer field obtained by the FLA, which is reflected in the fact that steeper gradients in the profiles are generally less well captured. Notwithstanding this, the overall qualitative trend for the spatial distribution of the momentum transfer term are captured very well, illustrating the usefulness of the FLA as a means of more quickly obtaining insight into the effect of the interphase coupling on the flow behaviour.

{\color{rev1}
Evaluation of the relative $l^2$-norm of the momentum transfer source term for the different FLA droplet seedings and injection intervals is displayed in Table \ref{table:l2-norm_noevap_Re20} for time $t = 50$.
\begin{table}[!ht]
	\begin{center}
		\begin{tabular}{cccc}
			\toprule
			& $NP = 50$ & $NP = 100$ & $NP = 200$ \\
			& $\Delta t_i = 0.2$ & $\Delta t_i = 0.1$ & $\Delta t_i = 0.05$ \\
			\midrule
			$\text{Error}_{l^2}[{S_{\text{Mom}}}_x]$ & 0.5388 & 0.3186 & 0.2459 \\
			$\text{Error}_{l^2}[{S_{\text{Mom}}}_y]$ & 0.5028 & 0.2828 & 0.2568 \\
			\bottomrule
		\end{tabular} 
	\end{center}
	\caption{{\color{rev1}Relative $l^2$-norm for the components of the momentum transfer source term $\bm{S}_{\text{Mom}}$ evaluated using different droplet seedings for $Re = 20$ and $St = 0.1$ without evaporation at time $t = 50$.}}
	\label{table:l2-norm_noevap_Re20}
\end{table}
This evidences that the reconstruction using $NP = 100$ reduces the error in the momentum transfer source term by a notable amount compared to the case for $NP = 50$. Additionally, although the reconstruction provided by $NP = 200$ has a lower error than that for $NP = 100$, the increase in accuracy is relatively minimal at $NP = 200$, signifying that in this case using $NP = 100$ achieves an ideal combination of accuracy and the efficiency that can be gained using the kernel regression procedure for the coupling source term reconstruction in steady-state flow.
}

\subsection{Transient non-evaporating flow}
\label{sec:transient-non-evap-flow}

\begin{figure}[!ht]
	\begin{subfigure}[c]{\columnwidth}
		\includegraphics[width=\columnwidth,trim={0 10 5 15},clip] {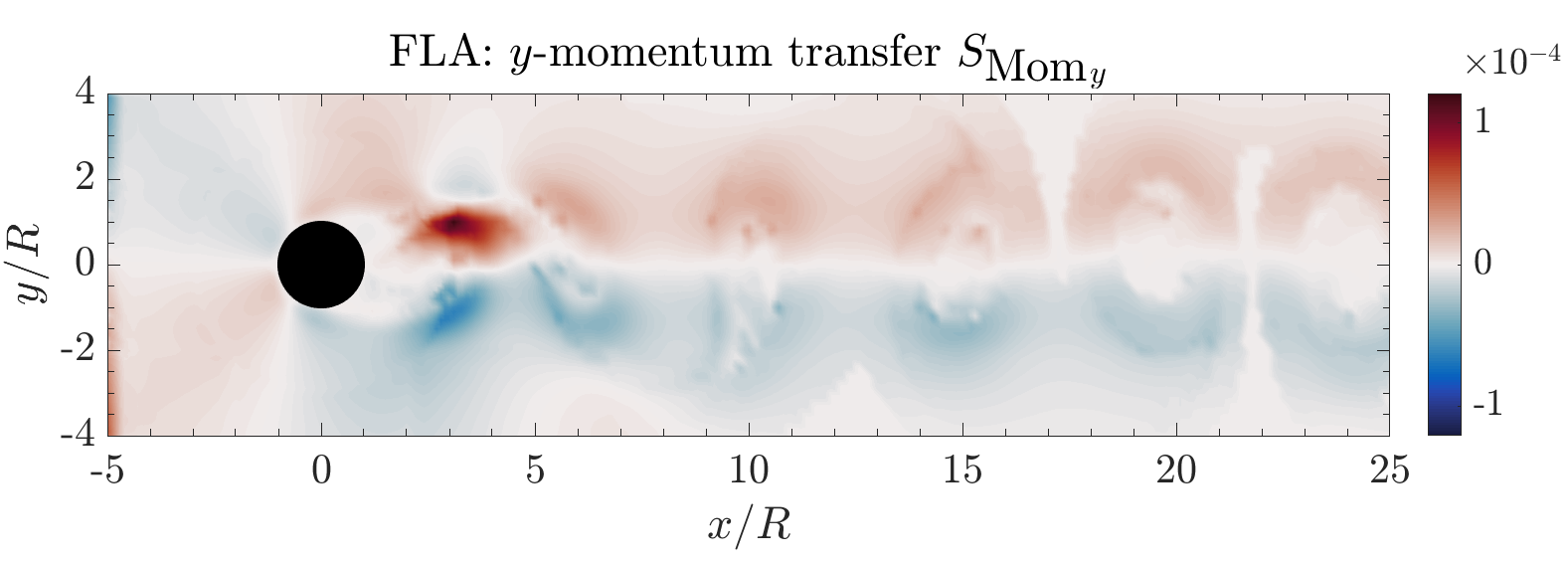}
		\caption{}
		\label{fig:fieldPlotFLA_nonevap_Re100_UTransY}
	\end{subfigure}
	\begin{subfigure}[c]{\columnwidth}
		\includegraphics[width=\columnwidth,trim={0 10 5 15},clip] {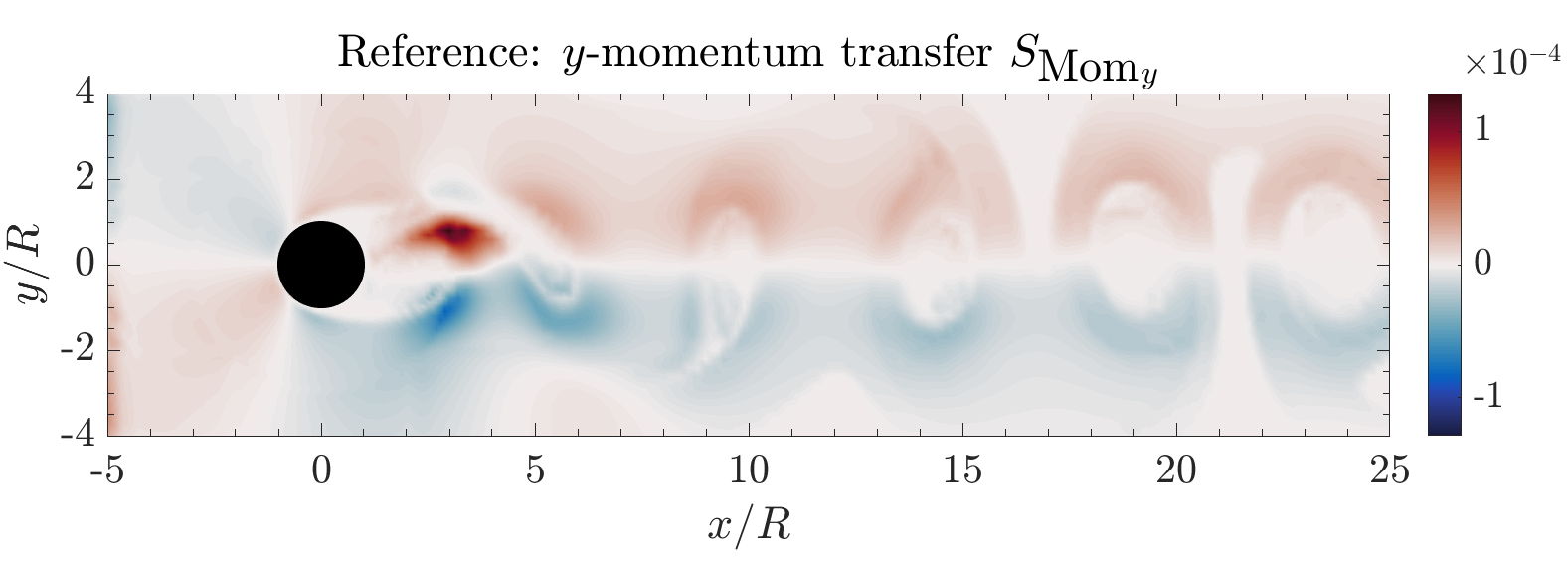}
		\caption{}
		\label{fig:fieldPlotCIC_nonevap_Re100_UTransY}
	\end{subfigure}
	\caption{Distribution of the $y-$component of the momentum transfer source term $\bm{S}_{\text{Mom}}$ for $Re = 100$ and $St = 0.1$ without evaporation at time $t = 50$ for: \eqref{fig:fieldPlotFLA_nonevap_Re100_UTransY} FLA solver; \eqref{fig:fieldPlotCIC_nonevap_Re100_UTransY} reference PSI-CELL solver.}
	\label{fig:fieldPlot_nonevap_Re100_UTransY}
\end{figure}

\begin{figure}[!ht]
	\begin{subfigure}[c]{\columnwidth}
		\includegraphics[width=\columnwidth,trim={0 0 0 0},clip] {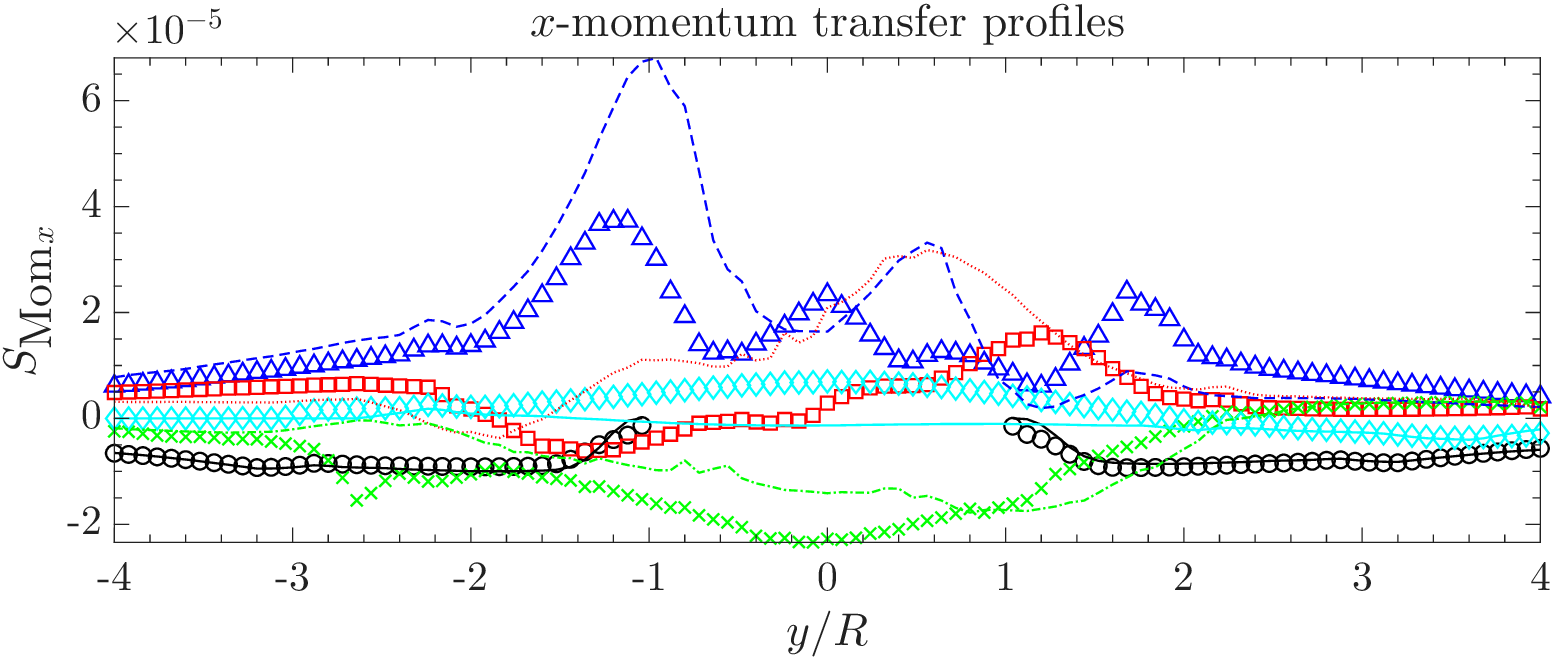}
		\caption{}
		\label{fig:plotprofCompFLACIC_nonevap_Re100_UTransX}
	\end{subfigure}
	\begin{subfigure}[c]{\columnwidth}
		\includegraphics[width=\columnwidth,trim={0 0 0 0},clip] {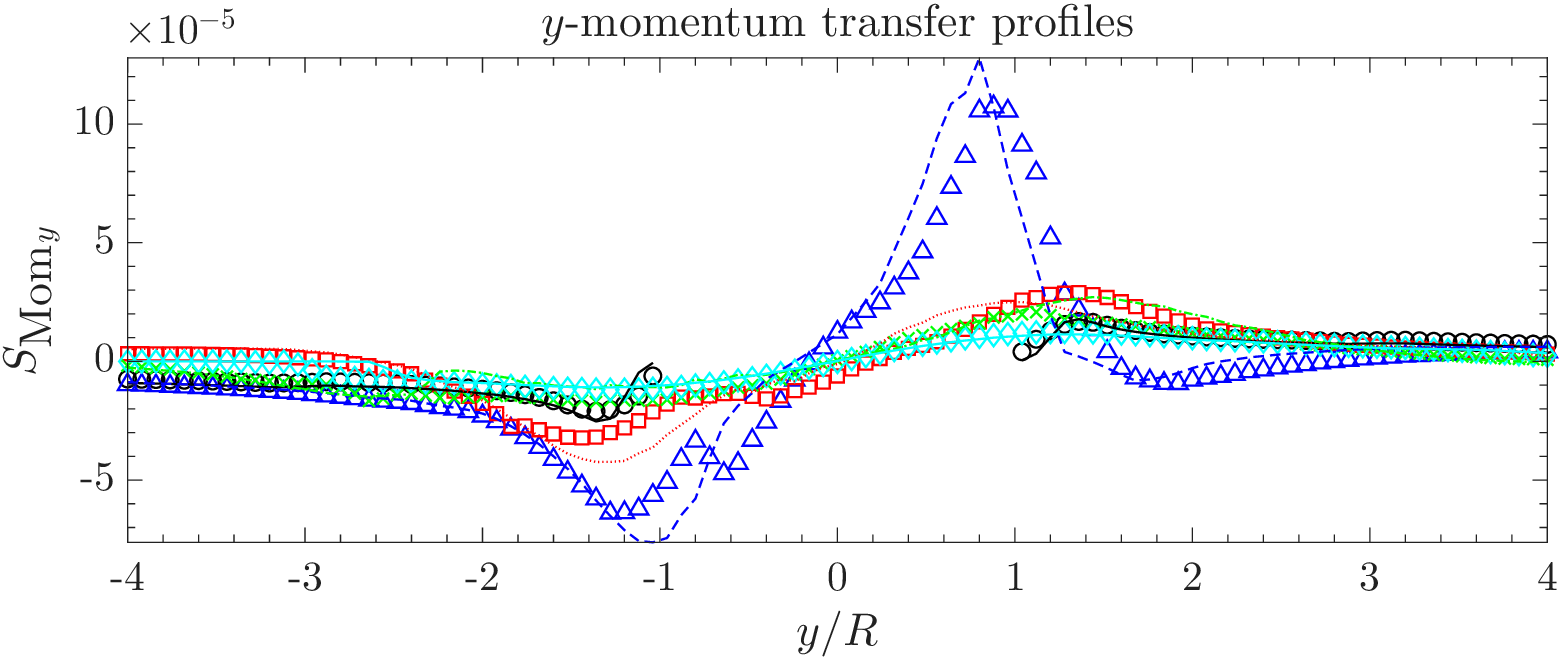}
		\caption{}
		\label{fig:plotprofCompFLACIC_nonevap_Re100_UTransY}
	\end{subfigure}
	\caption{Momentum transfer source term $\bm{S}_{\text{Mom}}$ profiles for the FLA solver (symbols) and the reference PSI-CELL solver (lines) at cross-sections
		{\color{black}$\boldsymbol{\bigcirc}$} ${x} / R = 0$, {\color{blue}$\boldsymbol{\bigtriangleup}$} ${x} / R = 3$,
		{\color{red}$\boldsymbol{\Box}$} ${x} / R = 6$,
		{\color{green}$\boldsymbol{\times}$} ${x} / R = 9$,
		{\color{cyan}$\boldsymbol{\diamond}$} ${x} / R = 12$
		for $Re = 100$ and $St = 0.1$ without evaporation at time $t = 50$ for: \eqref{fig:plotprofCompFLACIC_nonevap_Re100_UTransX} $x$-momentum; \eqref{fig:plotprofCompFLACIC_nonevap_Re100_UTransY} $y$-momentum.}
	\label{fig:plotprofCompFLACIC_nonevap_Re100}
\end{figure}

The next case retains non-evaporating droplets whilst focusing upon transient flow at $Re = 100$. In this scenario a periodic flow with asymmetric vortex shedding is created, and simulations were run to enable the vortex street to become fully developed before droplets were injected into the flow. The spatial distribution for the $y-$component of the momentum transfer source term at $t = 50$ are shown for this case in Figure \ref{fig:fieldPlot_nonevap_Re100_UTransY}.
As with the previous steady-state case, it is observed that there is a high level of qualitative agreement between the FLA reconstruction of the momentum transfer field in Figure \eqref{fig:fieldPlotFLA_nonevap_Re100_UTransY} and the reference PSI-CELL solver in Figure \eqref{fig:fieldPlotCIC_nonevap_Re100_UTransY}. This is more notable given the additional detail of the flow field structures involved in a transient flow, with both the shape and size of the vortices produced by the presence of the droplet phase being closely mirrored by the FLA. In particular, the only discrepancy is in the magnitude of ${S_{\text{Mom}}}_y$, for which it is seen that the FLA provides a slight over prediction for the area of high momentum transfer in the immediate wake of the cylinder at the formation of the vortex street. There is also a small amount of {\color{corr1}statistical} noise within the vortices in Figure \eqref{fig:fieldPlotFLA_nonevap_Re100_UTransY}, and this is due to the number density calculated via the FLA using Eq.~\eqref{eq:FLA-COM} becoming infinite as a result of the trajectories for the FLA seed droplets crossing \cite{ref:Osiptsov2000}. The kernel regression procedure is able to smoothen this effect to avoid the singularity that would otherwise appear in ${S_{\text{Mom}}}_y$ \cite{ref:StaffordRybdylova2022}, however the reconstructed field is not as smooth as the reference PSI-CELL case obtained with the higher droplet seeding in Figure \eqref{fig:fieldPlotCIC_nonevap_Re100_UTransY}. {\color{corr1}Furthermore, the numerical error incurred during the kernel regression procedure is larger in areas where there is a high number density gradient, and in particular on the boundaries of regions without droplets \cite{ref:StaffordRybdylova2022}.} These observations can be seen more clearly in Figure \ref{fig:plotprofCompFLACIC_nonevap_Re100}, which shows the cross-sectional profiles of both cases for the momentum transfer fields in the $x$ and $y$ directions.
For the $y-$components illustrated in Figure \ref{fig:fieldPlot_nonevap_Re100_UTransY}, Figure \eqref{fig:plotprofCompFLACIC_nonevap_Re100_UTransY} corroborates the close visual agreement, showing that the FLA profiles at all the selected cross sections agree in both trend and magnitude. In particular, the area of high momentum transfer at $x / R = 3$ is well represented, with both the peak and gradient of the field being adequately captured by the FLA. Notwithstanding this, the $x$-momentum transfer in Figure \eqref{fig:fieldPlotCIC_nonevap_Re20_UTransY} does however exhibit a higher level of difference between the profiles of the FLA and reference PSI-CELL cases, with the magnitude being inadequately represented in some locations. Despite this, it is seen that the FLA is generally able to provide a faithful representation of the true momentum transfer term across the flow using the reduced droplet seeding.

{\color{rev1}
The relative $l^2$-norm of the momentum transfer source term for the different FLA droplet seedings and injection intervals is displayed in Table \ref{table:l2-norm_noevap_Re100} for time $t = 50$.
\begin{table}[!ht]
	\begin{center}
		\begin{tabular}{cccc}
			\toprule
			& $NP = 50$ & $NP = 100$ & $NP = 200$ \\
			& $\Delta t_i = 0.2$ & $\Delta t_i = 0.1$ & $\Delta t_i = 0.05$ \\
			\midrule
			$\text{Error}_{l^2}[{S_{\text{Mom}}}_x]$ & 0.8137 & 0.6078 & 0.3703 \\
			$\text{Error}_{l^2}[{S_{\text{Mom}}}_y]$ & 0.5568 & 0.4319 & 0.2774 \\
			\bottomrule
		\end{tabular} 
	\end{center}
	\caption{{\color{rev1}Relative $l^2$-norm for the components of the momentum transfer source term $\bm{S}_{\text{Mom}}$ evaluated using different droplet seedings for $Re = 100$ and $St = 0.1$ without evaporation at time $t = 50$.}}
	\label{table:l2-norm_noevap_Re100}
\end{table}
As with the steady-state case in Section \ref{sec:steady-state-non-evap-flow}, it is seen that the increase in accuracy from a droplet seeding of $NP = 50$ to $NP = 100$ is notable, however for unsteady flow a similar increase in accuracy is also seen by increasing the droplet seeding to $NP = 200$. Therefore to reduce the error further it is appropriate to use the higher seeding of $NP = 200$ at $Re = 100$, however this does result in a computational cost that is 4 times higher than at $NP = 100$ due to the increased injection interval $\Delta t_i$.
}

\begin{figure}[!ht]
	\includegraphics[width=\columnwidth,trim={0 15 0 0},clip] {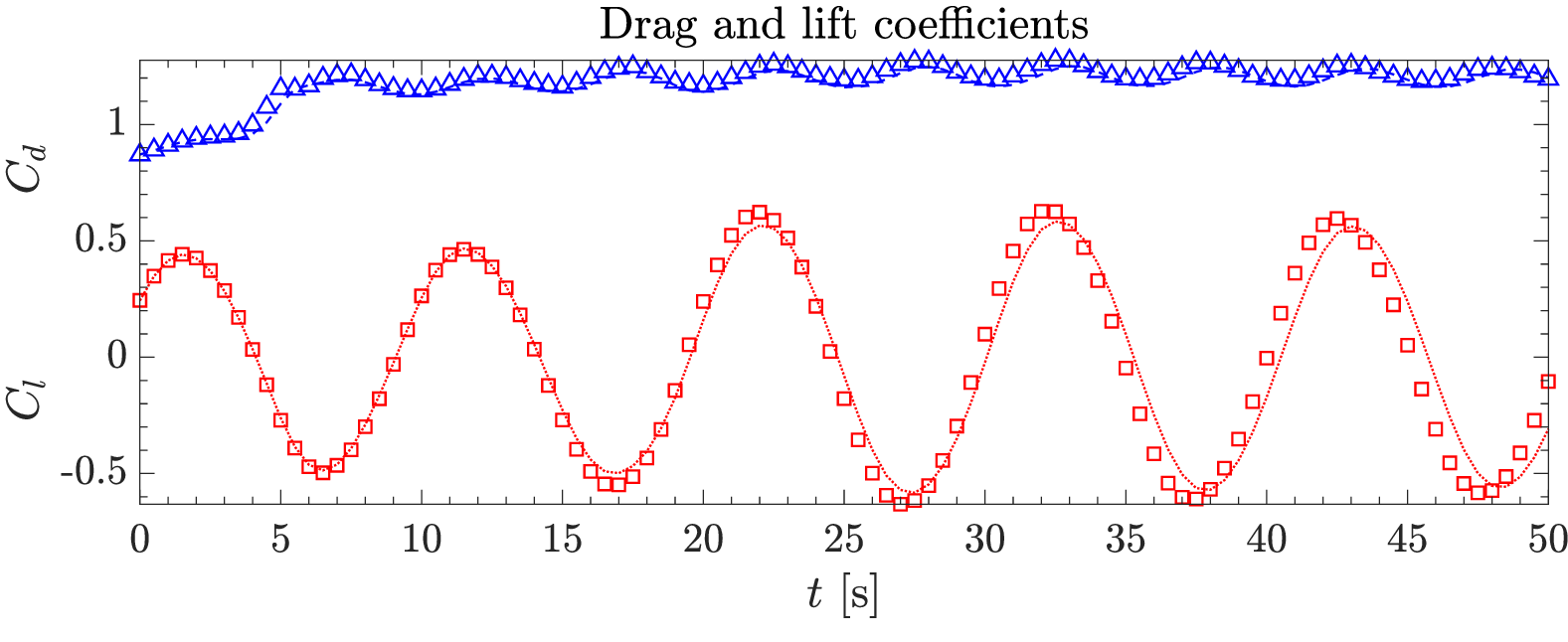}
	\caption{{\color{blue}$\boldsymbol{\bigtriangleup}$} Drag $C_d$ and {\color{red}$\boldsymbol{\Box}$} lift $C_l$ coefficients for the FLA solver (symbols) and the reference PSI-CELL solver (lines) for $Re = 100$ and $St = 0.1$ without evaporation.}
	\label{fig:plotforceCoeffsCompFLACIC_nonevap_Re100}
\end{figure}

Since the vortex street that develops in the flow is periodic, it is instructive to consider the associated period of the carrier flow field as the simulation evolves, since this will be directly affected by the momentum transfer from the droplet phase. This can be done by means of the drag $C_d$ and lift $C_l$ coefficients associated with the flow around a cylinder, defined respectively as \cite{ref:LiRybdylova2021}
\begin{subequations}
	\label{eq:force-coefficients}
	\begin{align}
		C_d (t) = \int_S \bm{e}_x \cdot \bm{P} \cdot \bm{n} \, dS \, , \label{eq:drag-coefficient} \\
		C_l (t) = \int_S \bm{e}_y \cdot \bm{P} \cdot \bm{n} \, dS \, , \label{eq:lift-coefficient}
	\end{align}
\end{subequations}
in which $\bm{P} = -p\bm{I} + (1 / Re) \left( \nabla \bm{u}  + \nabla \bm{u}^{\top} \right)$, $\bm{e}_x$ and $\bm{e}_y$ are the unit vectors in along the $x$ and $y$ coordinate axes respectively, and $\bm{n}$ is the unit normal to the circumference of the cylinder. Calculation of $C_d$ and $C_l$ for both the FLA and reference PSI-CELL cases is displayed in Figure \ref{fig:plotforceCoeffsCompFLACIC_nonevap_Re100}.
It is observed that as the flow field equilibrates once the droplet injections have started, both the FLA and reference PSI-CELL simulations adjust to steady-state values for the average drag coefficient and amplitude of the lift coefficient. The variation in the force coefficients of the reference PSI-CELL simulation is closely mirrored by those of the FLA in terms of both the amplitude and period, with the only notable difference being a slight decrease in the period of the flow in the FLA case by the end of the transient simulation periodic. This is as a result of the error in the momentum coupling source term depicted in Figure \ref{fig:plotprofCompFLACIC_nonevap_Re100} that arises due to the smoothing effect of the kernel regression procedure, which compounds as the transient simulation progresses, and eventually becomes notable in the period of the flow being affected. Despite the fact that this occurs, it is still notable that the FLA takes several periods of the flow to become even slightly different from the reference PSI-CELL case, and given the reduced droplet seeding used demonstrates the effectiveness of the FLA at reproducing the qualitative evolution of the transient flow at a much increased computational efficiency.

\subsection{Steady-state evaporating flow}
\label{sec:steady-state-evap-flow}

\begin{figure}[!ht]
	\begin{subfigure}[c]{\columnwidth}
		\includegraphics[width=\columnwidth,trim={0 10 5 15},clip] {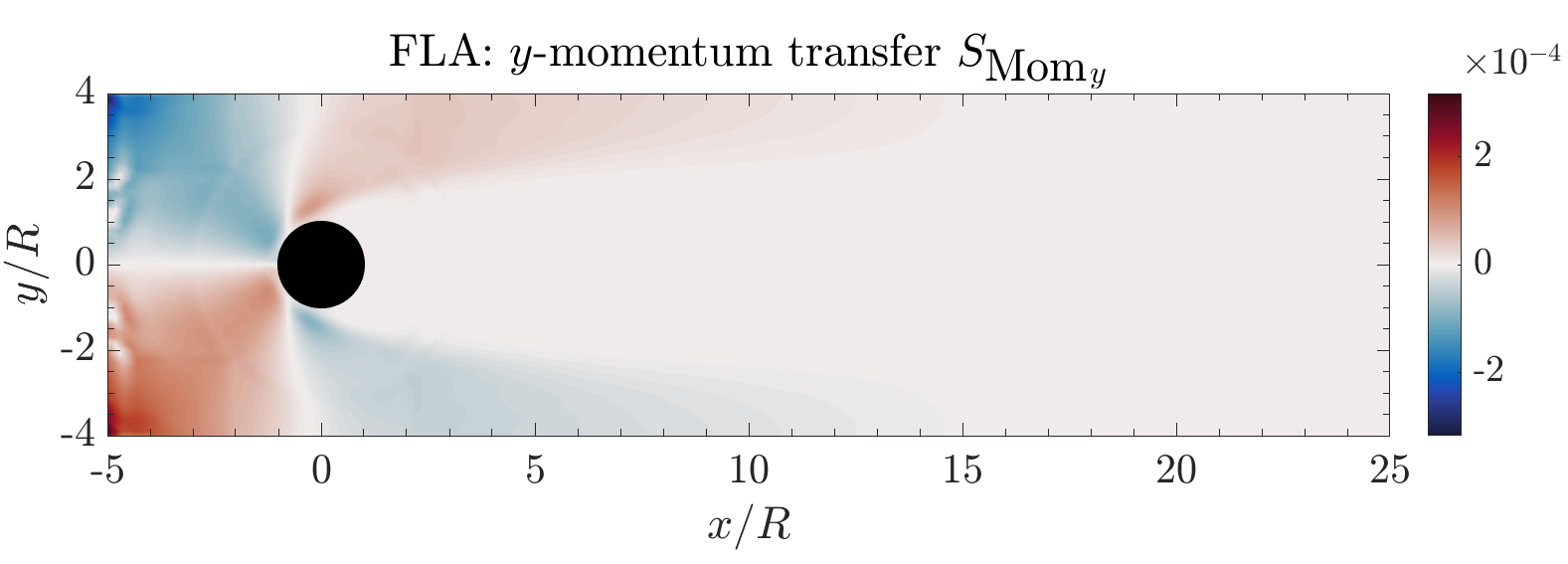}
		\caption{}
		\label{fig:fieldPlotFLA_evap_Re20_UTransY}
	\end{subfigure}
	\begin{subfigure}[c]{\columnwidth}
		\includegraphics[width=\columnwidth,trim={0 10 5 15},clip] {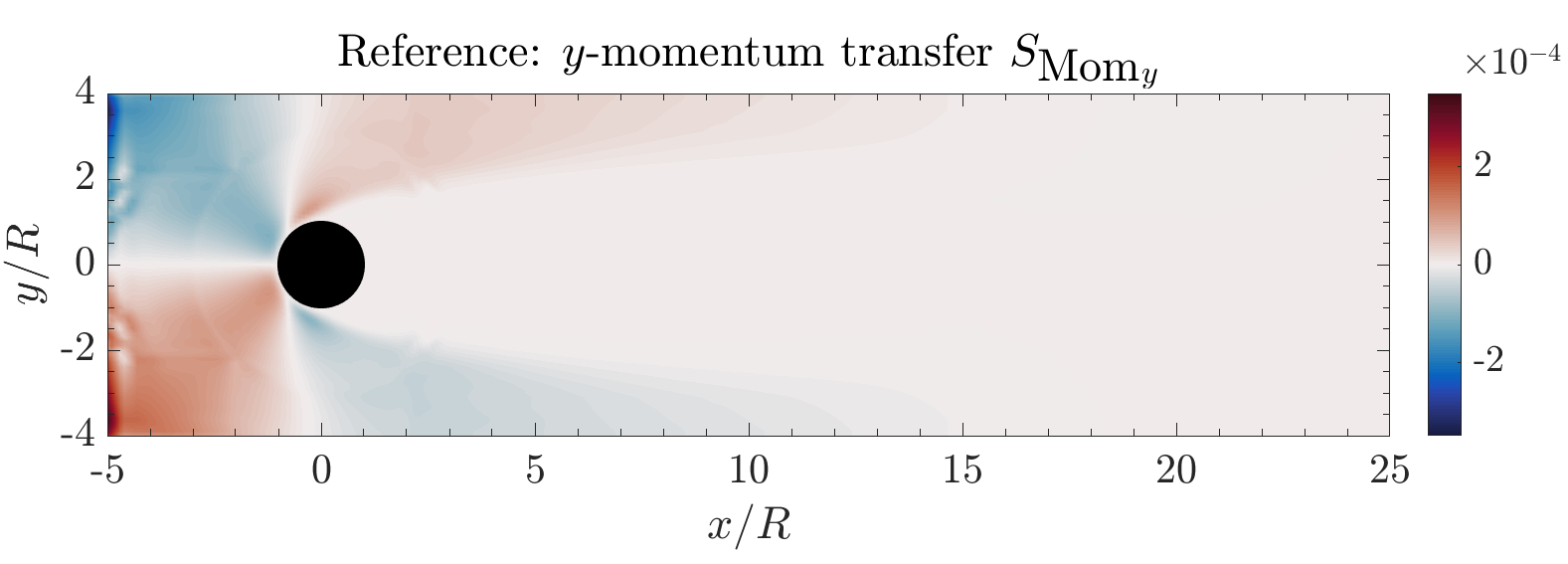}
		\caption{}
		\label{fig:fieldPlotCIC_evap_Re20_UTransY}
	\end{subfigure}
	\caption{Distribution of the $y-$component of the momentum transfer source term $\bm{S}_{\text{Mom}}$ for $Re = 20$ and $St = 0.1$ with evaporation at time $t = 50$ for: \eqref{fig:fieldPlotFLA_evap_Re20_UTransY} FLA solver; \eqref{fig:fieldPlotCIC_evap_Re20_UTransY} reference PSI-CELL solver.}
	\label{fig:fieldPlot_evap_Re20_UTransY}
\end{figure}

\begin{figure}[!ht]
	\begin{subfigure}[c]{\columnwidth}
		\includegraphics[width=\columnwidth,trim={0 0 0 0},clip] {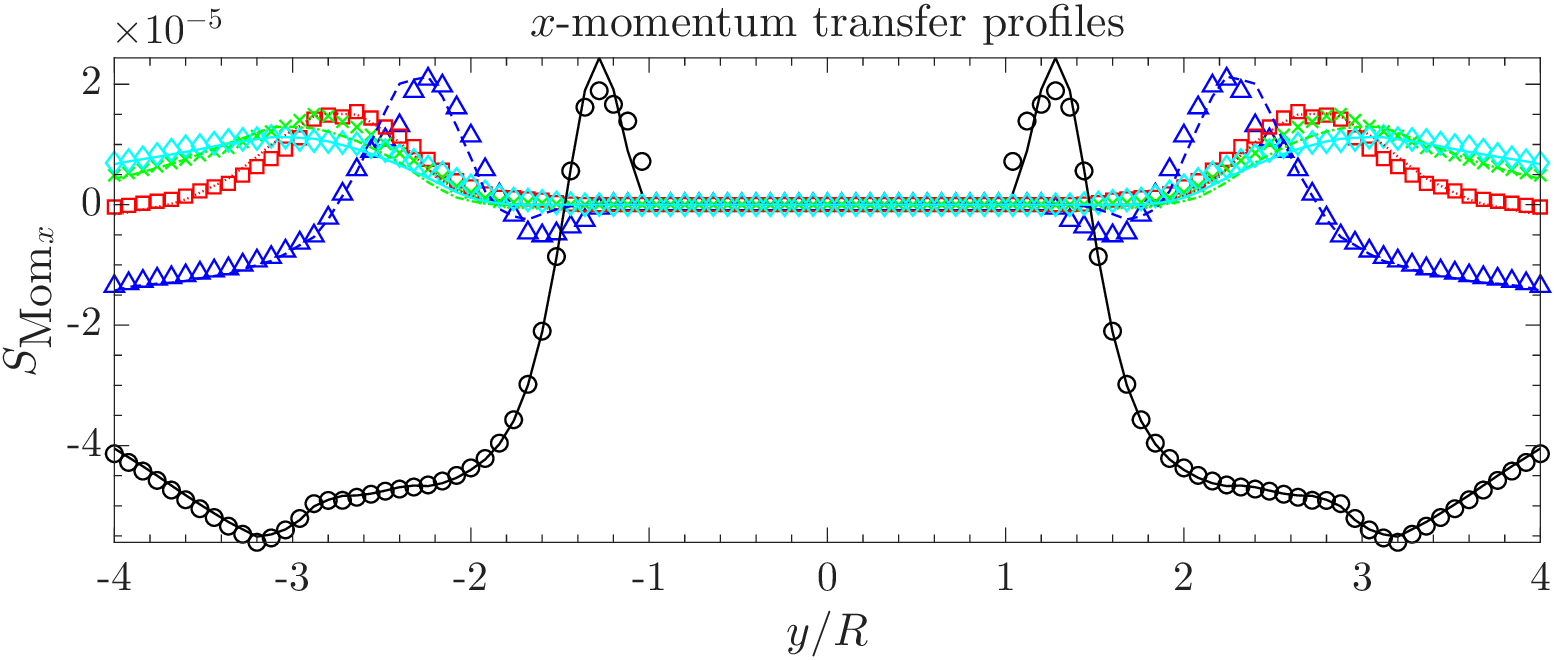}
		\caption{}
		\label{fig:plotprofCompFLACIC_evap_Re20_UTransX}
	\end{subfigure}
	\begin{subfigure}[c]{\columnwidth}
		\includegraphics[width=\columnwidth,trim={0 0 0 0},clip] {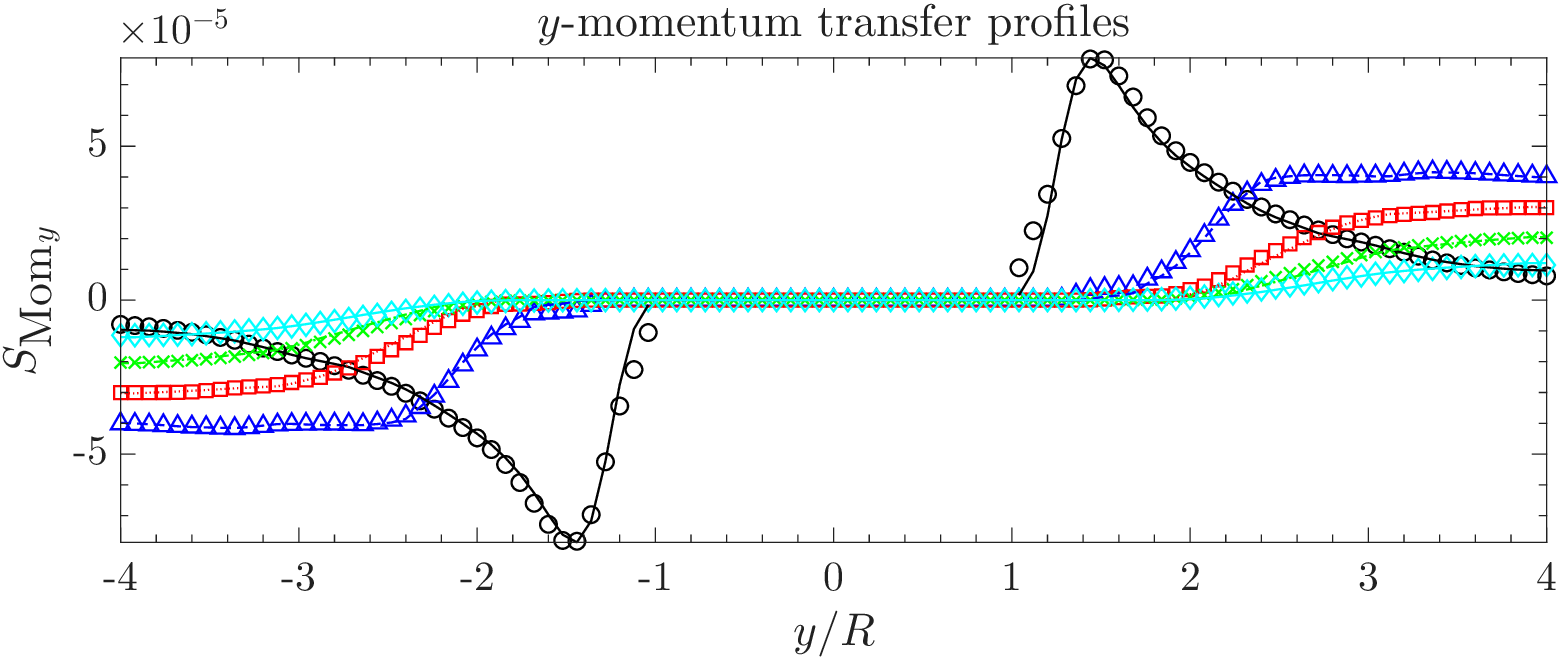}
		\caption{}
		\label{fig:plotprofCompFLACIC_evap_Re20_UTransY}
	\end{subfigure}
	\caption{Momentum transfer source term $\bm{S}_{\text{Mom}}$ profiles for the FLA solver (symbols) and the reference PSI-CELL solver (lines) at cross-sections
		{\color{black}$\boldsymbol{\bigcirc}$} ${x} / R = 0$, {\color{blue}$\boldsymbol{\bigtriangleup}$} ${x} / R = 3$,
		{\color{red}$\boldsymbol{\Box}$} ${x} / R = 6$,
		{\color{green}$\boldsymbol{\times}$} ${x} / R = 9$,
		{\color{cyan}$\boldsymbol{\diamond}$} ${x} / R = 12$
		for $Re = 20$ and $St = 0.1$ with evaporation at time $t = 50$ for: \eqref{fig:plotprofCompFLACIC_evap_Re20_UTransX} $x$-momentum; \eqref{fig:plotprofCompFLACIC_evap_Re20_UTransY} $y$-momentum.}
	\label{fig:plotprofCompFLACIC_evap_Re20_UTrans}
\end{figure}

To extend the capabilities of the FLA at handling interphase coupling in cases with {\color{corr1}additional} droplet physics, is it appropriate to consider the previous cases with the droplets now evaporating according to Eq.~\eqref{eq:evap-model}. This scenario then involves both mass and momentum coupling between the droplet phase and carrier flow, and accordingly requires that all parts of the source terms in Eqs.~\eqref{eq:coupling-terms-kernel} are calculated. To begin with, the interphase momentum transfer term is first focused upon as in the previous cases, with Figure \ref{fig:fieldPlot_evap_Re20_UTransY} displaying the distribution of the $y-$component for this term.
Whilst the carrier flow configuration is identical to that in Section \ref{sec:steady-state-non-evap-flow}, the effect of droplet evaporation causes the number density of droplets to decrease further away from the droplet injection inlet at $x / R = -5$, with the result being that at a certain distance from the inlet all droplets in the steady-state flow have evaporated. This is seen to occur by the location $x / R = 15$. Other than this change, the spatial distribution of the $y-$component of the momentum transfer is similar to that depicted for the non-evaporating case in Figure \ref{fig:fieldPlot_nonevap_Re20_UTransY}, and the difference between the distributions produced by the FLA and reference PSI-CELL simulations in Figures \eqref{fig:fieldPlotFLA_evap_Re20_UTransY} and \eqref{fig:fieldPlotCIC_nonevap_Re20_UTransY} respectively is indistinguishable to the eye. This is reflected in the cross-sectional profiles for these cases shown in Figure \ref{fig:plotprofCompFLACIC_evap_Re20_UTrans}.
It is seen that for both the $x-$components in Figure \eqref{fig:plotprofCompFLACIC_evap_Re20_UTransX} and $y-$component in Figure \eqref{fig:plotprofCompFLACIC_evap_Re20_UTransY}, the profiles show excellent quantitative agreement between the FLA case and the reference PSI-CELL solution, with the profiles downstream of the cylinder becoming increasing smaller in magnitude as the droplets evaporate. This highlights the efficacy of the FLA-based coupling procedure in this particularly stable scenario of steady-state flow with evaporation that minimizes transient effects.

\begin{figure}[!ht]
	\begin{subfigure}[c]{\columnwidth}
		\includegraphics[width=\columnwidth,trim={0 10 5 15},clip] {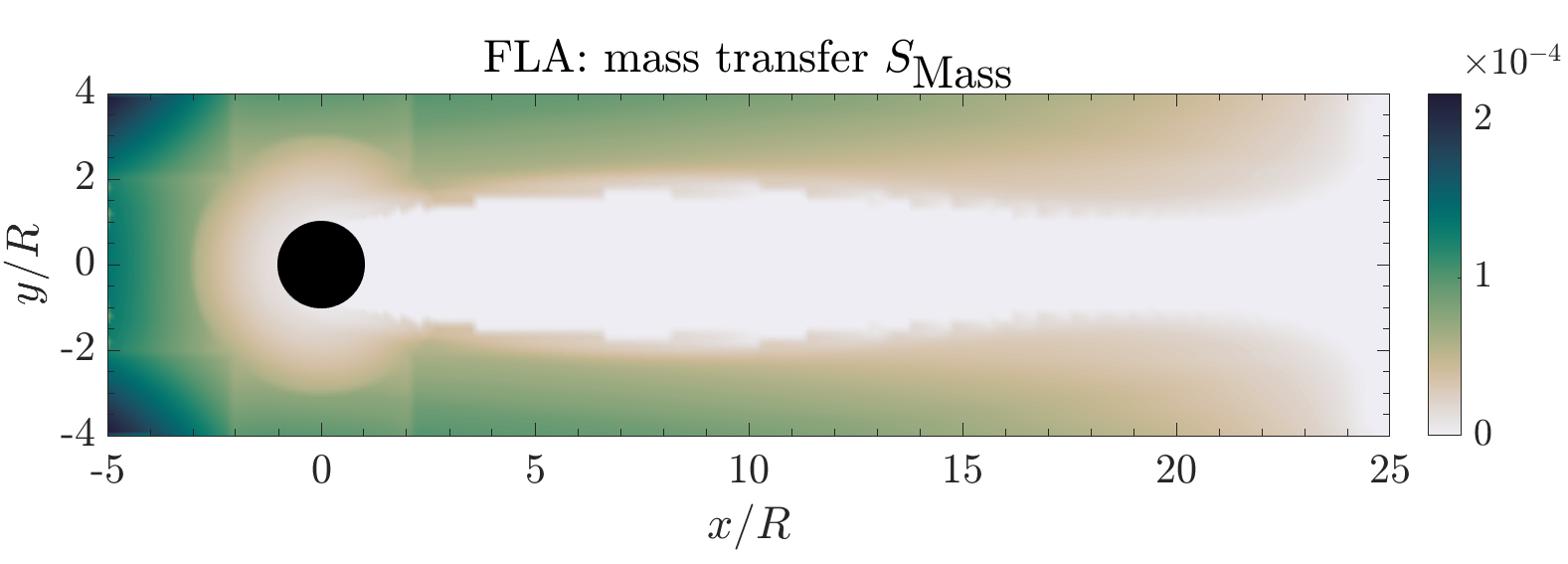}
		\caption{}
		\label{fig:fieldPlotFLA_evap_Re20_mTrans}
	\end{subfigure}
	\begin{subfigure}[c]{\columnwidth}
		\includegraphics[width=\columnwidth,trim={0 10 5 15},clip] {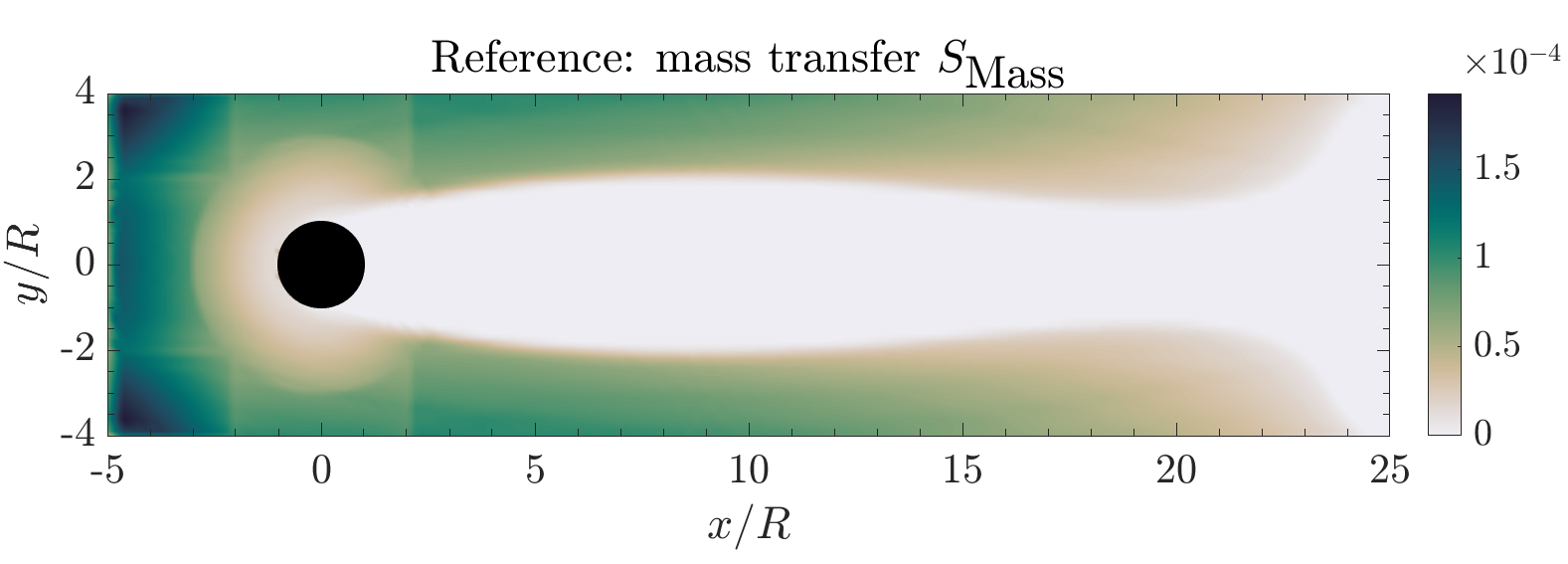}
		\caption{}
		\label{fig:fieldPlotCIC_evap_Re20_mTrans}
	\end{subfigure}
	\caption{Distribution of the mass transfer source term $S_{\text{Mass}}$ for $Re = 20$ and $St = 0.1$ with evaporation at time $t = 50$ for: \eqref{fig:fieldPlotFLA_evap_Re20_mTrans} FLA solver; \eqref{fig:fieldPlotCIC_evap_Re20_mTrans} reference PSI-CELL solver.}
	\label{fig:fieldPlot_evap_Re20_mTrans}
\end{figure}

\begin{figure}[!ht]
	\includegraphics[width=\columnwidth,trim={0 0 0 0},clip] {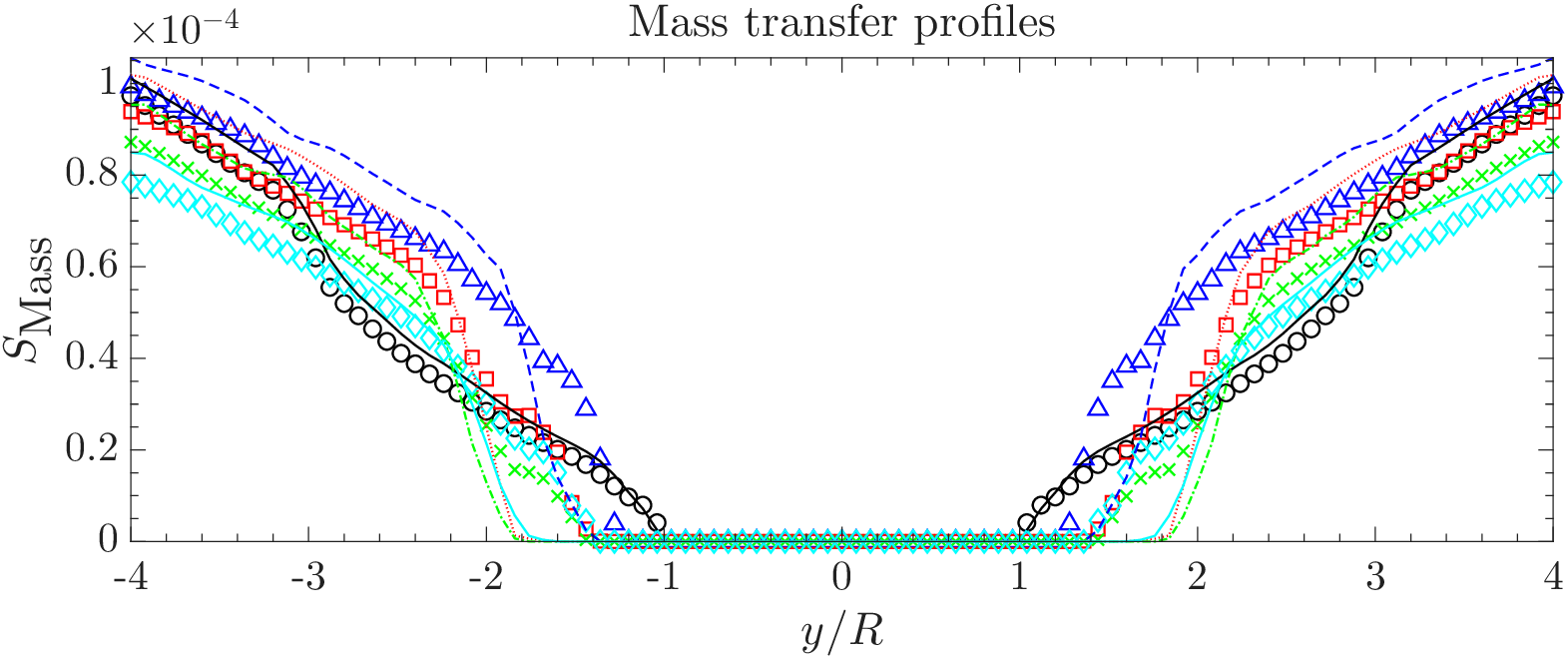}
	\caption{Mass transfer source term $S_{\text{Mass}}$ profiles for the FLA solver (symbols) and the reference PSI-CELL solver (lines) at cross-sections
		{\color{black}$\boldsymbol{\bigcirc}$} ${x} / R = 0$, {\color{blue}$\boldsymbol{\bigtriangleup}$} ${x} / R = 3$,
		{\color{red}$\boldsymbol{\Box}$} ${x} / R = 6$,
		{\color{green}$\boldsymbol{\times}$} ${x} / R = 9$,
		{\color{cyan}$\boldsymbol{\diamond}$} ${x} / R = 12$
		for $Re = 20$ and $St = 0.1$ with evaporation at time $t = 50$.}
	\label{fig:plotprofCompFLACIC_evap_Re20_mTrans}
\end{figure}

The novel information obtained in this case is through the interphase mass transfer term calculated through use of Eq.~\eqref{eq:mass-coupling_FLA}. Analogously to the momentum transfer term, the spatial distribution of $S_{\text{Mass}}$ is first examined in Figure \ref{fig:fieldPlotCIC_evap_Re20_mTrans}.
For both the FLA case in Figure \eqref{fig:fieldPlotFLA_evap_Re20_mTrans} and reference PSI-CELL solution in Figure \eqref{fig:fieldPlotCIC_evap_Re20_mTrans}, the interphase mass transfer is highest in the vicinity of the droplet injection at $x / R = -5$, and subsequently diminishes in magnitude as the droplets evaporate whilst travelling across the domain. This resembles the droplet number density distribution along the wake of the cylinder, although the mass transfer decreases close to the cylinder itself as the droplet field splits to pass around the cylinder. The increased droplet velocity in this region caused by the droplets having to flow around the obstacle means that they spent less time in the vicinity of the cylinder, and since the evaporation model in Eq.~\eqref{eq:evap-model} is constant in time, therefore resulting in a lower accumulation of mass transfer. The only apparent difference between the FLA and reference PSI-CELL simulations is along the edge of the droplet field, where the finite size of the kernel smoothing length causes the FLA mass transfer field to extend slightly further into the wake than in the reference PSI-CELL case. It should be noted however that this is for low values of the mass transfer term, and therefore does not contribute significantly to the overall mass coupling between the phases. This is exemplified in the cross-sectional profiles of the interphase mass transfer term illustrated in Figure \ref{fig:plotprofCompFLACIC_evap_Re20_mTrans}.
It can be seen that the mass transfer increases towards the edges of the domain, with there being no contribution in the region $y / R \in [-1,1]$ which constitutes the cylinder wake. The mass transfer is generally well captured by the FLA within the interior of the droplet field, with the majority of the error in the reconstruction arising along the edge of the droplet field around the cylinder wake. This is seen to occur within the regions $y \in \pm [1,2]$ and is characterised by the reference PSI-CELL solutions having a higher gradient and decreasing to zero more quickly than the FLA case, consistent with the interpretation of this effect arising from the kernel regression procedure which is unable to accurately represent high spatial gradients.

\begin{figure}[!ht]
	\begin{subfigure}[c]{\columnwidth}
		\includegraphics[width=\columnwidth,trim={0 10 5 15},clip] {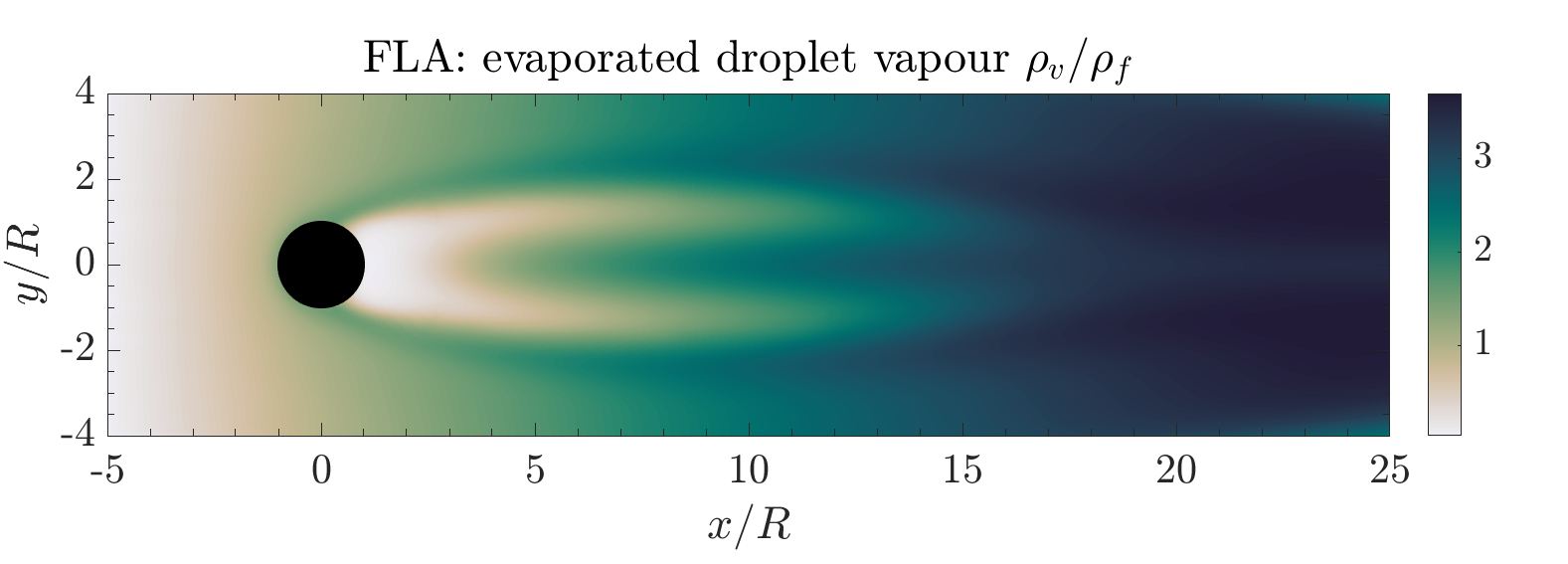}
		\caption{}
		\label{fig:fieldPlotFLA_evap_Re20_rhod}
	\end{subfigure}
	\begin{subfigure}[c]{\columnwidth}
		\includegraphics[width=\columnwidth,trim={0 10 5 15},clip] {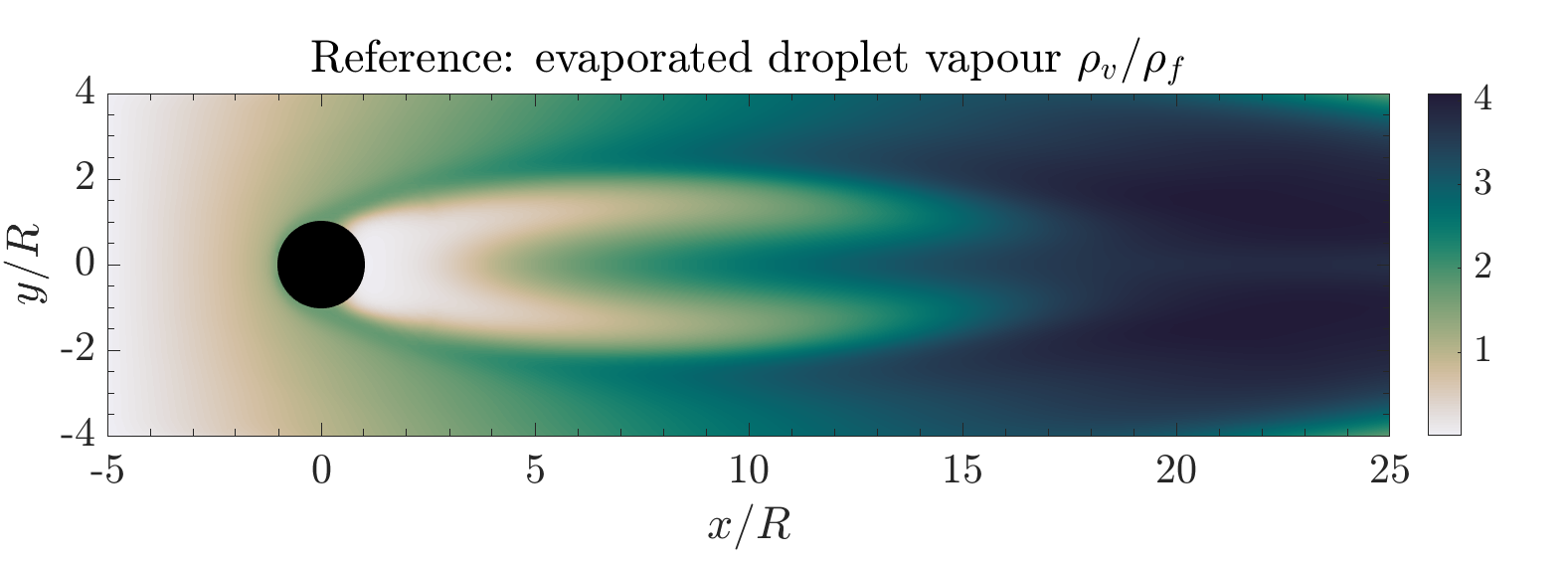}
		\caption{}
		\label{fig:fieldPlotCIC_evap_Re20_rhod}
	\end{subfigure}
	\caption{Distribution of the evaporated droplet vapour $\rho_v / \rho_f$ for $Re = 20$ and $St = 0.1$ with evaporation at time $t = 50$ for: \eqref{fig:fieldPlotFLA_evap_Re20_rhod} FLA solver; \eqref{fig:fieldPlotCIC_evap_Re20_rhod} reference PSI-CELL solver.}
	\label{fig:fieldPlot_evap_Re20_rhod}
\end{figure}

\begin{figure}[!ht]
	\includegraphics[width=\columnwidth,trim={0 0 0 0},clip] {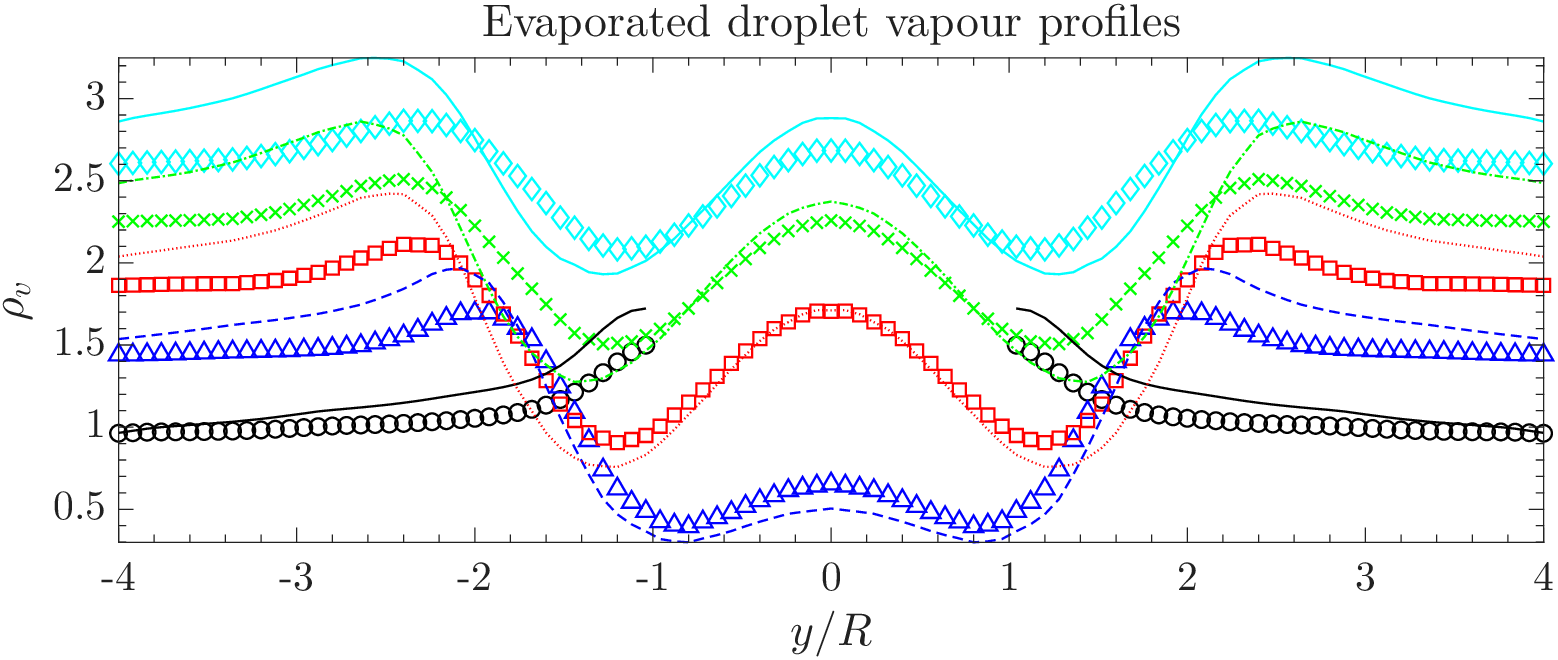}
	\caption{Evaporated droplet vapour $\rho_v$ profiles for the FLA solver (symbols) and the reference PSI-CELL solver (lines) at cross-sections
		{\color{black}$\boldsymbol{\bigcirc}$} ${x} / R = 0$, {\color{blue}$\boldsymbol{\bigtriangleup}$} ${x} / R = 3$,
		{\color{red}$\boldsymbol{\Box}$} ${x} / R = 6$,
		{\color{green}$\boldsymbol{\times}$} ${x} / R = 9$,
		{\color{cyan}$\boldsymbol{\diamond}$} ${x} / R = 12$
		for $Re = 20$ and $St = 0.1$ with evaporation at time $t = 50$.}
	\label{fig:plotprofCompFLACIC_evap_Re20_rhod}
\end{figure}

An additional consideration that can be made with mass coupling is for the evolution of the droplet vapour concentration field $\rho_v$ as governed by the scalar transport equation \eqref{eq:vapour-transport}, and this is depicted in Figure \ref{fig:fieldPlot_evap_Re20_rhod} relative to the density of the carrier flow $\rho_f$.
It is seen that with the droplet evaporation occurring as droplets travel across the domain, the resultant vapour accumulates such that the concentration generally increases with the distance from the injection interval at $x / R = -5$. The exception to this is the cylinder wake, for which the region immediately downstream of the cylinder remains devoid of any vapour, but the rest of the wake is affected by vapour diffusing and being advected into it. This is most notable along the centreline of the flow, where some backflow occurs once the separate regions of the droplet field have rejoined after $x / R \approx 20$. Despite the occurrence of this phenomenon, as with the distribution of the $y-$component of the interphase momentum transfer field for this case in Figure \ref{fig:fieldPlot_evap_Re20_UTransY}, the FLA and reference PSI-CELL simulations for the evaporated droplet vapour fields in Figures \eqref{fig:fieldPlotFLA_evap_Re20_rhod} and \eqref{fig:fieldPlotCIC_evap_Re20_rhod} show a high level of agreement. This is further exemplified through the corresponding cross-sectional profiles as illustrated in Figure \ref{fig:plotprofCompFLACIC_evap_Re20_rhod}.
The vapour transport in the wake region where no droplets exist is well predicted by the FLA, whilst the interior of the droplet field exhibits a degree of underprediction for the profiles further downstream of the cylinder. This is a direct result of the interphase mass transfer also being underpredicted by the FLA in this region, as observed in Figure \ref{fig:plotprofCompFLACIC_evap_Re20_mTrans}. Nonetheless, the qualitative behaviour remains well captured by the FLA-based coupling mechanism, and produces the correct spatial distribution of vapour as evidenced in Figure \ref{fig:fieldPlot_evap_Re20_rhod}.

{\color{rev1}
The relative $l^2$-norm of the momentum and mass transfer source terms and evaporated droplet vapour for the different FLA droplet seedings and injection intervals is displayed in Table \ref{table:l2-norm_evap_Re20} for time $t = 50$.
\begin{table}[!ht]
	\begin{center}
		\begin{tabular}{cccc}
			\toprule
			& $NP = 50$ & $NP = 100$ & $NP = 200$ \\
			& $\Delta t_i = 0.2$ & $\Delta t_i = 0.1$ & $\Delta t_i = 0.05$ \\
			\midrule
			$\text{Error}_{l^2}[{S_{\text{Mom}}}_x]$ & 0.4591 & 0.2114 & 0.2417 \\ 
			$\text{Error}_{l^2}[{S_{\text{Mom}}}_y]$ & 0.3674 & 0.2145 & 0.2140 \\
			$\text{Error}_{l^2}[S_{\text{Mass}}]$ & 0.3707 & 0.2218 & 0.1893 \\
			$\text{Error}_{l^2}[\rho_v / \rho_f]$ & 0.1481 & 0.1018 & 0.1156 \\
			\bottomrule
		\end{tabular} 
	\end{center}
	\caption{{\color{rev1}Relative $l^2$-norm for the components of the momentum transfer source term $\bm{S}_{\text{Mom}}$, mass transfer source term $S_{\text{Mass}}$, and  evaporated droplet vapour $\rho_v / \rho_f$ evaluated using different droplet seedings for $Re = 20$ and $St = 0.1$ with evaporation at time $t = 50$.}}
	\label{table:l2-norm_evap_Re20}
\end{table}
As with the non-evaporating steady state case in Section \ref{sec:steady-state-non-evap-flow}, a notable increase in accuracy is gained by increasing the injection droplet seeding from $NP = 50$ to $NP = 100$, however in the case of evaporating droplets it is seen here that further increasing the injection seeding to $NP = 200$ results in negligible additional reduction in error. This substantiates that $NP = 100$ is the ideal balance of accuracy and efficiency for applying the kernel regression procedure in steady-state flows.
}

\subsection{Transient evaporating flow}
\label{sec:transient-evap-flow}

\begin{figure}[!ht]
	\begin{subfigure}[c]{\columnwidth}
		\includegraphics[width=\columnwidth,trim={0 10 5 15},clip] {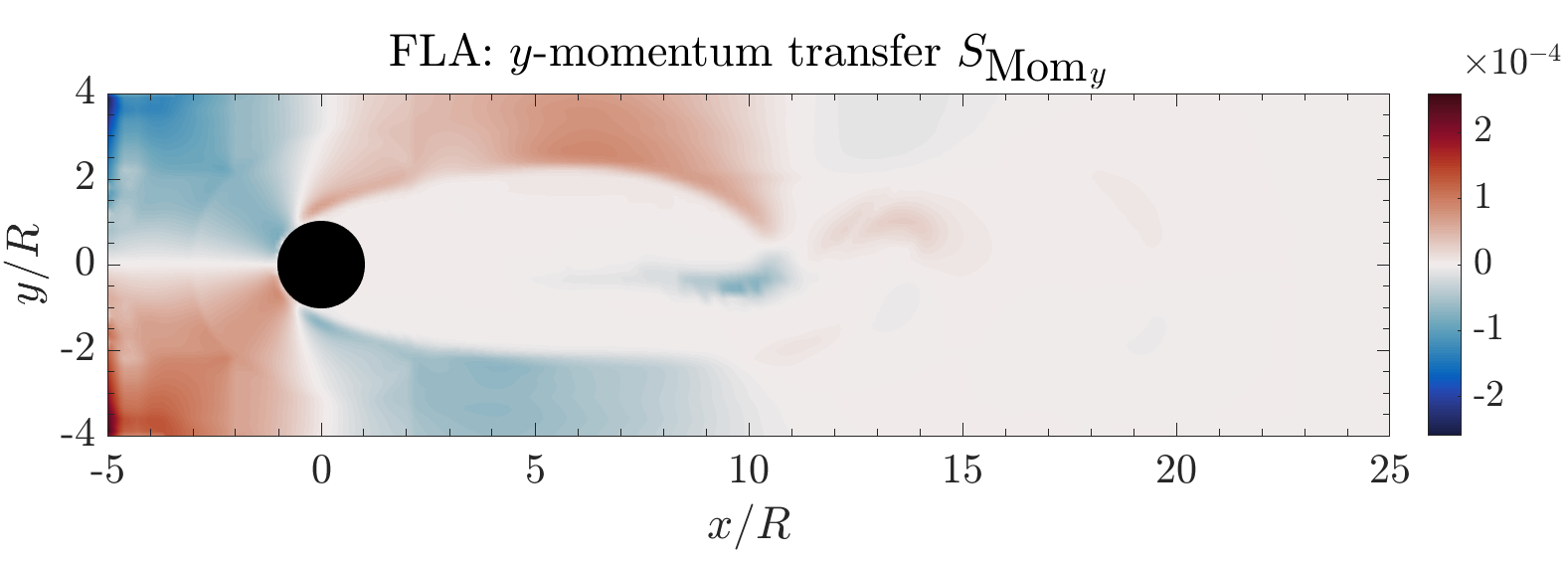}
		\caption{}
		\label{fig:fieldPlotFLA_evap_Re100_UTransY}
	\end{subfigure}
	\begin{subfigure}[c]{\columnwidth}
		\includegraphics[width=\columnwidth,trim={0 10 5 15},clip] {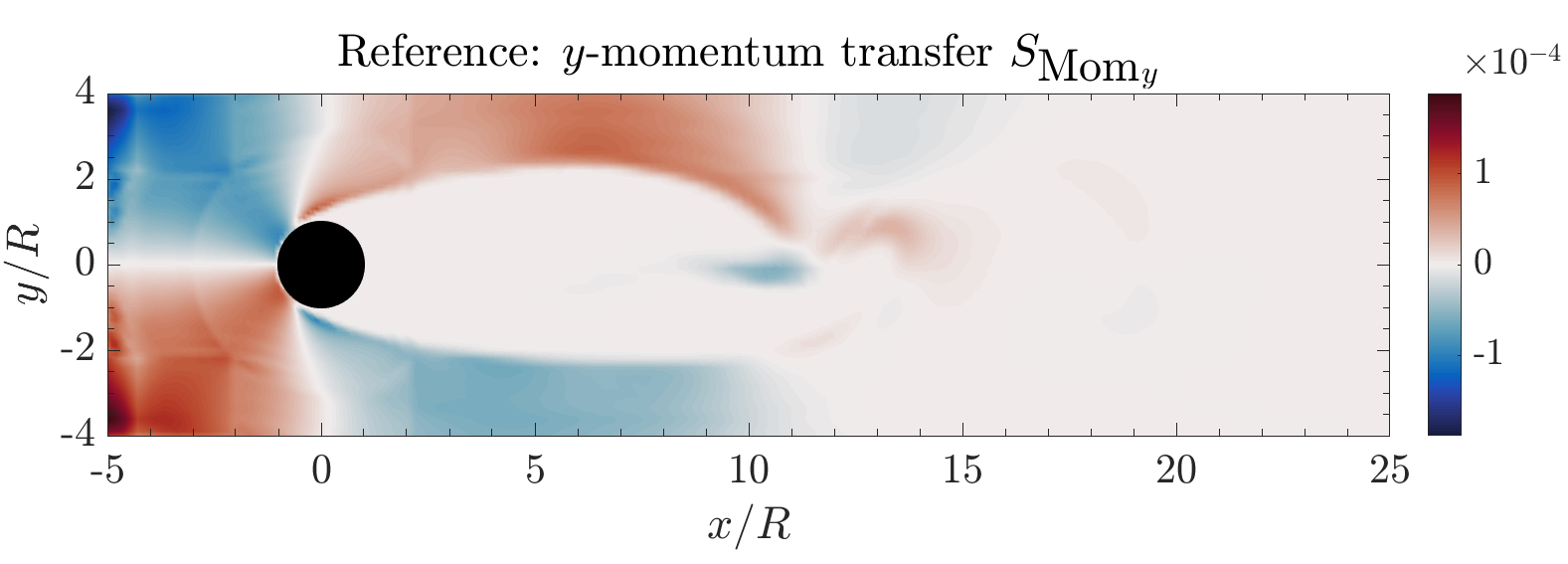}
		\caption{}
		\label{fig:fieldPlotCIC_evap_Re100_UTransY}
	\end{subfigure}
	\caption{Distribution of the $y-$component of the momentum transfer source term $S_{\text{Mom}}$ for $Re = 100$ and $St = 0.5$ with evaporation at time $t = 50$ for: \eqref{fig:fieldPlotFLA_evap_Re100_UTransY} FLA solver; \eqref{fig:fieldPlotCIC_evap_Re100_UTransY} reference PSI-CELL solver.}
	\label{fig:fieldPlot_evap_Re100_UTransY}
\end{figure}

\begin{figure}[!ht]
	\begin{subfigure}[c]{\columnwidth}
		\includegraphics[width=\columnwidth,trim={0 0 0 0},clip] {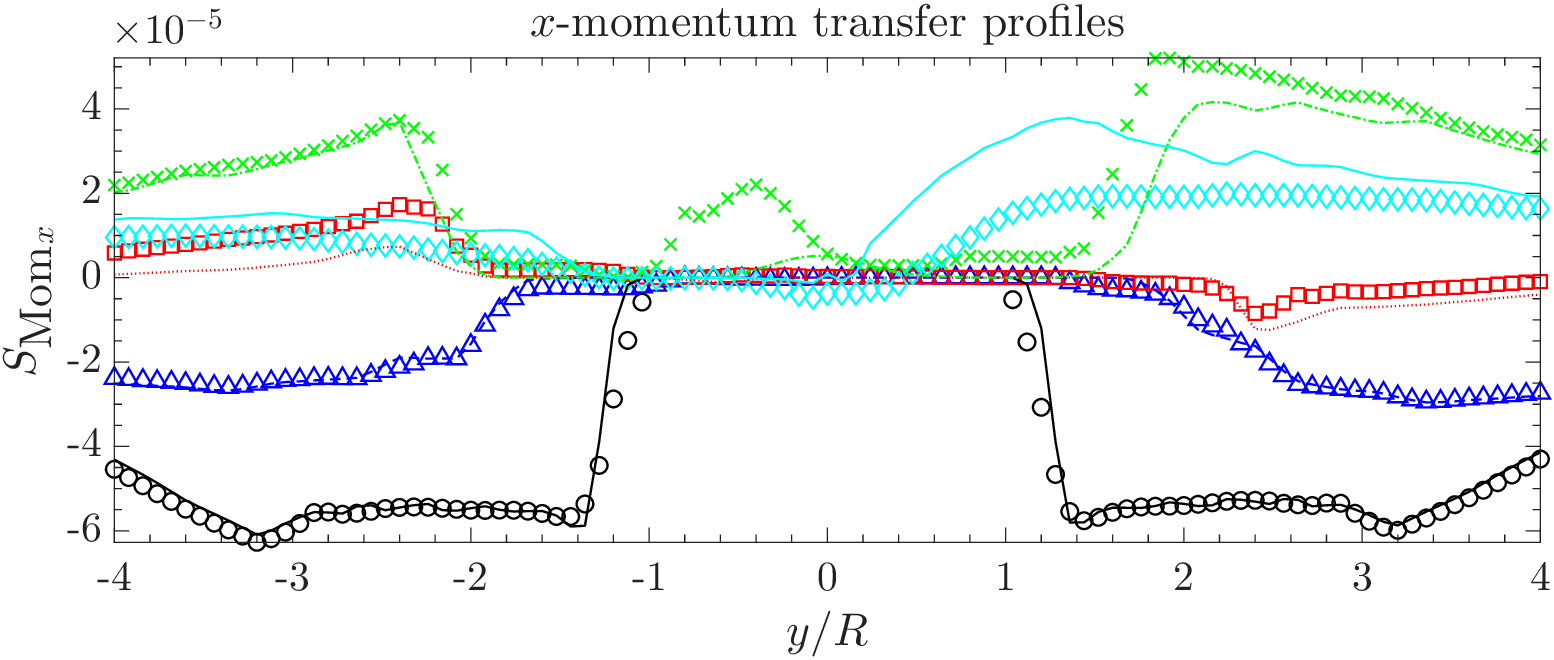}
		\caption{}
		\label{fig:plotprofCompFLACIC_evap_Re100_UTransX}
	\end{subfigure}
	\begin{subfigure}[c]{\columnwidth}
		\includegraphics[width=\columnwidth,trim={0 0 0 0},clip] {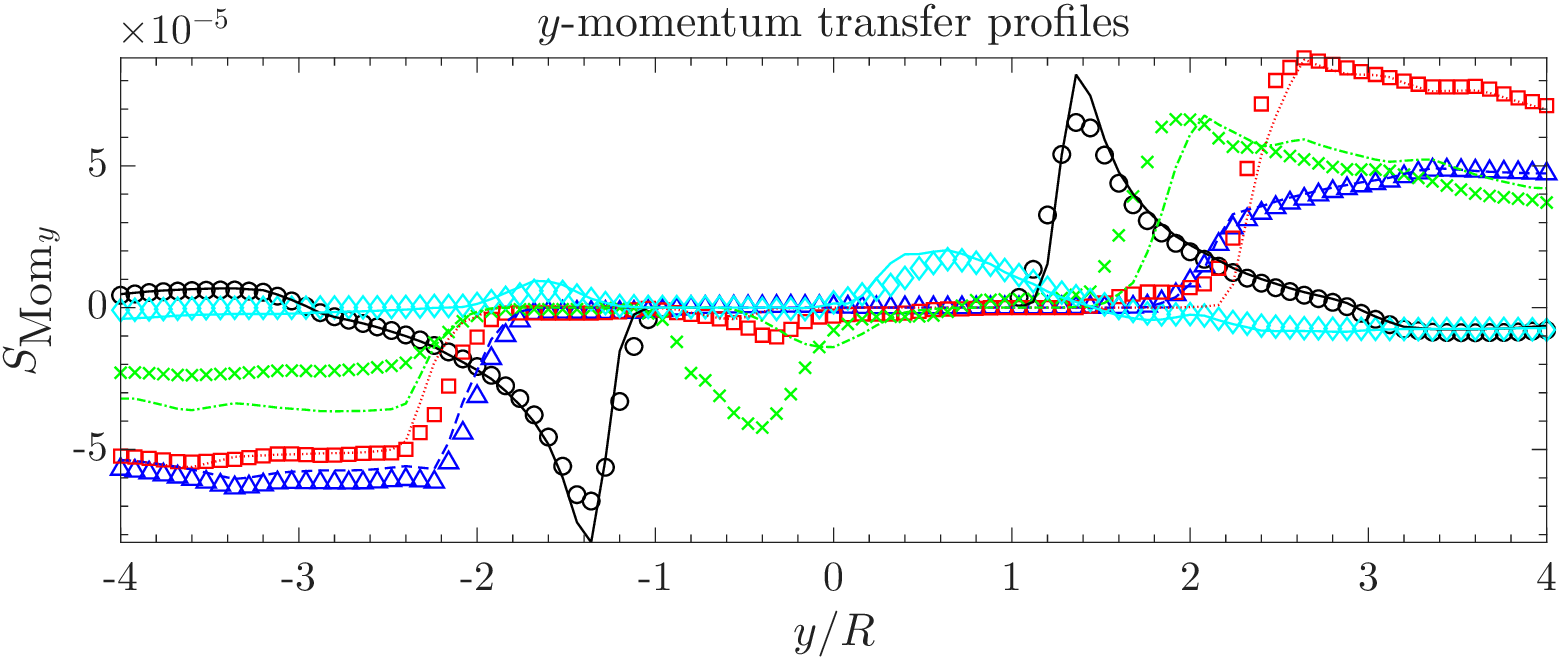}
		\caption{}
		\label{fig:plotprofCompFLACIC_evap_Re100_UTransY}
	\end{subfigure}
	\caption{Momentum transfer source term $S_{\text{Mom}}$ profiles for the FLA solver (symbols) and the reference PSI-CELL solver (lines) at cross-sections
		{\color{black}$\boldsymbol{\bigcirc}$} ${x} / R = 0$, {\color{blue}$\boldsymbol{\bigtriangleup}$} ${x} / R = 3$,
		{\color{red}$\boldsymbol{\Box}$} ${x} / R = 6$,
		{\color{green}$\boldsymbol{\times}$} ${x} / R = 9$,
		{\color{cyan}$\boldsymbol{\diamond}$} ${x} / R = 12$
		for $Re = 100$ and $St = 0.5$ with evaporation at time $t = 50$ for: \eqref{fig:plotprofCompFLACIC_evap_Re100_UTransX} $x$-momentum; \eqref{fig:plotprofCompFLACIC_evap_Re100_UTransY} $y$-momentum.}
	\label{fig:plotprofCompFLACIC_evap_Re100_UTrans}
\end{figure}

The final configuration to be considered is the most complex scenario of a transient carrier flow with droplet evaporation present. The combined effects of both mass and momentum transfer needing to be accounted for in an unsteady flow regime provides a stringent test of solver capabilities, and is most representative of the general configuration which is encountered in the majority of relevant fluid dynamical problems. In this case, the increased value of $St = 0.5$ is used to illustrate the effects of higher droplet inertia on the interphase coupling mechanism. As before, firstly the momentum coupling within this case is addressed. Figure \ref{fig:fieldPlot_evap_Re100_UTransY} displays the spatial distribution of the momentum transfer source term in the $y$ direction, and is depicted in Figure \eqref{fig:fieldPlotFLA_nonevap_Re100_UTransY} for the FLA procedure, and Figure \eqref{fig:fieldPlotCIC_evap_Re100_UTransY} for the reference PSI-CELL solver.
As with the steady-state evaporating case in Figure \ref{fig:fieldPlot_evap_Re20_UTransY}, it is observed that the evaporation model utilised causes all the droplets to evaporate by the location $x / R = 15$. Despite the Reynolds number being set at $Re = 100$ as in Section \ref{sec:transient-non-evap-flow}, the higher droplet inertia of $St = 0.5$ causes notably different behaviour than the previously observed vortex street, with the vortical structures being absent from Figure \ref{fig:fieldPlot_evap_Re100_UTransY}. Notwithstanding this, there are transient features which exist in the configuration, and the reconstruction provided by the FLA is seen to maintain a qualitatively accurate representation of the spatial distribution of the momentum coupling source term. The only observable difference is some slight diffusion of the droplet field along the centreline of the cylinder wake for the FLA case in Figure \eqref{fig:fieldPlotFLA_evap_Re100_UTransY}. This is detailed further in Figure \ref{fig:plotprofCompFLACIC_evap_Re100_UTrans}, with a comparison between the FLA and reference PSI-CELL simulations shown for the selected cross-sectional profiles for the $x-$ and $y-$components of the momentum transfer source term in Figures \eqref{fig:plotprofCompFLACIC_evap_Re100_UTransX} and \eqref{fig:plotprofCompFLACIC_evap_Re100_UTransY} respectively.
Whilst a fair degree of qualitative accuracy is maintained, it is seen that for the profiles which are a greater distance downstream from the cylinder a decreased level of accuracy in reconstructing the momentum transfer profile is achieved for the $x-$component, whilst the $y-$component generally maintains good quantitative accuracy. Although this emphasises the difficulties associated with capturing the full behaviour present in transient evaporating flows, it serves to illustrate that momentum coupling is nonetheless generally handled satisfactorily by the FLA framework even in this demanding case.

\begin{figure}[!ht]
	\begin{subfigure}[c]{\columnwidth}
		\includegraphics[width=\columnwidth,trim={0 10 5 15},clip] {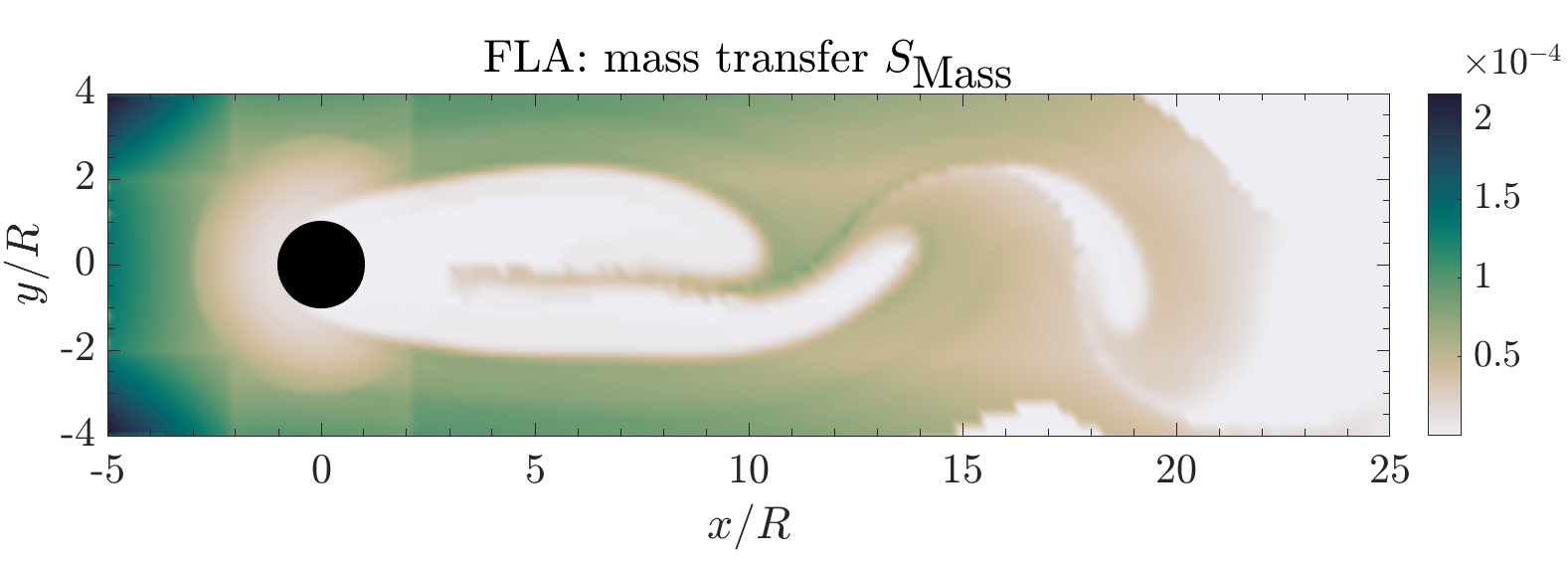}
		\caption{}
		\label{fig:fieldPlotFLA_evap_Re100_mTrans}
	\end{subfigure}
	\begin{subfigure}[c]{\columnwidth}
		\includegraphics[width=\columnwidth,trim={0 10 5 15},clip] {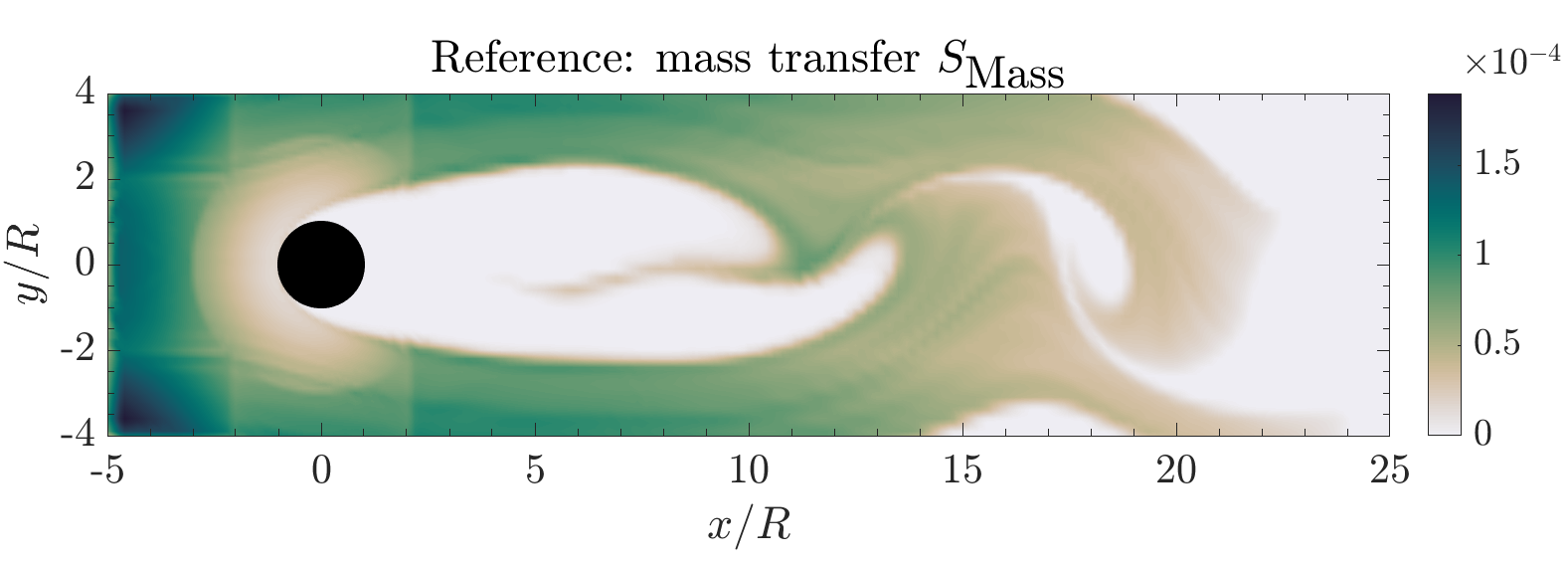}
		\caption{}
		\label{fig:fieldPlotCIC_evap_Re100_mTrans}
	\end{subfigure}
	\caption{Distribution of the mass transfer source term $S_{\text{Mass}}$ for $Re = 100$ and $St = 0.5$ with evaporation at time $t = 50$ for: \eqref{fig:fieldPlotFLA_evap_Re100_mTrans} FLA solver; \eqref{fig:fieldPlotCIC_evap_Re100_mTrans} reference PSI-CELL solver.}
	\label{fig:fieldPlot_evap_Re100_mTrans}
\end{figure}

\begin{figure}[!ht]
	\includegraphics[width=\columnwidth,trim={0 0 0 0},clip] {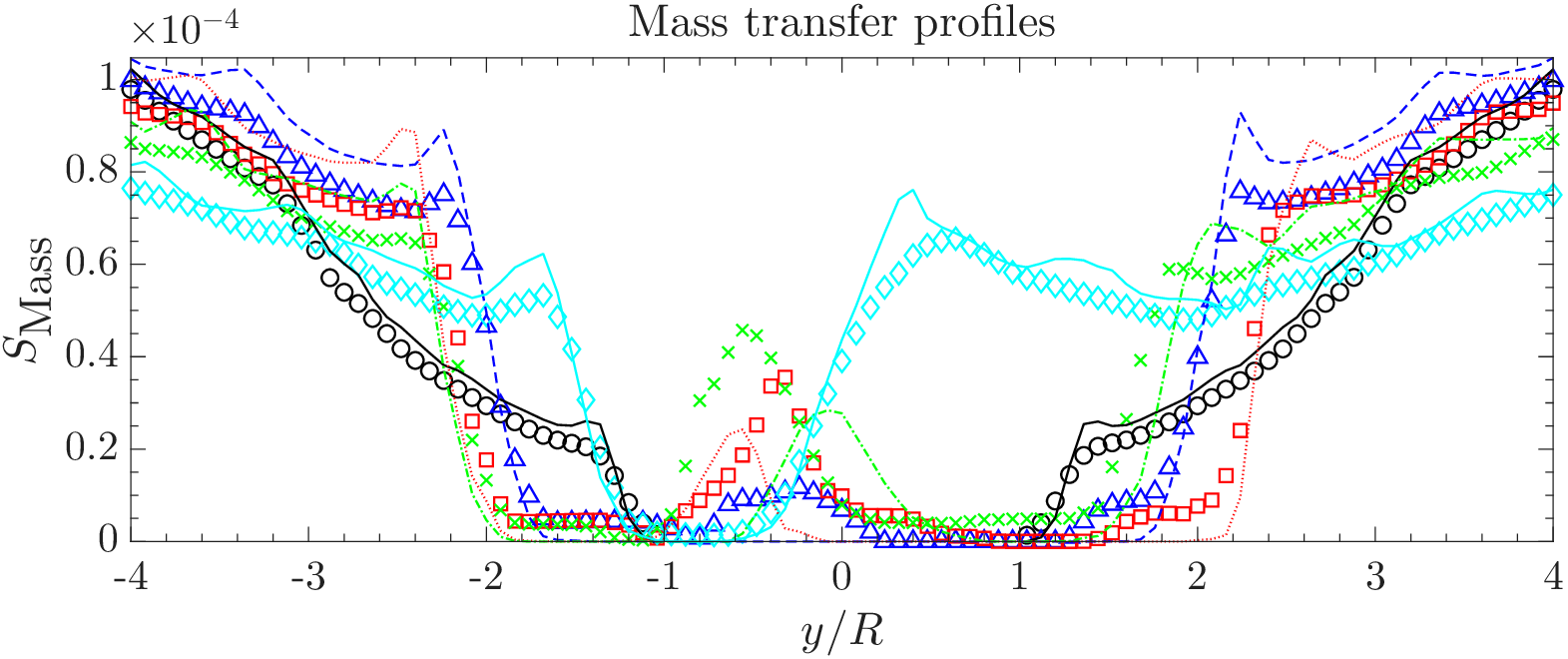}
	\caption{Mass transfer source term $S_{\text{Mass}}$ profiles for the FLA solver (symbols) and the reference PSI-CELL solver (lines) at cross-sections
		{\color{black}$\boldsymbol{\bigcirc}$} ${x} / R = 0$, {\color{blue}$\boldsymbol{\bigtriangleup}$} ${x} / R = 3$,
		{\color{red}$\boldsymbol{\Box}$} ${x} / R = 6$,
		{\color{green}$\boldsymbol{\times}$} ${x} / R = 9$,
		{\color{cyan}$\boldsymbol{\diamond}$} ${x} / R = 12$
		for $Re = 100$ and $St = 0.5$ with evaporation at time $t = 50$.}
	\label{fig:plotprofCompFLACIC_evap_Re100_mTrans}
\end{figure}

\begin{figure}[!ht]
	\begin{subfigure}[c]{\columnwidth}
		\includegraphics[width=\columnwidth,trim={0 10 5 15},clip] {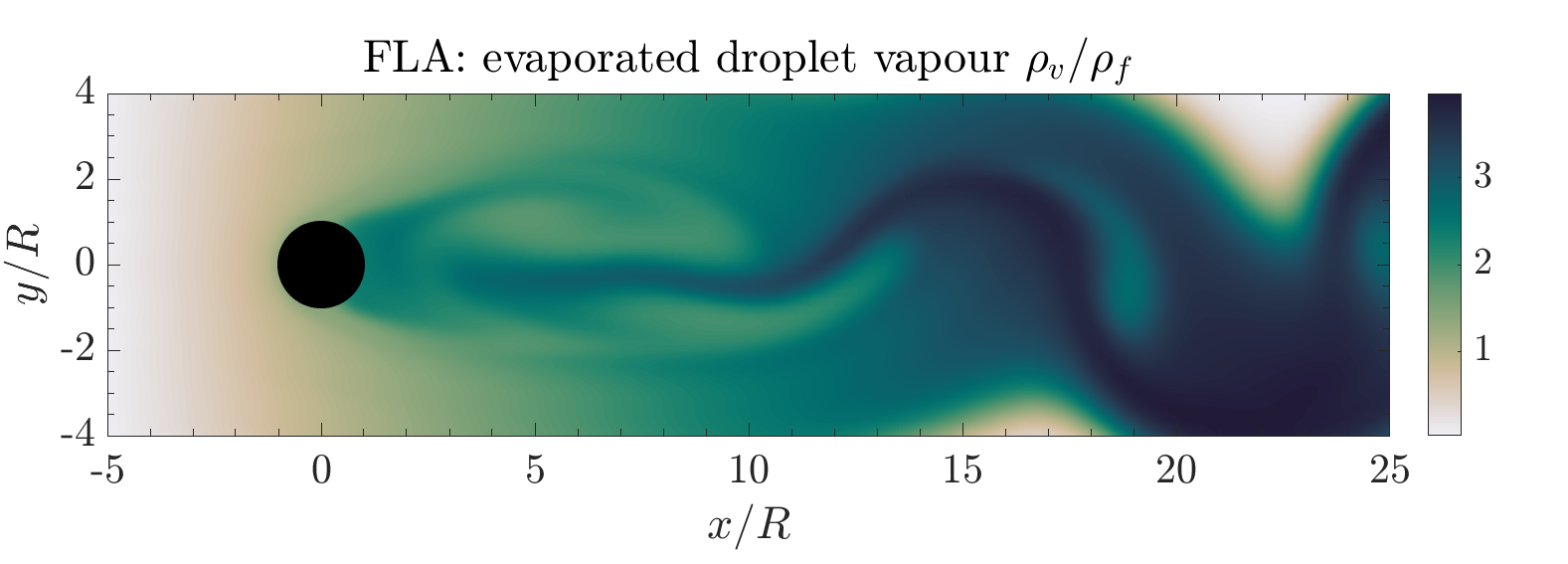}
		\caption{}
		\label{fig:fieldPlotFLA_evap_Re100_rhod}
	\end{subfigure}
	\begin{subfigure}[c]{\columnwidth}
		\includegraphics[width=\columnwidth,trim={0 10 5 15},clip] {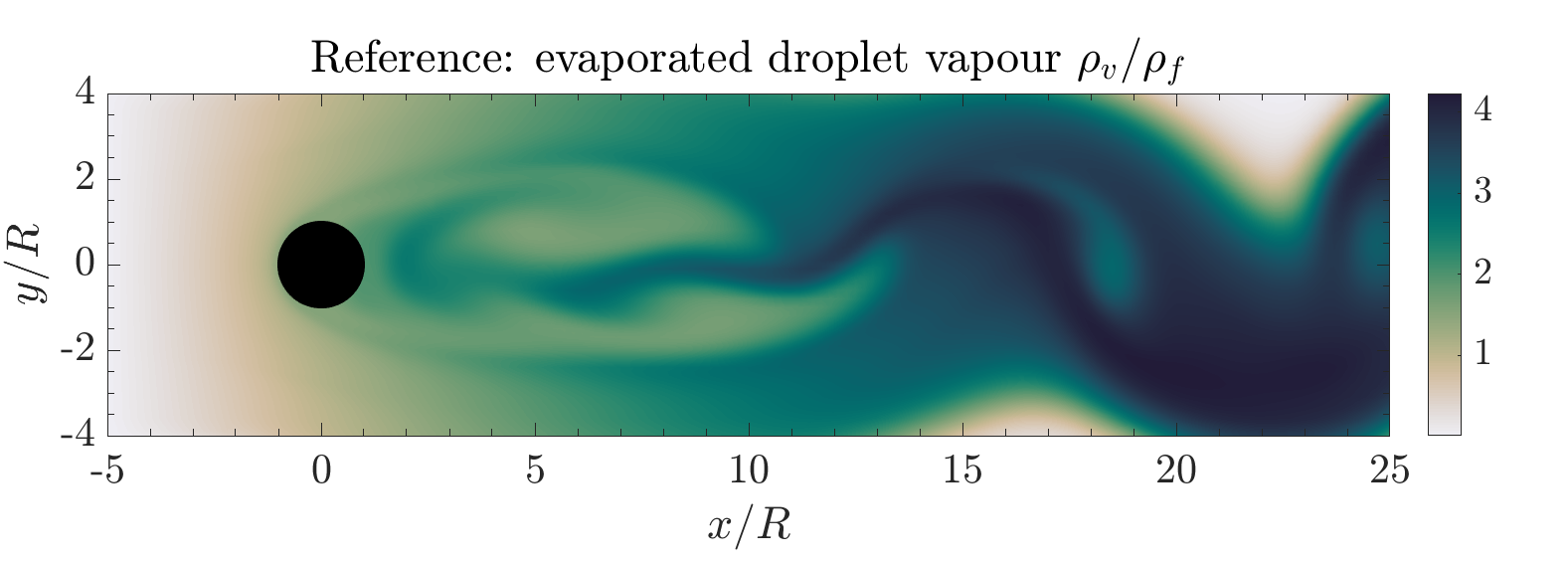}
		\caption{}
		\label{fig:fieldPlotCIC_evap_Re100_rhod}
	\end{subfigure}
	\caption{Distribution of the evaporated droplet vapour $\rho_v / \rho_f$ for $Re = 100$ and $St = 0.5$ with evaporation at time $t = 50$ for: \eqref{fig:fieldPlotFLA_evap_Re100_rhod} FLA solver; \eqref{fig:fieldPlotCIC_evap_Re100_rhod} reference PSI-CELL solver.}
	\label{fig:fieldPlot_evap_Re100_rhod}
\end{figure}

\begin{figure}[!ht]
	\includegraphics[width=\columnwidth,trim={0 0 0 0},clip] {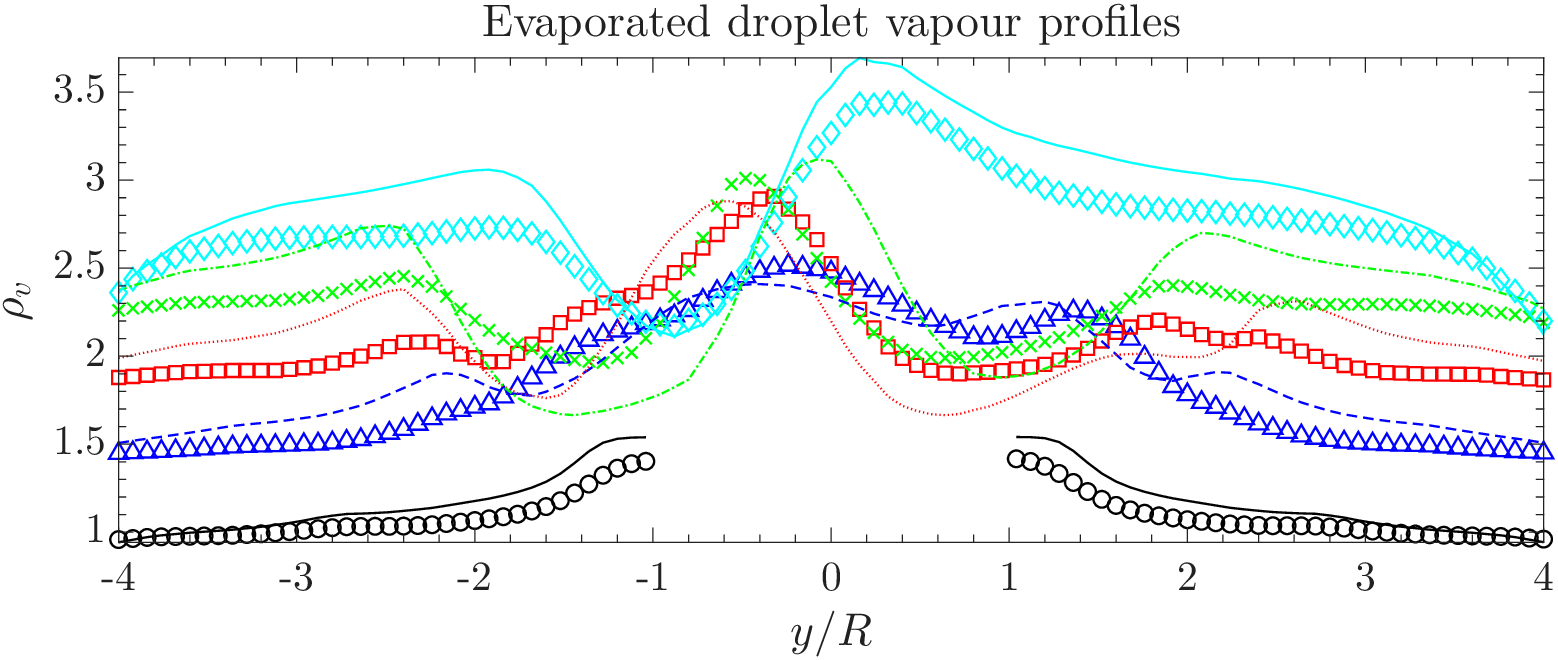}
	\caption{Evaporated droplet vapour $\rho_v$ profiles for the FLA solver (symbols) and the reference PSI-CELL solver (lines) at cross-sections
		{\color{black}$\boldsymbol{\bigcirc}$} ${x} / R = 0$, {\color{blue}$\boldsymbol{\bigtriangleup}$} ${x} / R = 3$,
		{\color{red}$\boldsymbol{\Box}$} ${x} / R = 6$,
		{\color{green}$\boldsymbol{\times}$} ${x} / R = 9$,
		{\color{cyan}$\boldsymbol{\diamond}$} ${x} / R = 12$
		for $Re = 100$ and $St = 0.5$ with evaporation at time $t = 50$.}
	\label{fig:plotprofCompFLACIC_evap_Re100_rhod}
\end{figure}

In terms of mass coupling, a corresponding analysis of the mass transfer source term is presented in Figure \ref{fig:fieldPlot_evap_Re100_mTrans}, where the associated spatial distributions are shown in Figure \eqref{fig:fieldPlotFLA_evap_Re100_mTrans} for the FLA solver and Figure \eqref{fig:fieldPlotCIC_evap_Re100_mTrans} for the reference PSI-CELL solver.
From these visualisations of the droplet field it is seen that again the FLA procedure provides a good qualitative transient reconstruction of the mass transfer. The notable exception to this is at the edge of the droplet field, where the finite size of the smoothing length associated with the kernel introduces an extended region in which mass transfer exists at a low magnitude in Figure \eqref{fig:fieldPlotFLA_evap_Re100_mTrans}. This is an artefact of the estimator in Eq.~\eqref{eq:mass-coupling_FLA} being an interpolation procedure, meaning that the droplet contributions are naturally distributed over a wider local region than in the corresponding reference PSI-CELL solver in Figure \eqref{fig:fieldPlotCIC_evap_Re100_mTrans}. Nonetheless, this is a recognised aspect of the kernel regression methodology \cite{ref:StaffordRybdylova2022}, and is acceptable since the additional regions of mass transfer are of such low magnitude that they do not influence the evolutionary behaviour of the transient flow. This is evidenced through the identical flow structures obtained in Figures \eqref{fig:fieldPlotFLA_evap_Re100_mTrans} and \eqref{fig:fieldPlotCIC_evap_Re100_mTrans}, demonstrating that such artefacts arising from the numerical methodology in the FLA solver do not have a significant effect. This is reinforced by the mass transfer profiles along the selected cross-sections displayed in Figure \eqref{fig:plotprofCompFLACIC_evap_Re100_mTrans}, which illustrates that across the majority of the flow domain the FLA produces a mass transfer field that is closely aligned with that from standard Lagrangian particle tracking simulations.
The evolution of the accumulated vapour distribution is shown in Figure \ref{fig:fieldPlot_evap_Re100_rhod}, with the result as obtained by the FLA solver depicted in Figure \eqref{fig:fieldPlotFLA_evap_Re100_rhod} and that from the reference PSI-CELL solver in Figure \eqref{fig:fieldPlotCIC_evap_Re100_rhod}.
As with the steady-state case in Figure \ref{fig:fieldPlot_evap_Re20_rhod}, since no evaporation has occurred across the injection inlet at $x/R = -5$, the vapour concentration is uniformly zero across this interval, and then increases in accordance with the evaporation as droplets advance along their trajectories through the flow field. As with the mass transfer source term, the vapour concentration distribution produced by the FLA is seen to offer a representation which accurately captures the transient structures of the flow field and spatial variation in the evaporation behaviour. Furthermore, for the given droplet loading, the presence of the vapour distribution downstream of the cylinder is highly significant relative to the carrier flow, meaning that that the effect of the interphase coupling is profound enough to alter the behaviour of the carrier flow. Selected $x$ cross-sectional profiles of the two approaches are contrasted in Figure \ref{fig:plotprofCompFLACIC_evap_Re100_rhod}, and it is seen that whilst the general trend of the vapour distribution is captured, some quantitative discrepancies do exist between the FLA and reference PSI-CELL solvers.
This is observed particularly at $x/R = 9$, and is again attributable to the kernel smoothing length reducing the fidelity of the reconstruction in this region of the flow field. Despite this, the transient evolution of the profiles is not affected, as evidenced by the furthest downstream profile at $x/R = 12$ being in good agreement with the reference PSI-CELL solver.

{\color{rev1}
The relative $l^2$-norm of the momentum and mass transfer source terms and evaporated droplet vapour for the different FLA droplet seedings and injection intervals is displayed in Table \ref{table:l2-norm_evap_Re100} for time $t = 50$.
\begin{table}[!ht]
	\begin{center}
		\begin{tabular}{cccc}
			\toprule
			& $NP = 50$ & $NP = 100$ & $NP = 200$ \\
			& $\Delta t_i = 0.2$ & $\Delta t_i = 0.1$ & $\Delta t_i = 0.05$ \\
			\midrule
			$\text{Error}_{l^2}[{S_{\text{Mom}}}_x]$ & 0.4817 & 0.3178 & 0.2627 \\ 
			$\text{Error}_{l^2}[{S_{\text{Mom}}}_y]$ & 0.4542 & 0.2363 & 0.2199 \\ 
			$\text{Error}_{l^2}[S_{\text{Mass}}]$ & 0.4284 & 0.2551 & 0.2406 \\
			$\text{Error}_{l^2}[\rho_v / \rho_f]$ & 0.2409 & 0.1374 & 0.1294 \\
			\bottomrule
		\end{tabular} 
	\end{center}
	\caption{{\color{rev1}Relative $l^2$-norm for the components of the momentum transfer source term $\bm{S}_{\text{Mom}}$, mass transfer source term $S_{\text{Mass}}$, and  evaporated droplet vapour $\rho_v / \rho_f$ evaluated using different droplet seedings for $Re = 100$ and $St = 0.5$ with evaporation at time $t = 50$.}}
	\label{table:l2-norm_evap_Re100}
\end{table}
As with the cases considered in Sections \ref{sec:steady-state-non-evap-flow} to \ref{sec:steady-state-evap-flow}, increasing the injection droplet seeding from $NP = 50$ to $NP = 100$ provides a notable reduction in error across all the coupling source term fields. However, unlike the non-evaporating unsteady case in Section \ref{sec:transient-non-evap-flow}, further increasing the injection droplet seeding to $NP = 200$ results in only a minor additional reduction in error. This suggests that for both steady-state and unsteady flow with evaporating droplets, an injection seeding of $NP = 100$ is the ideal balance between accuracy and efficiency for using the kernel regression procedure to reconstruct the coupling source terms.
}

As with the transient non-evaporating case in Section \ref{sec:transient-non-evap-flow}, the associated periodicity remains inherent in the carrier flow, and therefore it is appropriate to analyse this further. It is already evident from the spatial distribution of the momentum transfer in Figure \ref{fig:fieldPlot_evap_Re100_UTransY}, mass transfer in Figure \ref{fig:fieldPlot_evap_Re100_mTrans}, and droplet vapour in Figure \ref{fig:fieldPlot_evap_Re100_rhod} that the characteristic vortex street observed in the corresponding non-evaporating case in Figure \ref{fig:fieldPlot_nonevap_Re100_UTransY} is markedly altered. This is principally due to the higher droplet inertia used in this case with $St = 0.5$, and the effect this higher mass loading has on the mass and momentum transfer back to the carrier flow. Rather than seeing the development of a vortex street at $t = 50$, in this situation the increased mass loading in the droplet field as it parts to pass around the cylinder prevents the formation of vortices in the immediate trailing vicinity of the cylinder, with the carrier flow {\color{corr1}experiencing a delay in the onset of vortex shedding} instead. This is a result of the additional momentum transfer being sufficient enough to increase the effective viscosity of the carrier flow, meaning that the flow initially behaves as in the laminar regime. However, as the droplets evaporate whilst travelling across the domain, the mass loading decreases accordingly, meaning that the effective viscosity becomes lower, and the flow is able to transition to vortical structures after a certain distance. The evaporation ensures however that the droplet phase does not exist for long enough to allow the vortex street to become fully developed. The resulting spatial distribution is therefore a unique combination of the parameters chosen for the droplet inertia and evaporation model in this case, allowing a demonstration of the carrier flow modulation that such a configuration can produce.

\begin{figure}[!ht]
	\includegraphics[width=\columnwidth,trim={0 15 0 0},clip] {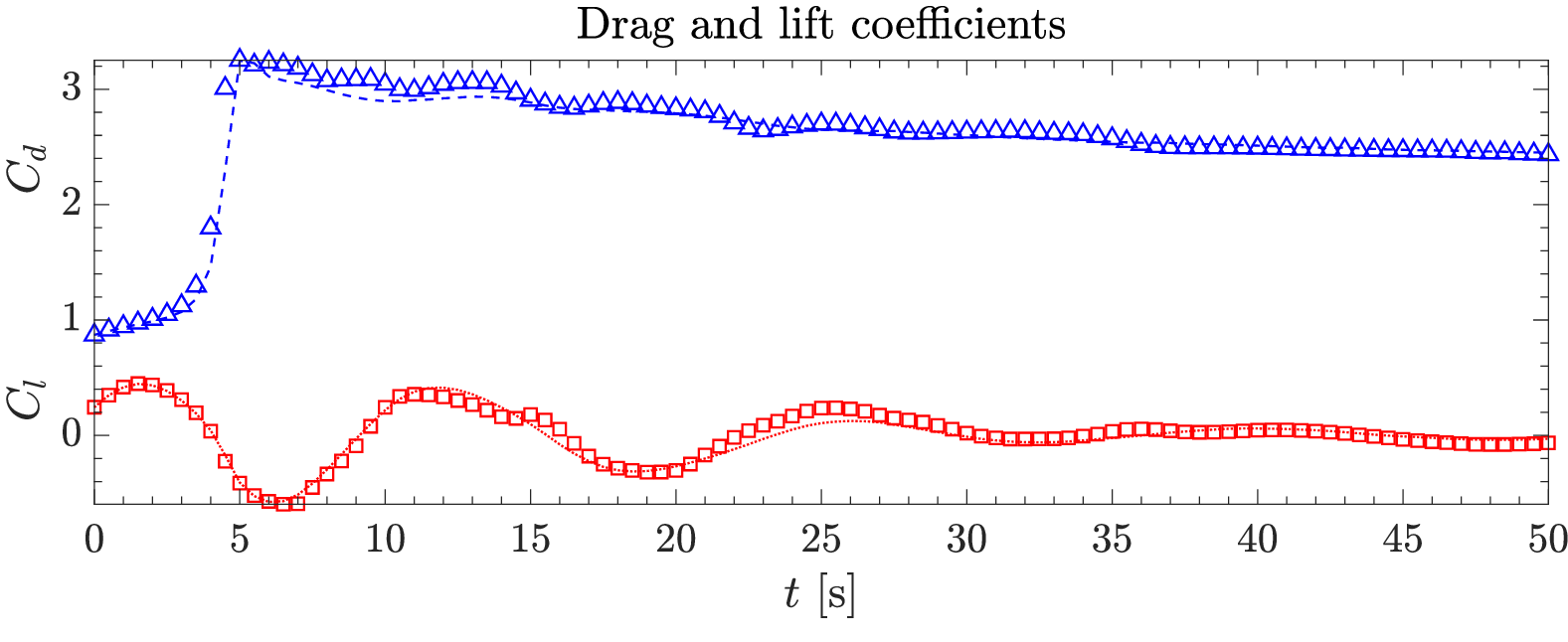}
	\caption{{\color{blue}$\boldsymbol{\bigtriangleup}$} Drag $C_d$ and {\color{red}$\boldsymbol{\Box}$} lift $C_l$ coefficients for the FLA solver (symbols) and the reference PSI-CELL solver (lines) for $Re = 100$ and $St = 0.5$ with evaporation.}
	\label{fig:plotforceCoeffsCompFLACIC_evap_Re100}
\end{figure}

Additional insight into this behaviour can be gained by considering the drag $C_d$ and lift $C_l$ coefficients defined in Eqs.~\eqref{eq:force-coefficients}. Calculation of these for the FLA and reference PSI-CELL simulations is illustrated in Figure \ref{fig:plotforceCoeffsCompFLACIC_evap_Re100}.
It is seen that initially the flow retains the periodic state it exhibits at the start of the droplet injection, which takes place into the fully developed vortex street. However, as time progresses, both the mean value of $C_d$ and the amplitude of $C_l$ decrease, reflecting the {\color{corr1}delay in the onset of vortex shedding experienced by} the flow. {\color{corr1}This can therefore be considered as the effect of the droplet phase being to dampen the oscillations of the carrier flow.} By the end of the simulation at $t = 50$, the amplitude of $C_l$ has reduced to zero, indicating the absence of any periodic structures in the flow at this point. Similarly, the mean value of $C_d$ has become steady and no longer fluctuates. 	 This behaviour conclusively demonstrates the effect of interphase coupling on a transient evaporating flow configuration, and the close agreement between the FLA and reference PSI-CELL data evidences the applicability of the FLA procedure at capturing the detail inherent in this behaviour with the reduced droplet seeding used.

The foregoing results illustrate the efficacy of the FLA when utilised as the core of a Lagrangian solver for dealing with the evolution of the droplet phase in simulations. Combined with the computational saving offered by the reduced droplet seeding that is afforded by the treatment of the droplet phase as a continuum, the procedure offers an attractive approach for investigating the transient behaviour of gas-droplet flows, with only a small degree of smoothing observed in the spatial distribution of the interphase coupling mass and momentum transfer fields when compared to conventional Lagrangian particle tracking simulations. This serves to highlight the potential of the FLA for investigation of more complex flow configurations involving physical boundaries, turbulence and multi-component droplets, opening up the possibility of advancing the modelling capabilities of simulations across a range of applications.

\section{Conclusions}
\label{sec:conclusions}

A novel procedure for two-way coupled evaporating spray simulations that utilises the fully Lagrangian approach in conjunction with a kernel regression procedure has been developed and implemented in OpenFOAM. The performance of the new model has been investigated on the benchmark configuration of a two-dimensional flow around a cylinder, and compared with a reference PSI-CELL solver that uses a standard Lagrangian particle tracking approach. The accuracy of the FLA methodology for predicting the transient flow behaviour is {\color{rev1}qualitatively} very good, and highlights its ability to resolve the distinct spatial structures that occur in the droplet phase, with the error that exists being attributable to the statistical nature of the kernel regression procedure. This level of detail is obtained {\color{rev1}notably} faster by the FLA when compared to the reference PSI-CELL solver for accumulating the source term contributions, with a computational speedup factor of around 100 times {\color{rev1}for the specific benchmark configuration tested with a droplet seeding at injection of $NP = 100$ and injection time interval of $\Delta t_i = 0.1$}. {\color{rev1}The FLA is the central factor to this computational efficiency advantage, both due to providing the number density along Lagrangian droplet trajectories, and also since it is incorporated into the kernel regression procedure so that the kernel size is adaptively scaled during the coupling source term reconstruction.} This establishes the FLA as a promising means of incorporating interphase coupling within gas-droplet flow simulations.

In the proposed procedure for the interphase coupling source term reconstruction, the FLA number density acts as a {\color{corr1}scaling factor} to the seed droplet contributions, and incorporates both the constant mass scaling coefficient needed to make simulations comparable to reference PSI-CELL simulations with a higher droplet seeding (\ref{sec:initial-mass-loading}), and also the adaptive volume scaling factor that represents the local droplet concentration on non-uniform computational meshes as the flow evolves (\ref{sec:cell-volume-scaling}). The suggestion of using variable statistical weights in previous work \cite{Garg2009} is remarkably similar in nature to the role played by the FLA number density within the proposed accumulation procedure in the present work. The advantage of the FLA-based formulation in this respect is that it uses an evolution equation which is physically consistent with the description of the droplet phase as a continuum to accurately describe the number density across the flow field, rather than an empirical measure which is unable to reflect more extreme variations in number density.

It is recognised that the interpretation of the momentum source term within the fluid phase momentum equation constitutes a significant problem in the context of a dispersed phase solver which uses Lagrangian particle tracking with the point-force particle assumption, whereby this term should be zero everywhere except for a delta function at the location of each droplet \cite{Eaton2009}. To that end, representation of the droplet phase as a continuum goes some way towards mitigating this problem, and the FLA provides the ideal means through which this can be achieved.

It has been seen that the more involved scenario of transient evaporating flows is handled well by the FLA, notably recovering the coupling-induced {\color{corr1}delay in the onset of vortex shedding that occurs in the carrier flow}. That this effect is not a numerical artefact induced by the kernel regression treatment applied to the FLA is seen in the reference PSI-CELL simulations, which display the same behaviour. This demonstrates that the FLA is able to consistently reproduce a high level of qualitative agreement in capturing the transient flow physics despite the reduced droplet seeding, and highlights the efficacy of this method when combined with kernel regression at obtaining a high level of detail from simulations within a notably reduced timeframe.

{\color{rev1}
Looking forward, this work demonstrates the potential for using the FLA as the core of a numerical solver for dispersed multiphase flows, and the accompanying gain in computational efficiency which can be achieved. This would be of particular utility in the treatment of sprays, which are characterised by polydisperse evaporating droplets along with the additional complexity of turbulence in the carrier flow. Such developments would require further fundamental modelling using the FLA framework, in order to account for effects such as breakup, agglomeration and segregation within sprays.

Generalising the coupling term reconstruction procedure to polydisperse droplet flows requires the additional consideration of accumulating contributions across the droplet size distribution. This framework has been developed by applying the kernel regression procedure to the gFLA \cite{ref:StaffordRybdylova2022}, and requires a number of additional Eulerian fields to store fixed droplet size classes in simulations. Implementation of this procedure in a computationally efficient manner presents the main challenge involved with this.

The FLA is straightforward to extend to more complex evaporation models which depend on local gas-phase conditions, with corresponding Jacobian governing equations being derivable. Due to the differing evaporation rates of droplets in such a model, this naturally only pertains to a polydisperse droplet suspension, and therefore the kernel regression procedure must be applied to the gFLA framework to properly capture such evaporation behaviour.
}

\section*{Acknowledgements}
The authors are grateful to the UKRI (grant MR/T043326/1) for their financial support and to the University of Brighton for the access to the university’s High Performance Computing Cluster.

For the purpose of open access, the authors have applied a Creative Commons Attribution (CC-BY) license to any Author Accepted Manuscript version arising from this submission.

\section*{Conflict of Interest}

The authors have no conflicts to disclose.

\section*{CRediT Authorship Contribution Statement}

\textbf{C. P. Stafford}: Conceptualization (equal); Data curation; Formal analysis (equal); Investigation (equal); Methodology (equal); Software (lead); Writing – original draft. \textbf{O. Rybdylova}: Conceptualization (equal); Formal analysis (equal); Funding acquisition; Investigation (equal); Methodology (equal); Software (supporting); Supervision; Project administration; Writing – review \& editing.

\section*{Data Availability}

The data that support the findings of this study are available from the corresponding author upon reasonable request.

\appendix
\section{FLA number density specification}
\label{sec:initial-number-density}

\subsection{Initial mass loading}
\label{sec:initial-mass-loading}

To ensure that the reduced droplet seeding used by the FLA is representative of the seeding used in the reference PSI-CELL solver, the procedure for two-way coupling using the FLA must be implemented so as to respect the difference in mass loading between these two cases. In a PSI-CELL-based simulation, the number of droplets injected must be calculated to directly satisfy the desired mass loading. In contrast, in an FLA-based simulation the user has the licence to choose the number of droplets to inject for the reduced droplet seeding. The scaling factor $M_{\text{scal}}$ between the mass loading of the reduced FLA seeding and the true mass loading used in the PSI-CELL solver is then used to determine the initial FLA number density $n_d^0 = n_d (\bm{x}_d,0)$ defined in Eq.~\eqref{eq:FLA-COM}.

Specification of $M_{\text{scal}}$ requires knowledge of the injection rates for both the FLA seed parcels and droplets in the associated PSI-CELL simulation. The mass loading of the droplet phase at the point of injection is determined as
\begin{equation} \label{eq:mass-flow}
	\dot{M} = \text{Mass introduced per injection } \times \text{ Injection rate} \, ,
\end{equation}
where $\dot{M}$ is the mass flow rate specified at the droplet inlet. The mass introduced per injection is simply
\begin{equation} \label{eq:mass}
	\text{Mass introduced} = \text{Number of droplets } \times \text{ mass per droplet} \, ,
\end{equation}
whilst the injection rate is specified as
\begin{equation} \label{eq:injection-rate}
	\text{Injection rate} = \frac{1}{\text{Injection interval}} \, ,
\end{equation}
where the injection interval is the time period between droplet injections into the domain.

Combining Eqs.~\eqref{eq:mass-flow}, \eqref{eq:mass} and \eqref{eq:injection-rate}, then for the reference PSI-CELL solver, we have
\begin{equation} \label{eq:mass-flow-ref}
	\dot{M}^{\text{ref}} = \frac{N^{\text{ref}} \, m}{\tau_{\text{inj}}^{\text{ref}}} \, ,
\end{equation}
where $N^{\text{ref}}$ is the number of droplets injected into the domain for the PSI-CELL solver, $m$ is the droplet mass, and $\tau_{\text{inj}}^{\text{ref}}$ is the time interval between injections for the PSI-CELL solver. Correspondingly, for the FLA solver, we have
\begin{equation} \label{eq:mass-flow-FLA}
	\dot{M}^{\text{FLA}} = \frac{N^{\text{FLA}} \, m}{\tau_{\text{inj}}^{\text{FLA}}} \, ,
\end{equation}
where $N^{\text{FLA}}$ is the number of droplets injected into the domain for the FLA solver, and $\tau_{\text{inj}}^{\text{FLA}}$ is the time interval between injections for the FLA solver. The mass scaling factor $M_{\text{scal}}$ can then be determined as the ratio of Eq.~\eqref{eq:mass-flow-ref} to Eq.~\eqref{eq:mass-flow-FLA}, yielding
\begin{equation} \label{eq:mass-load-scaling}
	M_{\text{scal}} = \frac{N^{\text{ref}} \, \tau_{\text{inj}}^{\text{FLA}}}{N^{\text{FLA}} \, \tau_{\text{inj}}^{\text{ref}}} \, .
\end{equation}
Thus it is seen from Eq.~\eqref{eq:mass-load-scaling} that the droplet mass $m$ cancels since both the FLA and PSI-CELL solvers
use identical droplets in corresponding simulations, with the number of droplets and injection interval used for each solver then determining the ratio between the respective mass loadings of the parcels used to represent the droplet phase in each case.

\subsection{Cell volume scaling}
\label{sec:cell-volume-scaling}

Since the FLA number density $n_d$ is a measure of the number of droplets per unit volume, the correct specification of the reference volume associated with each droplet is essential to being able to interpret the true mass loading associated with an FLA-based simulation. Then to use the FLA for reconstruction of two-way coupling source terms on a computational mesh, the FLA number density $n_d$ at a given time along a droplet trajectory must be renormalised with respect to the volume of the cell in which that droplet currently resides. This is necessary, as even though the kernel based interpolation procedure used for accumulating the source terms is meshfree \cite{ref:StaffordRybdylova2022}, it must be reinterpreted on the computational mesh in order to be coupled with the Eulerian solver for the carrier flow. To that end, it is sufficient to multiply $n_d$ by the volume scaling factor $\mathcal{V}_{\text{scal}}$ defined by
\begin{equation} \label{eq:vscal-def}
	\mathcal{V}_{\text{scal}} = \frac{\mathcal{V}_{\text{cell}}}{\mathcal{V}^{\text{FLA}}} \, ,
\end{equation}
where $\mathcal{V}_{\text{cell}}$ is the volume of the cell in which a droplet resides at a given time, and $\mathcal{V}^{\text{FLA}}$ is the reference volume associated with an individual droplet that has an initial FLA number density $n_d^0 = n_d (\bm{x}_d,0)$ prescribed at the point of droplet injection. The quantity $\mathcal{V}_{\text{cell}}$ is obtained from the simulation at runtime as a property of the computational mesh, whilst $\mathcal{V}^{\text{FLA}}$ can be found from the specifics of the droplet injection. Consider the definition of $n_d^0$ as
\begin{equation} \label{eq:nd0}
	n_d^0 = \frac{N^{\text{FLA}}}{\mathcal{V}_{\text{inj}}} \, .
\end{equation}
where $\mathcal{V}_{\text{inj}}$ is the volume associated with a given injection of droplets. This is applicable for a general injection in which $N^{\text{FLA}}$ droplets are injected into the domain through a face with cross-sectional area $\mathcal{A}_{\text{inj}}$. Then the reference volume $\mathcal{V}^{\text{FLA}}$ associated with a droplet with initial number density $n_d^0$ is given by the reciprocal of Eq.~\eqref{eq:nd0},
\begin{equation} \label{eq:vfla}
	\mathcal{V}^{\text{FLA}} = \frac{\mathcal{V}_{\text{inj}}}{N^{\text{FLA}}} \, .
\end{equation}
The injection volume $\mathcal{V}_{\text{inj}}$ can be defined as
\begin{equation} \label{eq:vinj}
	\mathcal{V}_{\text{inj}} = \mathcal{A}_{\text{inj}} \mathcal{D}_{\text{inj}} \, ,
\end{equation}
where $\mathcal{D}_{\text{inj}}$ is the average depth into the domain the droplets travel during the time interval between injections $\tau_{\text{inj}}^{\text{FLA}}$. This can be specified using the average droplet injection speed $v_{\text{inj}}$ into the domain as a reference velocity scale, giving
\begin{equation} \label{eq:dinj}
	\mathcal{D}_{\text{inj}} = v_{\text{inj}} \tau_{\text{inj}}^{\text{FLA}}
\end{equation}
which can be interpreted as the characteristic lengthscale of the droplet injections. Combining Eqs.~\eqref{eq:vscal-def}, \eqref{eq:vfla}, \eqref{eq:vinj}, and \eqref{eq:dinj} gives the required volume scaling factor as
\begin{equation} \label{eq:vscal}
	\mathcal{V}_{\text{scal}} = \frac{\mathcal{V}_{\text{cell}} N^{\text{FLA}}}{\mathcal{A}_{\text{inj}} v_{\text{inj}} \tau_{\text{inj}}^{\text{FLA}}}
\end{equation}
Eq.~\eqref{eq:vscal} provides the quantification needed to convert the FLA number density from a carried variable along droplet trajectories to produce a consistent interpretation on the computational mesh for an arbitrary grid cell of volume $\mathcal{V}_{\text{cell}}$.

With this interpretation, the reference volume $\mathcal{V}^{\text{FLA}}$ associated with each droplet must geometrically tile the cross-sectional area $\mathcal{A}_{\text{inj}}$ through which droplets are injected in order for the FLA to accurately represent the associated continuum description of the droplet field. This necessitates that $\mathcal{A}_{\text{inj}}$ is effectively discretised into uniform {\color{corr1}cell face sub-areas corresponding to} volume $\mathcal{V}^{\text{FLA}}$, with the droplets then injected at the cell centre of each.

\subsection{Combining the mass and volume scaling factors}

The overall scaling factor needed to specify the FLA number density on the mesh at a given time in the simulation is found by multiplying the mass scaling factor \eqref{eq:mass-load-scaling} and cell volume scaling factor \eqref{eq:vscal} together, yielding the non-dimensional number density
\begin{equation} \label{eq:mv-scal}
	\hat{n}_d = \frac{\mathcal{V}_{\text{cell}} N^{\text{ref}}}{\mathcal{A}_{\text{inj}} v_{\text{inj}} \tau_{\text{inj}}^{\text{ref}}}
\end{equation}
It is notable that the FLA parameters $N^{\text{FLA}}$ and $\tau_{\text{inj}}^{\text{FLA}}$ cancel out in the combined expression, leaving only the dependence on the reference parameters from the PSI-CELL simulation. This is consistent with the mass loading of the system under consideration in reality, since the PSI-CELL simulation contains all the droplets needed to specify the desired mass flow rate, whilst the continuum description offered by the FLA only uses a fraction of this number of parcels to represent the droplet phase.

Whilst the average droplet injection speed $v_{\text{inj}}$ is most easily assigned as a uniform value across the entire droplet field for a given injection, it is still applicable for a distribution of droplet velocities with a {\color{corr1}global mean velocity} of $v_{\text{inj}}$. In this case the injection volume $\mathcal{V}_{\text{inj}}$ associated with the droplet field for the injection remains the same, despite the actual depth travelled into the domain during the injection interval $\tau_{\text{inj}}^{\text{FLA}}$ being different for every droplet. In a similar manner, the geometric tiling of the cross-sectional area $\mathcal{A}_{\text{inj}}$ by the individual droplet reference volumes $\mathcal{V}^{\text{FLA}}$ is most easily realised by discretising $\mathcal{A}_{\text{inj}}$ into a uniform grid to generate a tessellation of all the $\mathcal{V}^{\text{FLA}}$ over the injection volume $\mathcal{V}_{\text{inj}}$. The methodology is not however limited to this scenario, with a random spatial distribution of the same number of $N^{\text{FLA}}$ droplets across $\mathcal{A}_{\text{inj}}$ being permissible, since the reference volumes $\mathcal{V}^{\text{FLA}}$ still effectively tile $\mathcal{V}_{\text{inj}}$ and therefore provide a continuum representation of the droplet field. This can be interpreted as enforcing that all FLA seed droplets initially represent the same proportion of the continuum which they collectively model.

The remaining dependence of Eq.~\eqref{eq:mv-scal} on the reference PSI-CELL simulation parameters is through the ratio $N^{\text{ref}} / \tau_{\text{inj}}^{\text{ref}}$, which is just the number of droplets injected per second. As an input parameter it can be more physically relevant to specify this as a mass flow rate, which for the injection of monodisperse droplets under consideration is given by Eq.~\eqref{eq:mass-flow-ref}. Then Eq.~\eqref{eq:mv-scal} can be written as
\begin{equation} \label{eq:mv-scal-massflow}
	\hat{n}_d = \frac{\mathcal{V}_{\text{cell}} \dot{M}^{\text{ref}}}{\mathcal{A}_{\text{inj}} v_{\text{inj}} m}
\end{equation}
in which the sole input parameter needed from the reference PSI-CELL simulation is the mass flow rate.

Within the developed FLA-based solver, Eq.~\eqref{eq:mv-scal} is applied to the coupling terms $S_{\text{Mass}}^{\text{FLA}}$ and $\bm{S}_{\text{Mom}}^{\text{FLA}}$ as they are reconstructed using Eqs.~\eqref{eq:coupling-terms-kernel} by including the additional multiplicative factor $\hat{n}_d$ in the calculation.

\section{Solution procedure for the interphase coupling terms} \label{sec:solution-coupling}

The solution method for the reconstruction of the interphase mass and momentum transfer source terms at a given time step during the simulation can be summarised as follows:

\begin{enumerate}
	\item \textbf{Parcel level:} loop through all droplets and evaluation of individual droplet contributions.
	\begin{enumerate}
		\item Calculation of change in mass ($\dot{m}_d^i$), and momentum ($\dot{m}_d^i \bm{v}_d^i$ {\color{corr1}and} $\bm{F}_d^i$) {\color{corr1}contributions} from droplet $i$ at each substep of $\Delta t_d$ used for time integration of the Lagrangian droplet trajectories.
		\item Summation of mass {\color{corr1}and momentum} contributions for droplet $i$ over all droplet substeps within the carrier flow timestep $\Delta t$.
		\item Convolve mass, momentum and force contributions from droplet $i$ with the kernel \eqref{eq:kernel} and FLA number density $n_d^i$, and append these terms to the numerator of the kernel estimators in Eqs.~\eqref{eq:mass-coupling_FLA}, \eqref{eq:momentum-coupling_FLA}.
		\item Append the weight provided by the kernel  \eqref{eq:kernel} to the denominator of the kernel estimators in Eqs.~\eqref{eq:mass-coupling_FLA}, \eqref{eq:momentum-coupling_FLA}.
	\end{enumerate}
	\item \textbf{Cloud level:} accumulation of all droplet contributions onto the computational mesh.
	\begin{enumerate}
		\item Evaluate the kernel estimators by performing the division in Eqs.~\eqref{eq:mass-coupling_FLA}, \eqref{eq:momentum-coupling_FLA} to obtain the mass ($S_{\text{Mass}}^{\text{FLA}}$) and momentum ($\bm{S}_{\text{Mom}}^{\text{FLA}}$) coupling source terms.
		\item Compute the FLA scaling coefficient $\hat{n}_d$ using Eq.~\eqref{eq:mv-scal}.
		\item Multiply the reconstructed coupling source terms $S_{\text{Mass}}^{\text{FLA}}$ and $\bm{S}_{\text{Mom}}^{\text{FLA}}$ by $\hat{n}_d$.
	\end{enumerate}
	\item \textbf{Carrier flow solver:} append the interphase coupling contributions from the droplet cloud to the fluid transport equations.
	\begin{enumerate}
		\item Add the momentum source term $\bm{S}_{\text{Mom}}^{\text{FLA}}$ to the momentum transport equation \eqref{eq:fluid-momentum}.
		\item Add the mass source term $S_{\text{Mass}}^{\text{FLA}}$ to the scalar transport equation \eqref{eq:vapour-transport} for the droplet vapour concentration.
	\end{enumerate}
\end{enumerate}

\bibliographystyle{elsarticle-num} 
\bibliography{bibfile}

%
%

\end{document}